%% file: main.tex
\documentclass{article}
\usepackage[utf8]{inputenc}
\usepackage{algorithm}
\usepackage{algpseudocode}
\usepackage{subcaption}
\usepackage{multirow}
\usepackage{lscape} 
\usepackage{amsmath}
\usepackage{url}
\usepackage{balance}
\usepackage{paralist}
\usepackage{graphicx}
\usepackage{amsmath,amssymb}
\usepackage{rotating}

\providecommand{\keywords}[1]
{
  \small	
  \textbf{\textit{Keywords---}} #1
}

\author{Gorka Abad$^{1,2}$, Stjepan Picek$^{1}$, \\ Víctor Julio Ramírez-Durán$^{2}$, Aitor Urbieta$^{2}$   \\
        \small $^{1}$Radboud University, Nijmegen, The Netherlands. \\
        \small $^{2}$Ikerlan Technology Research Centre, Arrasate-Mondragón, Spain. \\
}

\begin{document}

\title{On the Security \& Privacy in Federated Learning}

\maketitle

\input{00-Abstract}


\input{01-Introduction}

\input{02-MachineLearning}

\input{03-FederatedLearning}

\input{04-ThreatModel}

\input{05-TrainingInAdversarialSettings}

\input{06-InferringInAdversarialSettings}

\input{07-TowardsSecureFL}

\input{08-Challenges}

\input{09-RelatedWork}

\input{10-Conclusions}

\section*{Acknowledgements}

The European commission financially supported this work through Horizon Europe program under the IDUNN project (grant agreement number 101021911). It was also partially supported by the Ayudas Cervera para Centros Tecnológicos grant of the Spanish Centre for the Development of Industrial Technology (CDTI) under the project EGIDA (CER-20191012), and by the Basque Country Government under the ELKARTEK program, project TRUSTIND - Creating Trust in the Industrial Digital Transformation (KK2020/00054).

\balance

\bibliographystyle{plain}
\bibliography{bibliography.bib}

\end{document}

%% file: 00-Abstract.tex
\begin{abstract}

Recent privacy awareness initiatives such as the EU General Data Protection Regulation subdued Machine Learning (ML) to privacy and security assessments. Federated Learning (FL) grants a privacy-driven, decentralized training scheme that improves ML models' security. The industry's fast-growing adaptation and security evaluations of FL technology exposed various vulnerabilities that threaten FL's confidentiality, integrity, or availability (CIA). This work assesses the CIA of FL by reviewing the state-of-the-art (SoTA) and creating a threat model that embraces the attack's surface, adversarial actors, capabilities, and goals. We propose the first unifying taxonomy for attacks and defenses and provide promising future research directions.

\end{abstract}

%% file: 01-Introduction.tex
\section{Introduction}

Recent improvements and research in Artificial Intelligence (AI) and Machine Learning (ML) opened many applications and opportunities~\cite{zhou2017machine}. More concretely, ML is applied to different tasks, from medical use~\cite{fatima2017survey} to autonomous driving~\cite{grigorescu2020survey} and cybersecurity~\cite{ucci2019survey}. What is more, ML can be used in both offensive and defensive tasks. For example, ML has a defensive role, determining if the network traffic is legitimate or a file contains malware. However, ML has also been applied in offensive tasks, e.g., generating malicious content to bypass such defenses~\cite{anderson2017evading}. 
In the same way, the usage of ML models in some critical scenarios, i.e., autonomous driving, requires safety measures for preventing accidents involving human lives~\cite{shafaei2018uncertainty}. Related to this, in the last three years, AI sector leaders, e.g., Google, Amazon, Microsoft, and Tesla, received attacks on their ML algorithms~\cite{kumar2020adversarial}, which are publicly available as Machine Learning as a Service for little cost.
As such, to improve the security and privacy of ML becomes an urgent and crucial task.

To enhance security and privacy, in 2016, Google presented a novel privacy-driven, decentralized ML scheme that enables training many devices towards a single-shared model without exposing the training data, named Federated Learning (FL)~\cite{mcmahan2017communicationefficient}. Similar to ML, FL has grown in popularity due to its broad applicability~\cite{niknam2020federated, li2019survey}. Since FL is privacy aimed~\cite{mcmahan2017communicationefficient}, legal requirements as the EU General Data Protection Regulation (GDPR)-- European Parliament and Council Regulation No 2016/679 boosted FL's progress and usage~\cite{yang2019federated}. The wide range in applicability and popularity placed FL in the scope of security and privacy assessments~\cite{li2020federated}. Recent privacy assessments concluded that privacy leaks occur in different segments of the FL procedure by various means~\cite{mothukuri2021survey}. 
Different research areas (cryptography, ML, distributed technologies) have already analyzed FL's security and privacy under their scope~\cite{lyu2020threats}. However, since FL's introduction in 2016, to the best of our knowledge, none have developed a unifying taxonomy for attacks and defenses that helps FL with its security and privacy assessment. Some work has been done respectively in specific areas of FL, e.g., privacy~\cite{yin2021comprehensive}, security~\cite{jere2020taxonomy}, edge computing~\cite{lim2020federated}, and suggesting open problems~\cite{kairouz2019advances}.

The rapid growth in proposals of attacks and defenses complicates determining the current state of FL security and privacy~\cite{kairouz2019advances}.
This paper systematizes the current state of security and privacy in FL by developing two unifying taxonomies. First, we analyze the state-of-the-art (SoTA), which consolidates a threat model. We further develop two holistic taxonomies for attacks and defenses by applying them to the reviewed papers. By this means, the proposed taxonomies solve the lack of consensus in attacks and defenses derived from the extensive attack surface and fast evolution of the technology. Therefore, our contributions are threefold:

\begin{compactitem}
   \item We create a threat model describing all the assets that pose a risk for FL. It identifies and makes understandable the FL's threats while considering its context.
   \item By applying the threat model, we develop two unifying novel taxonomies for attacks or defenses, which joins different lines of research regarding FL's security and privacy. The proposed taxonomies consider a wide range of parameters that unequivocally describe an attack or a defense.
   \item According to the proposed holistic taxonomies, we investigate the SoTA attacks and defenses and classify them accordingly. From there, we extract the relationships between attacks and defenses, metrics used for evaluation, and standard experimental setups.
\end{compactitem}

%% file: 02-MachineLearning.tex
\section{Machine Learning}
\label{sec:machine learning}

This section briefly explains the centralized ML procedure; a dataset is collected from various sources and kept in a single place for the training process. In particular, we describe the phases involving the ML procedure.

\subsection{Types of Learning}

ML (see Fig.~\ref{fig:ML}) is the automated process of accurately making predictions based on the existing data. ML algorithms adapt their structure to fit the provided data to maximize the outcome via an iterative process known as training~\cite{murphy2012machine}. 
Learning can be supervised, unsupervised, semi-supervised, and reinforcement~\cite{murphy2012machine, goodfellow2016deep}. 

We mainly focus on supervised learning for the rest of the paper.
In supervised learning, the goal is to perform classification (map an input to a label) or regression (map an input to a continuous output). The ML algorithm uses a dataset, a collection of labeled examples $\{(\textbf{x}_i,y_i)\}^\mathcal{N}_{i=1}$, where $\mathcal{N}$ is the size of the collection. Each sample $\textbf{x}_i$, named a feature vector, contains a label $y_i$ of class $c_n:\mathcal{C}=\{c_1,c_2,c_3,...,c_n\}$. 
Each dimension of the feature vector $\textbf{x}^{(j)}_i$ ($j=1,2,3,...$) describes (in a specific way) the sample. For instance, a feature $\textbf{x}^{(1)}_i$ may describe the height of a person and $\textbf{x}^{(2)}_i$ its weight. The goal of supervised learning is to create a model that, with an unseen feature vector $\textbf{x}$ as an input, can infer its label $y$.

\subsection{Machine Learning Phases}

The ML procedure is divided into two phases. The training phase describes where the model uses the dataset for learning. 
A model $f$ takes a feature vector~$\textbf{x}:f(\textbf{x})$ as input for describing its label $y \leftarrow f(\textbf{x})$. During the learning process, the model looks for an optimal vector of weights $\theta \in \Theta:f_\Theta$. $f_\Theta$ is set by an iterative learning process, aiming to minimize a loss function $\mathcal{L}(\mathbf{x}):\operatorname*{argmin}\limits_\Theta \mathcal{L}(\mathbf{x})$ while a disjoint test set is used for measuring the quality of the model.

The model is deployed after achieving convergence at the inference (test) phase, ready for performing a given task. A trained model $f_\Theta$ performs inference over a feature vector $y \leftarrow f_\Theta(\textbf{x})$ as the input.

%% file: 03-FederatedLearning.tex
\subsection{Federated Learning}
\label{sec:federated learning}

\begin{figure}[!htb]
    \centering
    \begin{subfigure}[b]{0.45\textwidth}
        \centering
        \includegraphics[width=\columnwidth]{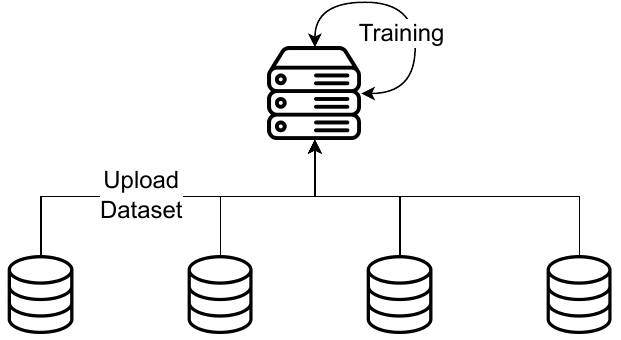}
        \caption{Centralized Machine Learning procedure.}
        \label{fig:ML}
    \end{subfigure}
    \hfill
    \begin{subfigure}[b]{0.3\textwidth}
        \centering
        \includegraphics[width=\columnwidth]{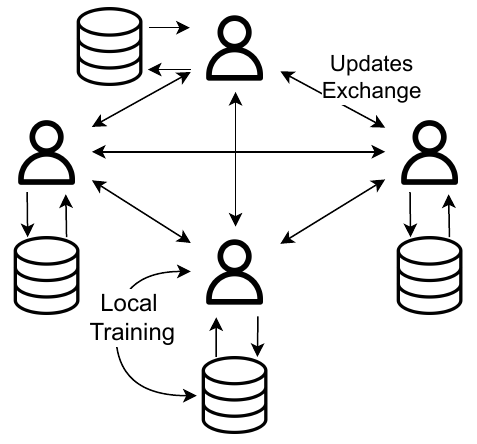}
        \caption{Peer to Peer FL procedure.}
        \label{fig:P2P}
    \end{subfigure}
    \hfill
    \begin{subfigure}[b]{0.45\textwidth}
        \centering
        \includegraphics[width=\columnwidth]{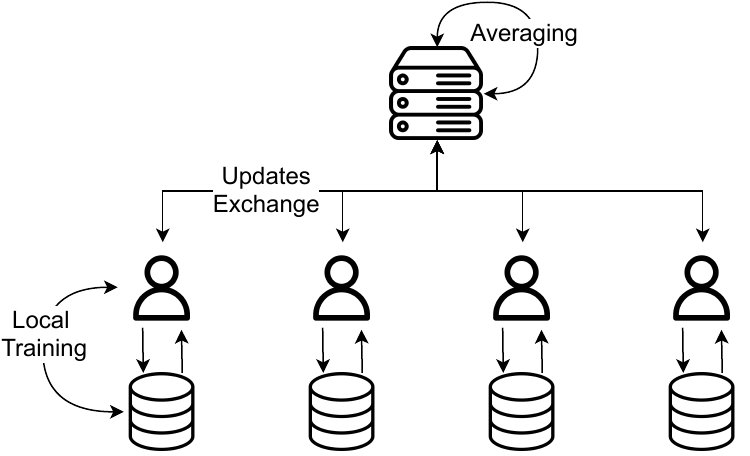}
        \caption{Federated Learning procedure.}
        \label{fig:federatedlearning}
    \end{subfigure}
    \caption{ML, FL, and P2P procedures.}
    \label{fig:MLandFL}
\end{figure}

FL is a collaborative, decentralized scheme for training privacy-driven ML models. The FL network is composed of: the clients, the aggregator, and their communications (see Fig.~\ref{fig:federatedlearning}). The clients, upon consensus, train the same ML algorithm. The training parameters, e.g., batch size and learning rate, are also chosen by agreement. Each client owns a disjoint dataset to locally train the model, which is afterward uploaded to the aggregator (the vector of weights representing the model). Later, the aggregator, following an aggregation algorithm, i.e., Federated Averaging (see Section~\ref{sec:FedAvg}), joins each client's model. After aggregation, the vector of weights is sent back to each client for another training round until convergence. 

This client-aggregator distribution is known as client-server or decentralized architecture. By removing the aggregator and performing the aggregation process among the clients in the network, a fully decentralized architecture is obtained (see Fig.~\ref{fig:P2P}), named peer-to-peer (P2P) FL or fully decentralized FL~\cite{rodriguez2020dynamic, vanhaesebrouck2017decentralized}.

\subsubsection{Federated Averaging}
\label{sec:FedAvg}

Stochastic Gradient Descent (SGD) is the most common optimization method for minimizing a loss function in ML~\cite{kiefer1952stochastic}. Since its release, the ML community has driven its research towards improving SGD~\cite{goodfellow2016deep}.
Federated SGD (FedSGD) is considered the baseline aggregation algorithm for developing an improved version of SGD~\cite{sattler2020}. It functions as follows: for each client, $k \in \mathcal{K}$ where $|\mathcal{K}| > 0$, it performs a local training procedure. The batch size $\beta$ is the entire dataset size, i.e., $\beta=\infty$. Thus, for each training round, each client $k$ sends the trained weights $\Theta$ to the server. Then, the server will, by averaging, aggregate those and send the newly joined model to each client $k$. 

Federated Averaging (FedAvg) improves the training procedure and leverages the decentralized scheme of FL. It is based on mini-batch training where $\beta$ is smaller than the dataset, i.e., $\beta \neq \infty$ for a single epoch, improving the training speed. For each training epoch with $\beta$ as batch size, the weights $\Theta$ are sent to the server. Then the server aggregates the updates and sends those back to each client $k$ for another epoch. FedAvg improves the generalization and accuracy of the overall model. However, it incurs noticeable communication overhead~\cite{mcmahan2017communicationefficient}. 

\subsection{Datasets \& Data Distribution}

Since datasets are owned by each client, their heterogeneity and data distribution directly influence the local model accuracy. Therefore, FL leverages the heterogeneity of clients' datasets to improve the global model quality. The data distribution where each client has different dataset lengths and data types is known as non-Independent and Identically Distributed (non-IID). Unlike non-IID, in Independent and Identically Distributed Data (IID), each client has the same dataset length without overlapping. For achieving a non-IID distribution from IID data, the Dirichlet distribution~\cite{minka2000estimating} is usually used. The data distribution directly affects the attacks and defenses performance, as demonstrated in~\cite{awan2021contra}. In FL real-world scenarios, data distribution will commonly follow a non-IID distribution~\cite{mcmahan2017communicationefficient, kairouz2019advances}. 

Additionally, data is partitioned into Vertical FL (VFL), Horizontal FL (HFL), Federated Transfer Learning (FTL), and hybrid FL. HFL (see Table~\ref{tab:HFL}) is a data distribution that shares the same features related to unique identities. The original use of Google's GBoard used HFL since clients make similar use of the keyboards, from which it extracted the same features for distinct users~\cite{hard2018federated}. This approach takes advantage of either the extracted features without identifying clients. As an example, an application collects usage data from clients worldwide. The collected data is transverse, but the clients' identity remains unknown. Datasets are horizontally partitioned when: let $i,j$ be two clients that share the same features $\mathcal{F}$ for a feature space $\mathcal{X}: \mathcal{X}^i_\mathcal{F} = \mathcal{X}^j_\mathcal{F}$. However, they do not share the same client IDs $ID: \mathcal{X}^i_{ID} \neq \mathcal{X}^j_{ID}$.

Contrary to HFL, VFL (see Table~\ref{tab:VFL}) is the data partitioning scheme that shares the same client identity for distinct features. Usually, the application of VFL requires a trusted third party and encryption~\cite{yang2019parallel}. For example, an application creates personalized advertisements. The application collects or buys information regarding a target client from different sources that enhance the experience of the target client. Datasets are vertically partitioned when: let $i,j$, be two clients that share unique features $\mathcal{F}$ for a feature space $\mathcal{X}: \mathcal{X}^i_\mathcal{F} \neq X^j_\mathcal{F}$. However, they share the same clients IDs $ID: \mathcal{X}^i_{ID} = \mathcal{X}^j_{ID}$.

FTL~\cite{chen2020fedhealth, Liu2020, pan2009survey}, adapted from classical ML scenarios, is a scheme that grants the ability to train a new requirement on an already trained model. The new requirement could solve a different problem from what the original was trained for by training on a similar dataset. Training on a pre-trained model enhances the overall model quality, which is easier than training from scratch~\cite{pan2009survey}. For example, a model that detects malicious network traffic from HTTP requests now detects malicious traffic over MQTT~\cite{mothukuri2021survey}. A model $f$ that was trained over a domain $\mathcal{D}$ containing a feature space $\mathcal{X}$ can predict a label $y$ on the label space $\mathcal{Y}:y\in \mathcal{Y}$ for an input $\textbf{x}:y \leftarrow f(\textbf{x})$. Considering the source domain $\mathcal{D}_\mathcal{S}$ and the target domain $\mathcal{D}_\mathcal{T}$ where $\mathcal{D}_\mathcal{S} \neq \mathcal{D}_\mathcal{T}$ the objective of FTL is to help improve the learning capabilities of the target predictive function $f_\mathcal{T}$ in $\mathcal{D}_\mathcal{T}$ using the knowledge of $\mathcal{D}_\mathcal{S}$.

Similarly, under hybrid settings, different clients share distinct features~\cite{yoshida2020hybrid}. Let $i,j$ be two clients that share the distinct features $\mathcal{F}$ and distinct IDs $ID$ for a feature space $\mathcal{X}: \mathcal{X}^i_\mathcal{F} \neq \mathcal{X}^j_\mathcal{F} \wedge \mathcal{X}^i_{ID} \neq \mathcal{X}^j_{ID}$.

\begin{table}[!htb]
\caption{HFL and VFL description. Let $f_N \in \mathcal{F}$ and $ID_M$ distinct data samples.}
\begin{subtable}[c]{0.45\textwidth}
\centering
    \subcaption{HFL}
    \begin{tabular}{|c||c|c|c|c|c|}
    \hline
           & $f_1$ & $f_2$ & $f_3$        & ... & $f_N$ \\ \hline \hline
    $ID_1$ & $i$   & $i$   & $i$          &     & $i$   \\ \hline
    $ID_2$ & $i$   & $i$   & $i$          &     & $i$   \\ \hline
    $ID_3$ & $j$   & $j$   & $j$          &     & $j$   \\ \hline
    ...    &       &       &              &     &       \\ \hline
    $ID_M$ & $j$   & $j$   & $j$          &     & $j$   \\ \hline
    \end{tabular}
    \label{tab:HFL}
\end{subtable}
\begin{subtable}[c]{0.45\textwidth}
\centering
    \caption{VFL}
    \begin{tabular}{|c||c|c|c|c|c|}
    \hline
           & $f_1$ & $f_2$ & $f_3$ & ... & $f_N$ \\ \hline   \hline
    $ID_1$ & $i$   & $i$   & $j$   &     & $j$   \\ \hline
    $ID_2$ & $i$   & $i$   & $j$   &     & $j$   \\ \hline
    $ID_3$ & $i$   & $i$   & $j$   &     & $j$   \\ \hline
    ...    &       &       &       &     &       \\ \hline
    $ID_M$ & $i$   & $i$   & $j$   &     & $j$   \\ \hline
    \end{tabular}
    \label{tab:VFL}
\end{subtable}
\end{table}

Most of the studied papers partitioned the data according to the HFL scheme. Similarly, every paper used a standard server-client architecture while only one used a P2P.

Further, depending on the number of clients and their capabilities, the FL network is divided into cross-silo or cross-device settings. Clients are high-end powerful devices or parties in cross-silo settings, e.g., datacenters~\cite{kairouz2019advances}. Under this set-up, the clients are connection reliable and submit their model when required~\cite{yang2019federated}. Usually, the network comprises 2-100 clients, and data is split either horizontally or vertically. 

However, clients are not that powerful or connection reliable in cross-device scenarios. The network contains many low computational powered devices, as in Google's GBoard scenario~\cite{mcmahan2017communicationefficient, kairouz2019advances, konevcny2016federated}. Usually, the clients are IoT devices or smartphones~\cite{hard2018federated}. Clients have slow connections, usually via Wi-Fi, which may cause connection dropouts or be unavailable for submitting the model when required. Therefore, only a partition of the whole network will be available, around 5\%. The number of clients, contrary to cross-silo settings, is more significant, up to $10^{10}$ clients. In this case, the data distribution is always partitioned horizontally.

\subsection{Methodology}

This research was conducted by manually searching through two of the most known search engines, dblp (\url{https://dblp.org/}) and Google Scholar (\url{https://scholar.google.es/}).
In order to choose the papers for review, we have used some keywords in our search, i.e., adversarial federated learning, poisoning federated learning, backdoor federated learning, attack federated learning, threat federated learning, and defense federated learning. From the results of the search, we set a paper inclusion criteria. 

\begin{enumerate}
    \item The paper should be written in English.
    \item The paper should introduce a novel study in the research of the security of FL. Mainly it should cover either attacks or defenses. For example, proposals related to the efficiency of FL or learning proposals are not considered.
    \item Also, surveys papers are considered and evaluated as a baseline with our proposal.
    \item The paper proposal should be evaluated with experimentation.
\end{enumerate}

Table~\ref{tab:search} shows the finding for each search query, which we then evaluated through the paper, without duplicates.

\begin{table}[!htb]
\centering
\caption{The number of paper findings for each query.}
\begin{tabular}{|c||c|c|c|c||c|}
\hline
Query                                                                    & Attacks & Defenses & Survey & Unrelated & Total \\ \hline \hline
\begin{tabular}[c]{@{}c@{}}Adversarial\\ Federated Learning\end{tabular} & 5       & 13       & 0      & 7         & 25    \\ \hline
\begin{tabular}[c]{@{}c@{}}Poisoning\\ Federated Learning\end{tabular}   & 9       & 11       & 1      & 3         & 24    \\ \hline
\begin{tabular}[c]{@{}c@{}}Backdoor\\ Federated Learning\end{tabular}    & 2       & 8        & 2      & 6         & 18    \\ \hline
\begin{tabular}[c]{@{}c@{}}Attack\\ Federated Learning\end{tabular}      & 17      & 13       & 4      & 10        & 44    \\ \hline
\begin{tabular}[c]{@{}c@{}}Threat\\ Federated Learning\end{tabular}      & 2       & 1        & 3      & 2         & 8     \\ \hline
\begin{tabular}[c]{@{}c@{}}Defense\\ Federated Learning\end{tabular}     & 1       & 6        & 2      & 2         & 11    \\ \hline \hline
Total                                                                    & 36      & 52       & 12     & 30        & \textbf{130}   \\ \hline
\end{tabular}
\label{tab:search}
\end{table}

Also, the rest of the selected papers are relevant SoTA papers in centralized ML and other pertinent areas required to introduce background knowledge.

%% file: 04-ThreatModel.tex
\section{Threat model}
\label{sec:threat model}

In this section, we analyze and structurally represent the information that affects the security and privacy of FL. We first investigate the attack surface, i.e., what sections of FL are vulnerable in what phases. Second, we study the different actors involved in the FL process and how they may threaten the FL environment. Third, we analyze the adversarial capabilities and classify them as white-box and black-box. Finally, we review the adversarial goals and map them according to confidentiality, integrity, and availability (CIA).

\subsection{FL Attack Surface}

During the training phase, clients train their datasets locally. A poisoned dataset, e.g., gathered from untrusted sources, will contaminate the client's model, polluting the aggregated model after submission. Furthermore, an external attacker may compromise model exchanges by intercepting, eavesdropping, interrupting, or modifying the communications. 

Since FL is an iterative process, models are not required to be near convergence for using them. Hence, models can be threatened during the test phase by querying adversarial inputs, i.e., Adversarial Examples, or extracting private information, i.e., Model Inversion, by observing the model's outputs.
Similarly, external adversaries might threaten the aggregator that mangled the aggregation process.

\subsection{Adversarial Actors}

Two actors participate in the FL process, the clients and the aggregator. Under adversarial settings, a third one, an outsider as an external actor, has to be considered. The trust level for the FL environment is directly affected by the trust levels of the actors. Whether the adversarial actors are part of the network or outsiders, they are classified as insiders or outsiders. An insider is an actor that actively participates in the FL process. The aggregator and the clients belong to this group. The client has access to the model, its weights, the learning procedure, the dataset, and the communications. Likewise, the aggregator has access to each client model's weights, the communications, and the aggregation algorithm. Usually, insiders are categorized as \textit{honest-but-curious} and interact passively with the capabilities mentioned above.
The outsider has a low level of knowledge which depending on its capabilities, can access the communications or query the model at the test phase.

\subsection{Adversarial Capabilities}

As previously introduced, outsiders and insiders have different levels of knowledge and capabilities. This section defines and categorizes adversarial capabilities accordingly.

Since clients have unlimited access to the model's snapshots, they could leverage certain information, such as the local dataset, local learning rate, number of local epochs, local batch size, model architecture, communications, or its weights for empowering the attack. Similarly, the aggregator has access to a snapshot of the model, leveraging each client update, communications, or aggregation algorithm to improve attacks capabilities. On the contrary, an outsider could leverage information regarding communications and oracle access to the model to enhance its attacks.

Thus, we consider a \textbf{black-box} attack when an adversary can query the model $f(\textbf{x})$ with any arbitrary input $\textbf{x}$ and obtain the result. However, the adversary cannot access any inner computation of the model or the training procedure. 
In a \textbf{white-box} setting, the attacker may leverage or modify any of the above information to empower the attack~\cite{nasr2019comprehensive}. Because of the FL nature, most of the attacks are white-box.

\subsection{Adversarial Goals}

Adversarial actors may have different aims and approaches for attacking certain parts of the FL surface. We classify the adversarial goals concerning their capabilities, the FL phase, and the CIA triad.
The attack aim is categorized as:

\begin{compactitem}
    \item \textbf{Targeted:} the adversarial aim is to map chosen inputs to desired outputs or predictions.
    \item \textbf{Untargeted:} the adversarial goal is to degrade the primary task performance, so the model does not achieve near-optimal performance.
\end{compactitem}

Targeted and untargeted attacks are achieved independent of the FL phase or the adversarial knowledge. However, training phase attacks target the CIA triad of the model. For example, Poisoning attacks target the integrity and availability of the model, while Inference attacks target confidentiality. Test-time attacks mainly target confidentiality but could also target the model's integrity as with Adversarial Examples. Likewise, untargeted attacks, whose primary goal is to degrade the model's performance, can lead to a denial of service, causing unavailability. 

The details of the attacks and CIA triad mapping are explained in Sections~\ref{sec:trainigAdvSettings} and~\ref{sec:infAdvSettings}. In Sections~\ref{sec:trainigAdvSettings},~\ref{sec:infAdvSettings}, and~\ref{sec:secureFL}, we study the threat model and the SoTA countermeasures.

%% file: 05-TrainingInAdversarialSettings.tex
\section{Training in Adversarial Settings} 
\label{sec:trainigAdvSettings}

\begin{figure}[!htb]
    \centering
    \includegraphics[width=0.7\columnwidth]{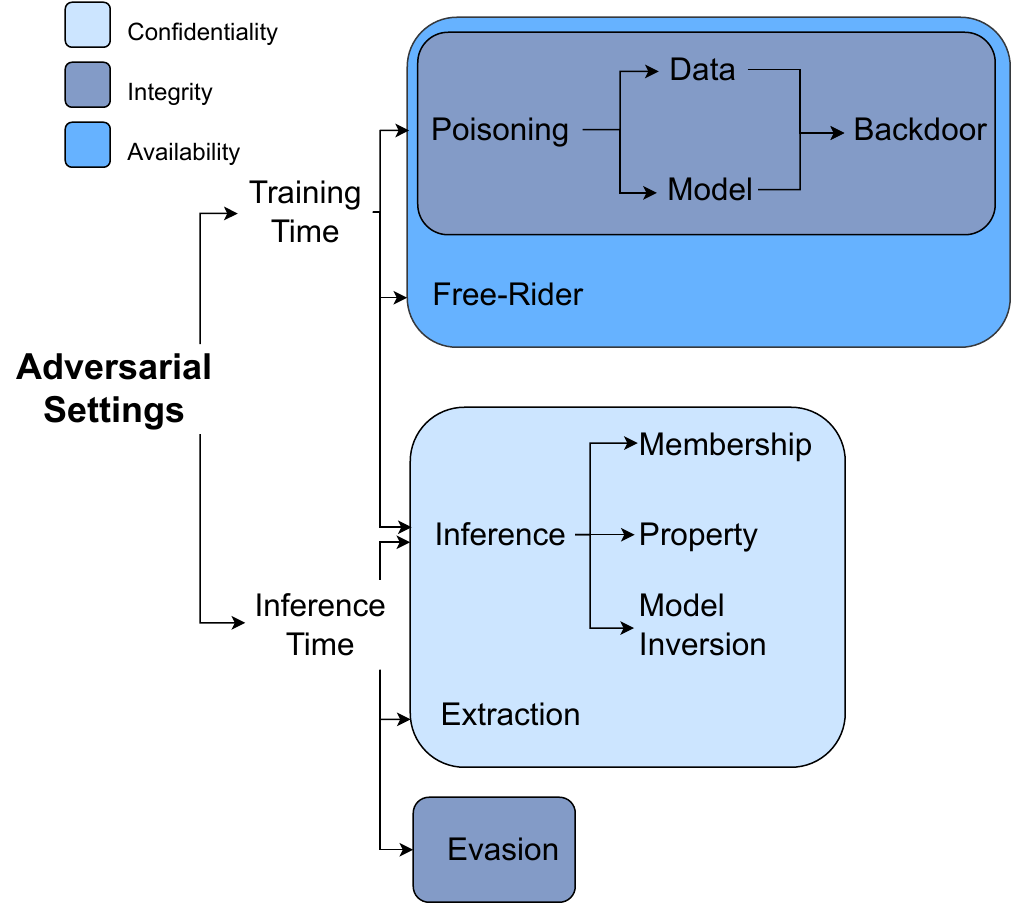}
    \caption{Proposed taxonomy of attacks.}
    \label{fig:taxAttacks}
\end{figure}

This section categorizes and explains the training-time attacks concerning the CIA. We further propose a taxonomy for each attack type by analyzing the SoTA attacks presented in Table~\ref{tab:attacks2} and considering the adversarial goals and capabilities shown in Fig.~\ref{fig:taxAttacks}.

\subsection{Targeting Integrity \& Availability}

\subsubsection{Poisoning Attacks}

Under FL, each client owns a dataset that could be modified for malicious purposes. As insiders, attackers have access to the training procedure, the model, and the dataset. An attacker leverages those capabilities to perform Poisoning attacks. Untargeted Poisoning attacks reduce the overall performance of the model. The attack ranges from degrading the main task where the model does not achieve a near-optimal performance to increasing the convergence time leading to a Denial of Service (DoS)~\cite{Fang2020}. However, targeted Poisoning attacks modify the model to achieve the desired output.

Once the joined model is poisoned, the aggregator shares it with each client in the network, thus poisoning every local model. The adversarial objective is achievable under two techniques:
\begin{compactitem}
	\item \textbf{Single-shot:} poison a model in a single FL round with a single adversarial client~\cite{Bagdasaryan2020}.
	\item \textbf{Multiple-shot:} poison a model in several FL rounds with $1\leq$ adversarial clients~\cite{Bagdasaryan2020}.
\end{compactitem}

Furthermore, depending on the capabilities of the attackers, attacks are further categorized into Data Poisoning, Model Poisoning, and Backdoors attacks. 

\paragraph{Data Poisoning}

Data Poisoning attacks are most common in ML but not very effective in FL because updates from benign clients tend to reduce the negative effect of the malicious dataset during the aggregation process~\cite{Tolpegin, qureshi2021performance}. As the attackers modify the dataset, Data Poisoning is thus considered a white-box attack.

Data Poisoning attacks rely on dataset modification during local training to degrade the overall model performance achieved by label-flipping~\cite{xiao2012adversarial}.
The intuition behind label-flipping is to alter the labels of the local dataset.
An attacker owns a dataset made of $\{(\textbf{x}_i,y_i)\}^\mathcal{N}_{i=1}$, a set of vectors $\textbf{x}_i$ and its corresponding class $y_i$, where $\mathcal{N}$ is the size of the dataset. The attack aims to change the ground truth class $y_i$ of $\textbf{x}_i$, to an adversarial class $a$. Then, the dataset gets poisoned by an adversarial entry $\{(\textbf{x}_i,a)\}$. For example, for the MNIST dataset, an attacker will swap the ``1'' data samples to ``7'', namely 1 to 7 attack~\cite{biggio2012poisoning}. For the general case $\{(\textbf{x}_i,y_i)\} \rightarrow \{(\textbf{x}_i,a)\}$ and for this example $\{(\textbf{x}_i,1)\} \rightarrow \{(\textbf{x}_i,7)\}$. 

For example, the authors in~\cite{DucNguyen} presented an anomaly detection model for IoT using FL, where they gradually injected malicious traffic in the dataset using label-flipping, successfully bypassing defensive clustering techniques. Following the same technique, authors~\cite{Tolpegin} noticeably reduced the model's accuracy.

Label-flipping is further categorized regarding the existence of a source of label verifier, which proves the correctness of the chosen label as a pre-processing step before training:
\begin{compactitem}
\item \textbf{Clean-label}~\cite{shafahi2018poison,munoz2017towards,turner2019label}: the intuition behind clean-label attacks is that they do not require control over the labeling function. The poisoned dataset seems correct to the eyes of an expert label-verifier identity. These types of attacks are difficult to detect and usually occur when data is gathered from untrusted sources to construct a dataset. An attacker may place a poisoned instance online and wait for it to be gathered, labeled, and included in the dataset. Since the data piece seems to be accurately labeled and used for training, the model is trained over a poisoned dataset and, therefore, contaminated.
  
\item \textbf{Dirty-label}~\cite{Chen2017}: dirty-label consists of modifying the dataset by changing the label of a data piece to the desired one (targeted) or to a random one (untargeted), e.g., changing all planes labels into birds that the final model classifies planes images as birds (targeted) or arbitrarily assigning labels to the training samples.
\end{compactitem}

\paragraph{Model Poisoning}

Data Poisoning attacks have shown poor performance in real-life scenarios~\cite{Bagdasaryan2020}. Thus, a new alternative has emerged to manipulate model weights directly: Model Poisoning attacks~\cite{Bhagoji}. They overcome the low success of Data Poisoning attacks by using Boosting. Boosting is a technique that enlarges clients' contributions by multiplying the vector weights (the update) by a scaling factor. Similarly, clients could modify any additional parameter from the training procedure for malicious purposes.
After the training procedure over a poisoned dataset, the attacker has a model $f$ represented by a vector of weights $\textbf{w}$. The attacker boosts its model by multiplying it by a Boosting factor $\lambda: \lambda \textbf{w}$.

Authors in~\cite{Bhagoji} further optimized the Model Poisoning attack by applying SGD to the scaling factor for finding its optimal value. An adversarial client could change the local training rate, the weights, the training epochs, or submit arbitrary updates. Since this attack modified the vector of weights directly, it is a white-box attack.
A boosted model has more influence over the global model, achieving successful single-shot attacks. However, defenses have arisen towards this Boosting-based model replacement attacks because it increases the $l_n$ norm of the vector of weights (see Section~\ref{sec:secureFL}). Additionally, several proposed stealthy techniques evaded such defenses~\cite{WeiCovert, shejwalkar2021manipulating, Shejwalkar, Tomsett, Huang, hossain2021desmp}.
Furthermore, authors~\cite{douceur2002sybil} proposed Sybil attacks, which evaded and improved attacks quality. They are coordinated attacks among various adversarial clients. A single attacker could control more than a single client, allowing an adversary to smartly distribute its attack, improving stealthiness and its adversarial accuracy. Sybils are mainly used in Model Poisoning but are also present in other types of adversarial threats~\cite{Mallah2021}.
Likewise, the authors in~\cite{Fang2020} evaluated Model Poisoning attacks against Byzantine-robust FL. They tested Model Poisoning against Krum~\cite{blanchard2017machine}, Bulyan~\cite{guerraoui2018hidden}, trimmed mean~\cite{yin2018byzantine}, and mean defensive mechanisms. None of them could stop the attack. The authors further investigated adapting Krum and Reject on Negative Impact (RONI)~\cite{barreno2010security} for defending Model Poisoning attacks, but with not enough success. Other approaches~\cite{Zhang2020a, Zhang2019} leveraged Generative Adversarial Networks (GANs)~\cite{goodfellow2014generative} for generating poisoned data. The created poisoned dataset is then used to train an adversarial model that reduced the attack's assumptions and increased the Model Poisoning attack's feasibility in real-world scenarios.

\paragraph{Backdoor Attacks} 
Backdoors~\cite{Gu2019} are a particular type of Poisoning attack, also named Trojans. Authors in~\cite{Bagdasaryan2020} investigated the possibility of performing a Data Poisoning attack on chosen labels while working non-adversarially on the rest. They leveraged specific data of the dataset that contained certain information, i.e., green cars with white stripes, and then applied label-flipping to those. Thus, the resulting model performed adversarially over green cars with white stripes as input. However, Backdoors need Boosting to become effective~\cite{Bagdasaryan2020}.
A model $f$ is trained on backdoored data samples and represented by $\textbf{w}$ vector of weights. The attacker may (or may not) want to boost its model by an scaling factor $\lambda: \lambda \textbf{w}$. Furthermore, the attacker may leverage information about an anomaly detection $\mathcal{L}_{ano}$ for evading aggregator's defensive techniques. By controlling the importance of evading the defense $\alpha$, the attacker may change the loss function $\mathcal{L}$ from its training procedure, $\mathcal{L} = \alpha \mathcal{L} + (1 - \alpha) \mathcal{L}_{ano}$~\cite{Bagdasaryan2020}.

Regarding the trigger used, they are further classified into:
\begin{compactitem}
    \item \textbf{In-Backdoor}~\cite{Bhagoji}: when the trigger label exists in at least one of the datasets of any client in the network. For a set of datasets $\mathcal{S}$ where $\mathcal{S}=\{\mathcal{D}_1,\mathcal{D}_2,\mathcal{D}_3,...\mathcal{D}_n\}$, let $n$ be the number of clients in the network. $\mathcal{D}_n$ is made of data samples $\{(\textbf{x}_i,y_i)\}^\mathcal{N}_{i=1}$. At least one sample's class $y_i$ of at least one party in $\mathcal{S}$ does belong to the adversarial class $a: \{(\textbf{x}_i,a)\} \in \mathcal{S}$.
  
    \item \textbf{Out-Backdoor}~\cite{Kerkouche}: when the trigger class does not exist in any of the datasets for any client in the network, i.e., $\{(\textbf{x}_i,a)\} \notin \mathcal{S}$.

\end{compactitem}

\begin{figure}[!htb]
    \centering
    \includegraphics[width=0.7\textwidth]{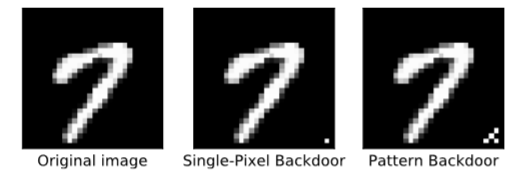}
    \caption{Single pixel and pixel pattern backdoors. Figure extracted form~\cite{Gu2019}.}
    \label{fig:backdoor}
\end{figure}

Depending on the trigger type, Backdoor attacks are categorized into:
\begin{compactitem}

\item \textbf{Edge Backdoor:} the authors in~\cite{Wang} suggested a type of Backdoor attack that relied on rare data pieces, i.e., edge samples of the dataset with adversarial labels, for local training, so the resulting model will only misclassify those rare inputs. One benefit of this method is that gradient-based defense techniques are unlikely to detect this attack~\cite{Wang}. The attacker uses an edge case dataset $\mathcal{D'}$ extracted from $\mathcal{D}$. Then it performs standard local training aiming to maximize the accuracy. However, Server Cleaning techniques (see Section~\ref{sec:secureFL}) removed such a simple attack because the data distribution of $\mathcal{D'}$ is largely different from $\mathcal{D}$. For solving it, they further proposed a stealthier Projected Gradient Descent (PGD) attack. The attacker applied PGD to maximize the similarity of updates from benign clients. To improve the attack, the authors used Boosting for achieving a model replacement. 
  
\item \textbf{Semantic Backdoor:} a type of attack that does not modify the input but the labels, e.g., all the cars with white stripes are labeled birds~\cite{Bagdasaryan2020}. It is compulsory to use Boosting to ensure that the aggregation survives the averaging, and the Backdoor is thus successfully transferred to the global model.
  
\item \textbf{Pixel-pattern Backdoor:} primarily used in image recognition, and in contrast with Semantic Backdoors, pixel-pattern attacks~\cite{Gu2019, Zhou2021} inject a pattern in a particular area of the input space (see Fig.~\ref{fig:backdoor}). After altering some local data, the attacker performs local training with a mixture of adversarial and benign data to follow a similar non-Backdoored data distribution. The resulting model misclassified the inputs that contained the trigger pattern while correctly classifying harmless inputs, e.g., some pixel pattern has been injected into several car images and labeled as a bird. After local training and global model aggregation, the resulting model will misclassify every car input containing the trigger, predicting them as birds, while a benign user gets a correct classification for a non-Backdoored car input. For improving the stealthiness, the authors in~\cite{Wang2020} evaluated pixel-pattern Backdoors with no scaling factor where they evaded most of the clipping defensive mechanisms. As a particular case of pixel-pattern attacks, single-pixel attacks do not necessarily use large patterns to trigger a Backdoor. Instead, they only require modifying a single pixel in the input space~\cite{moosavi2017analysis}.
\end{compactitem}

Since Backdoors development, the research drifted towards testing them in different environments~\cite{Nuding, chen2020backdoor} and improving Backdoor techniques as dynamically changing Backdoors~\cite{Huang}. In addition,~\cite{Sun} investigated Backdoors in real-world scenarios. They performed the research over non-IID data distribution and various datasets. They also analyzed norm-clipping and Differential Privacy defenses, with almost no model performance degradation. Authors in~\cite{Xie, Chen2021} proposed evading techniques, namely distributed Backdoor attacks, where the Backdoor got split among adversarial clients. Each client performed a Backdoor in coordination, improving stealthiness, and after aggregation, the split Backdoor gets joined.

Backdoor attacks are recognized as Data and Model Poisoning, depending on the altered factor. Solely altering the dataset, i.e., a Backdoor without Boosting is considered a Data Poisoning-based Backdoor. However, when using Boosting techniques, Backdoors are considered Model Poisoning-based Backdoors~\cite{goldblum2020dataset}.

\subsection{Targeting Availability}

\subsubsection{Free-riders}

Free-riders~\cite{fraboni2021free, lin2019free} are clients that do not contribute to the network by different means. The motivation behind the attack is that the Free-rider client does not want to collaborate in the network or does not have enough resources for training the local model, e.g., enough computational power or a sufficiently large dataset. The attacker disguises itself by sending random updates similar to the joined model to the aggregator downgrading its overall quality. Commonly the adversary does not require any extra information for sending random updates.
An attacker submits a random vector of weights $\textbf{w}_{rand}$ to the aggregator. Or the attackers send an update that resemblances to the aggregated model $\textbf{w}_{agg}:\textbf{w}_{rand} \simeq \textbf{w}_{agg}$.

When the server implements defense mechanisms~\cite{lin2019free}, the attacker might leverage additional information for evading them in a white-box manner. SoTA Free-rider attacks aimed to bypass defenses implemented by the aggregator~\cite{Wan2021}. The attacker trains the model locally with a small dataset for saving resources. Since clients control the training procedure, the attacker claims to use a large dataset. The attack reduced the model convergence time by 50\%. The authors in~\cite{fraboni2021free} analyzed the ease of detection for plain Free-rider attacks. Thus, they developed a stealthy Free-rider attack, which added noise to the crafted updates. Then, SGD is applied to maximize the resemblance with the updates of other clients. Since some servers offered rewards for contributing clients, a Free-rider might change the weights via Boosting for achieving higher representation on the global server and own a more significant part of the reward~\cite{Wan2021}.

%% file: 06-InferringInAdversarialSettings.tex
\section{Inferring in Adversarial Settings}
\label{sec:infAdvSettings}

This section explains inference-time attacks and maps them according to the CIA triad.

\subsection{Targeting Integrity}

\subsubsection{Evasion Attacks}

\begin{figure}[!htb]
    \centering
    \includegraphics[width=0.8\textwidth]{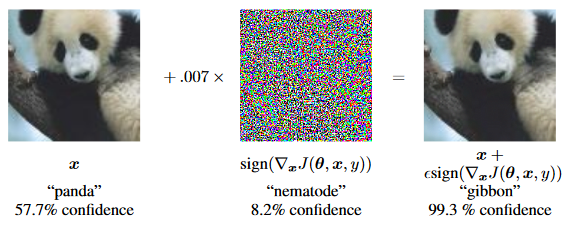}
    \caption{Adversarial examples created by the Fast Gradient Sign Method. Figure extracted from~\cite{goodfellow2014explaining}.}
    \label{fig:FGSM}
\end{figure}

Mostly used in ML, Adversarial Examples~\cite{carlini2017towards, goodfellow2014explaining} leverage a precisely crafted input to misclassify the model at inference-time (see Fig.~\ref{fig:FGSM}). Since the model is unknown to the attacker, the attack is a black-box. However, under FL, where the model is known to an adversarial inside the network, they may leverage that extra information for performing a more powerful attack. Thus it would be considered a white-box. The intuition behind the attack is: from an input $\textbf{x}$ and a crafted noise $\epsilon$ fool the model $f$ to misclassify it in a targeted or untargeted manner. A crafted input $\textbf{x} + \epsilon$ and its ground truth label $y$ fool the model as $y \nleftarrow f(\textbf{x} + \epsilon)$.
Similarly, Evasion attacks are also developed in a physical context~\cite{sharif2016accessorize, wang2019advpattern}, where the crafted noise is included physically in the real world rather than embedded via software, e.g., evading face detection systems~\cite{sharif2016accessorize}.

\subsection{Targeting Confidentiality}

\subsubsection{Inference Attacks}

Inference attacks use different knowledge to extract private information. Depending on the adversarial capabilities and the phase attacked, the attacker leverages oracle access to the model, its inner computations, or the updates~\cite{Zhu}. Depending on the attacker's activities, Inference attacks are classified as:
\begin{compactitem}
    \item \textbf{Passive}~\cite{Zhang2020,nasr2019comprehensive}: the attacker, \textit{honest-but-curious}, could only read or eavesdrop on communications, the local model, and the dataset.
   
    \item \textbf{Active}~\cite{nasr2019comprehensive}: more effective than passive attacks but less stealthy, leverages the extra information obtained by mangling the local model, local training procedure, or its parameters, local dataset, or communications.
\end{compactitem}

Furthermore, depending on the target information, Inference attacks are further classified as:
\begin{compactitem}
   
\item \textbf{Membership Inference}~\cite{nasr2019comprehensive}: the goal is to infer whether some data belongs to the training dataset. Under FL, it is even possible to suggest which user owns the dataset. The attacker aim is to infer if some data piece $\{(\textbf{x},y)\}$ belongs to a local dataset $\mathcal{D}_n: \{(\textbf{x},y)\} \in \mathcal{D}_n$ where $n$ is the number of clients in the network. Or in the general case whether $\{(\textbf{x},y)\} \in \mathcal{S}$ belongs to the dataset $\mathcal{S}=\{\mathcal{D}_1,\mathcal{D}_2,\mathcal{D}_3,...,\mathcal{D}_n\}$.
   
\item \textbf{Property Inference}~\cite{Melis2019, lyu2021novel}: this attack leverages model snapshot updates concerning a dataset without the objective property and another one containing it. Afterward, a trained binary classifier can relate the property with an unseen update, e.g., an update got by eavesdropping, to conclude whether the update owns the objective property. For a data piece $\textbf{x}$ containing the property $p:\textbf{x}^{(p)}$ belonging to the dataset $\mathcal{D}_n$ being $n$ the number of clients, the attacker infers if the property belongs to the dataset $\textbf{x}^{(p)} \in \mathcal{D}_n$.
   
\item \textbf{Model Inversion}~\cite{fredrikson2015model, wu2020plfg}: 
an ML model is trained to infer a label for a given input. As demonstrated by~\cite{fredrikson2014privacy} ML models tend to remember training data. The same authors inverted the ML inference-time process by first providing the label backward to the model while maximizing the likelihood of the target label. The result is a piece of data similar to the one used during training. Even authors in~\cite{fredrikson2015model} achieved pixel-wise accuracy on reconstructing images and token-accurate on text-based models~\cite{carlini2020extracting}. Model Inversion attacks can be categorized depending if the attacker can solely query the model, i.e., black-box, or has access to the model parameters, i.e., white-box. From a model $f$ trained on the dataset $\mathcal{D}$, the Model Inversion attack tries to reconstruct a piece of data $\hat{\textbf{x}} \simeq \textbf{x} \in \mathcal{D}$ by querying $f$.
\end{compactitem}

Regarding the SoTA, the authors in~\cite{nasr2019comprehensive} proposed a white-box Membership Inference attack. They leveraged the attacker's position inside the network to acquire the model. The model obtained every layer's intermediate computation on an input. They then trained a Convolutional Neural Network (CNN) with the gained weights, distinguishing if the input is present in some client dataset. Extending this attack, the authors in~\cite{Chen} successfully performed user-level Membership Inference mapping the input to a specific client's dataset. Leveraging the inferred information, the authors~\cite{hu2021source} performed an improved Model Inversion attack.
Similarly, the authors in~\cite{Zhang2020} used a GAN for empowering their attack. The GAN generated data samples that followed the same data distribution as the real one. Then this data was passed to the global model for setting a label and merging the generated data with actual data. Given that the generated data were not present in none of the clients' datasets, the authors trained a model to infer its membership. For Model Inversion attacks, the authors in~\cite{Song2020, sun2021information} also used GANs. With a malicious server, they deployed a GAN trained on the updates of a victim client. Levering an auxiliary dataset, the model could create samples similar to the input. Improving this approach, authors~\cite{Ren} developed a Generative Recurrent Neural Networks that removed the need for an auxiliary dataset. Contrary, in~\cite{Luo}, the authors considered VFL instead of HFL. They researched the effects of Membership Inference in VFL for Logistic Regression and Decision Trees, widely used models in the industry. The authors in~\cite{Li2021a} investigated privacy threats to Linear Regression, Logistic Regression, and Decision Tree while considering Model Inversion attacks.

\subsubsection{Model Extraction Attacks}

Model Extraction attacks try to mimic or fully copy a target model. Leveraging oracle access to a model $f$, Model Extraction attacks try to reconstruct it with oracle access $\textbf{x}:f(\textbf{x})\simeq \hat{f}(\textbf{x})$. The model $\hat{f}$ acquired by approximation relies on creating a model that, by iterative adjustments, performs similar to the original model. However, it is not architecturally the same. For reconstructing a floating-point accurate model, an analytical approach is needed, named direct analysis~\cite{Carlini, jagielski2020high}. Because of the gradient descent procedure during training, which implies second derivatives computation, the model holds some critical points where the derivative equals 0. Meeting those points leads to floating-point accurate weights recovery.

Extraction attacks are common in ML, where an adversary does not know details about the model rather than its architecture, i.e., black-box settings. However, clients of the FL network already have access to the model. Nevertheless, an outsider may perform a Model Extraction attack for owning the joined model.

\begin{sidewaystable}[!htbp]
    \caption{Comparison of attacks regarding different features.~\dag ~denotes a backdoor attack.}
    \label{tab:attacks2}
    \resizebox{\textwidth}{!}{

    \begin{tabular}{llllllllll}

        \hline
        
        & \begin{tabular}[c]{@{}l@{}} Attack   \\ \fbox{\tiny{1}} \fbox{\tiny{2}} \fbox{\tiny{3}} \fbox{\tiny{4}} \fbox{\tiny{5}} \end{tabular}
  
        & \begin{tabular}[c]{@{}l@{}} Type \\ \fbox{\tiny{1}} \fbox{\tiny{2}} \end{tabular}    
        
        & \begin{tabular}[c]{@{}l@{}} Data distribution   \\ \fbox{\tiny{1}} \fbox{\tiny{2}} \fbox{\tiny{3}} \end{tabular}
    
        & \begin{tabular}[c]{@{}l@{}}  Dataset \\ \fbox{\tiny{1}} \fbox{\tiny{2}} \fbox{\tiny{3}}\end{tabular} 
        
        & \begin{tabular}[c]{@{}l@{}} Evaluation metrics   \\ \fbox{\tiny{1}} \fbox{\tiny{2}} \fbox{\tiny{3}} \fbox{\tiny{4}} \fbox{\tiny{5}}\end{tabular}
            
        & \begin{tabular}[c]{@{}l@{}} Defences  \\ \fbox{\tiny{1}} \fbox{\tiny{2}} \fbox{\tiny{3}} \fbox{\tiny{4}} \fbox{\tiny{5}} \fbox{\tiny{6}} \fbox{\tiny{7}} \fbox{\tiny{8}} \end{tabular} 
        
        & \begin{tabular}[c]{@{}l@{}} Source code \\ \fbox{\tiny{1}} \fbox{\tiny{2}} \end{tabular}  
        
        & \begin{tabular}[c]{@{}l@{}} FL algorithm  \\ \fbox{\tiny{1}} \fbox{\tiny{2}} \fbox{\tiny{3}} \fbox{\tiny{4}} \fbox{\tiny{5}} \fbox{\tiny{6}} \fbox{\tiny{7}} \fbox{\tiny{8}} \fbox{\tiny{9}} \fbox{\tiny{10}} \end{tabular}
        
        & \begin{tabular}[c]{@{}l@{}} Framework  \\ \fbox{\tiny{1}} \fbox{\tiny{2}} \fbox{\tiny{3}} \fbox{\tiny{4}} \fbox{\tiny{5}} \fbox{\tiny{6}} \end{tabular} \\

        \hline
            
        & \begin{tabular}[c]{@{}l@{}} 1. Data Poisoning\\ 2. Model Poisoning\\ 
        3. Inference\\ 4. Model Inversion\\ 5. Free-rider\end{tabular}
        
        & \begin{tabular}[c]{@{}l@{}} 1. White-Box\\ 2. Black-Box\\\end{tabular}
        
        & \begin{tabular}[c]{@{}l@{}}1. IID\\ 2. Non-IID \\3. Not defined\end{tabular}
        
        & \begin{tabular}[c]{@{}l@{}} 1. Image\\ 2. Multivariate\\ 3. Text\end{tabular} 
        
        & \begin{tabular}[c]{@{}l@{}} 1. Confusion matrix\\ 2. ASR\\ 3. MSE\\ 4. $l_n$-norm \\ 5. Others\end{tabular}
        
        & \begin{tabular}[c]{@{}l@{}}1. Pruning \\ 2. Server Cleaning \\ 3. Cryptographic Methods \\ 4. Adversarial Training \\ 5. Differential Privacy  \\ 6. Robust FL Aggregation \\ 7. GAN-based \\ 8. None \end{tabular}
        
        & \begin{tabular}[c]{@{}l@{}} 1. Yes \\ 2. No \end{tabular}
        
        & \begin{tabular}[c]{@{}l@{}} 1. CNN \\ 2. ResNet \\ 3. LeNet \\ 4. LSTM \\ 5. LR \\ 6. AlexNet \\ 7. DenseNet \\ 8. VGG \\ 9. Others \\ 10. Not defined \end{tabular} 

        & \begin{tabular}[c]{@{}l@{}} 1. PyTorch \\ 2. TensorFlow \\ 3. Scikit-Learn \\ 4. PySyft \\ 5. FATE \\ 7. Not defined \end{tabular} \\

        \hline

        \begin{tabular}[c]{@{}l@{}} Wang \textit{et al.}~\cite{Wang2020}\dag \end{tabular}
        &
        $\square$\hspace{5pt}$\blacksquare$\hspace{5pt}$\square$\hspace{5pt}$\square$\hspace{5pt}$\square$& 
        
        $\blacksquare$\hspace{5pt}$\square$& 
        
        $\blacksquare$\hspace{5pt}$\blacksquare$\hspace{5pt}$\square$
        & 
        
        $\blacksquare$\hspace{5pt}$\square$\hspace{5pt}$\square$& 
        
        $\blacksquare$\hspace{5pt}$\square$\hspace{5pt}$\square$\hspace{5pt}$\blacksquare$\hspace{5pt}$\blacksquare$\hspace{5pt} & 
        
        $\square$\hspace{5pt}$\square$\hspace{5pt}$\square$\hspace{5pt}$\square$\hspace{5pt}$\square$\hspace{5pt}$\square$\hspace{5pt}$\square$\hspace{5pt}$\blacksquare$ & 
        
        $\square$\hspace{5pt}$\blacksquare$ & 
        
        $\square$\hspace{5pt}$\square$\hspace{5pt}$\square$\hspace{5pt}$\square$\hspace{5pt}$\square$\hspace{5pt}$\square$\hspace{5pt}$\square$\hspace{5pt}$\square$\hspace{5pt}$\square$\hspace{5pt}$\blacksquare$ &
        
        $\square$\hspace{5pt}$\square$\hspace{5pt}$\square$\hspace{5pt}$\square$\hspace{5pt}$\square$\hspace{5pt}$\blacksquare$

        \\
        
        \begin{tabular}[c]{@{}l@{}} Nguyen \textit{et al.} ~\cite{DucNguyen} \end{tabular}
        &
        $\blacksquare$\hspace{5pt}$\square$\hspace{5pt}$\square$\hspace{5pt}$\square$\hspace{5pt}$\square$& 
        
        $\blacksquare$\hspace{5pt}$\square$& 
        
        $\square$\hspace{5pt}$\square$\hspace{5pt}$\blacksquare$  & 
        
        $\square$\hspace{5pt}$\blacksquare$\hspace{5pt}$\square$& 
        
        $\blacksquare$\hspace{5pt}$\square$\hspace{5pt}$\square$\hspace{5pt}$\square$\hspace{5pt}$\square$\hspace{5pt} & 
        
        $\square$\hspace{5pt}$\blacksquare$\hspace{5pt}$\square$\hspace{5pt}$\square$\hspace{5pt}$\blacksquare$\hspace{5pt}$\square$\hspace{5pt}$\square$\hspace{5pt}$\square$ & 
        
        $\square$\hspace{5pt}$\blacksquare$\hspace{5pt} & 
        
        $\square$\hspace{5pt}$\square$\hspace{5pt}$\square$\hspace{5pt}$\square$\hspace{5pt}$\square$\hspace{5pt}$\square$\hspace{5pt}$\square$\hspace{5pt}$\square$\hspace{5pt}$\blacksquare$\hspace{5pt}$\square$  &      
        
        $\blacksquare$\hspace{5pt}$\square$\hspace{5pt}$\square$\hspace{5pt}$\square$\hspace{5pt}$\square$\hspace{5pt}$\square$

        \\

        \begin{tabular}[c]{@{}l@{}}  Tolpegin \textit{et al.} ~\cite{Tolpegin} \end{tabular}
        &
        $\blacksquare$\hspace{5pt}$\square$\hspace{5pt}$\square$\hspace{5pt}$\square$\hspace{5pt}$\square$& 
        
        $\blacksquare$\hspace{5pt}$\square$& 
        
        $\blacksquare$\hspace{5pt}$\square$\hspace{5pt}$\square$  & 
        
        $\blacksquare$\hspace{5pt}$\square$\hspace{5pt}$\square$& 
        
        $\blacksquare$\hspace{5pt}$\square$\hspace{5pt}$\square$\hspace{5pt}$\square$\hspace{5pt}$\square$\hspace{5pt} & 
        
        $\square$\hspace{5pt}$\blacksquare$\hspace{5pt}$\square$\hspace{5pt}$\square$\hspace{5pt}$\square$\hspace{5pt}$\square$\hspace{5pt}$\square$\hspace{5pt}$\square$ & 
        
        $\blacksquare$\hspace{5pt}$\square$\hspace{5pt} & 
        
        $\blacksquare$\hspace{5pt}$\square$\hspace{5pt}$\square$\hspace{5pt}$\square$\hspace{5pt}$\square$\hspace{5pt}$\square$\hspace{5pt}$\square$\hspace{5pt}$\square$\hspace{5pt}$\square$\hspace{5pt}$\square$ &
        
        $\blacksquare$\hspace{5pt}$\square$\hspace{5pt}$\square$\hspace{5pt}$\square$\hspace{5pt}$\square$\hspace{5pt}$\square$

        \\
        
        \begin{tabular}[c]{@{}l@{}} Wang \textit{et al.} ~\cite{Wang}\dag \end{tabular}
        &
        $\blacksquare$\hspace{5pt}$\blacksquare$\hspace{5pt}$\square$\hspace{5pt}$\square$\hspace{5pt}$\square$& 
        
        $\blacksquare$\hspace{5pt}$\square$& 
        
        $\blacksquare$\hspace{5pt}$\square$\hspace{5pt}$\square$  & 
        
        $\blacksquare$\hspace{5pt}$\blacksquare$\hspace{5pt}$\blacksquare$& 
        
        $\blacksquare$\hspace{5pt}$\square$\hspace{5pt}$\square$\hspace{5pt}$\square$\hspace{5pt}$\square$\hspace{5pt} & 
        
        $\square$\hspace{5pt}$\blacksquare$\hspace{5pt}$\square$\hspace{5pt}$\square$\hspace{5pt}$\blacksquare$\hspace{5pt}$\blacksquare$\hspace{5pt}$\square$\hspace{5pt}$\square$ & 
        
        $\blacksquare$\hspace{5pt}$\square$ & 
        
        $\square$\hspace{5pt}$\square$\hspace{5pt}$\blacksquare$\hspace{5pt}$\blacksquare$\hspace{5pt}$\square$\hspace{5pt}$\square$\hspace{5pt}$\square$\hspace{5pt}$\blacksquare$\hspace{5pt}$\square$\hspace{5pt}$\square$ &
        
        $\blacksquare$\hspace{5pt}$\square$\hspace{5pt}$\square$\hspace{5pt}$\square$\hspace{5pt}$\square$\hspace{5pt}$\square$

        \\
        
        \begin{tabular}[c]{@{}l@{}} Chen \textit{et al.} ~\cite{chen2020backdoor}\dag \end{tabular}
        &
        $\blacksquare$\hspace{5pt}$\blacksquare$\hspace{5pt}$\square$\hspace{5pt}$\square$\hspace{5pt}$\square$& 
        
        $\blacksquare$\hspace{5pt}$\square$& 
        
        $\blacksquare$\hspace{5pt}$\square$\hspace{5pt}$\square$  & 
        
        $\blacksquare$\hspace{5pt}$\square$\hspace{5pt}$\square$& 
        
        $\blacksquare$\hspace{5pt}$\square$\hspace{5pt}$\square$\hspace{5pt}$\square$\hspace{5pt}$\square$\hspace{5pt} & 
        
        $\square$\hspace{5pt}$\blacksquare$\hspace{5pt}$\square$\hspace{5pt}$\square$\hspace{5pt}$\blacksquare$\hspace{5pt}$\square$\hspace{5pt}$\square$\hspace{5pt}$\square$ & 
        
        $\square$\hspace{5pt}$\blacksquare$ & 
        
        $\blacksquare$\hspace{5pt}$\square$\hspace{5pt}$\square$\hspace{5pt}$\blacksquare$\hspace{5pt}$\square$\hspace{5pt}$\square$\hspace{5pt}$\square$\hspace{5pt}$\blacksquare$\hspace{5pt}$\square$\hspace{5pt}$\square$ &
        
        $\square$\hspace{5pt}$\square$\hspace{5pt}$\square$\hspace{5pt}$\square$\hspace{5pt}$\square$\hspace{5pt}$\blacksquare$

        \\
        
        \begin{tabular}[c]{@{}l@{}} Zhang \textit{et al.} ~\cite{Zhang2019} \end{tabular}
        &
        $\square$\hspace{5pt}$\blacksquare$\hspace{5pt}$\square$\hspace{5pt}$\square$\hspace{5pt}$\square$& 
        
        $\blacksquare$\hspace{5pt}$\square$& 
        
        $\square$\hspace{5pt}$\blacksquare$\hspace{5pt}$\square$  & 
        
        $\blacksquare$\hspace{5pt}$\square$\hspace{5pt}$\square$& 
        
        $\blacksquare$\hspace{5pt}$\square$\hspace{5pt}$\square$\hspace{5pt}$\square$\hspace{5pt}$\square$\hspace{5pt} & 
        
        $\square$\hspace{5pt}$\square$\hspace{5pt}$\square$\hspace{5pt}$\square$\hspace{5pt}$\square$\hspace{5pt}$\square$\hspace{5pt}$\square$\hspace{5pt}$\blacksquare$ & 
        
        $\square$\hspace{5pt}$\blacksquare$\hspace{5pt} & 
        
        $\blacksquare$\hspace{5pt}$\square$\hspace{5pt}$\square$\hspace{5pt}$\square$\hspace{5pt}$\square$\hspace{5pt}$\square$\hspace{5pt}$\square$\hspace{5pt}$\square$\hspace{5pt}$\square$\hspace{5pt}$\square$ &
        
        $\blacksquare$\hspace{5pt}$\square$\hspace{5pt}$\square$\hspace{5pt}$\square$\hspace{5pt}$\square$\hspace{5pt}$\square$

        \\

        \begin{tabular}[c]{@{}l@{}} Bhagoji \textit{et al.} ~\cite{Bhagoji} \end{tabular}
        &
        $\square$\hspace{5pt}$\blacksquare$\hspace{5pt}$\square$\hspace{5pt}$\square$\hspace{5pt}$\square$& 
        
        $\blacksquare$\hspace{5pt}$\square$& 
        
        $\blacksquare$\hspace{5pt}$\square$\hspace{5pt}$\square$  & 
        
        $\blacksquare$\hspace{5pt}$\square$\hspace{5pt}$\square$& 
        
        $\blacksquare$\hspace{5pt}$\square$\hspace{5pt}$\square$\hspace{5pt}$\square$\hspace{5pt}$\square$\hspace{5pt} & 
        
        $\square$\hspace{5pt}$\square$\hspace{5pt}$\square$\hspace{5pt}$\square$\hspace{5pt}$\square$\hspace{5pt}$\blacksquare$\hspace{5pt}$\square$\hspace{5pt}$\square$ & 
        
        $\blacksquare$\hspace{5pt}$\square$\hspace{5pt} & 
        
        $\blacksquare$\hspace{5pt}$\square$\hspace{5pt}$\square$\hspace{5pt}$\square$\hspace{5pt}$\square$\hspace{5pt}$\square$\hspace{5pt}$\square$\hspace{5pt}$\square$\hspace{5pt}$\blacksquare$\hspace{5pt}$\square$ &
        
        $\square$\hspace{5pt}$\blacksquare$\hspace{5pt}$\blacksquare$\hspace{5pt}$\square$\hspace{5pt}$\square$\hspace{5pt}$\square$         
        
        \\
        
        \begin{tabular}[c]{@{}l@{}} Wei \textit{et al.} ~\cite{WeiCovert} \end{tabular}
        &
        $\square$\hspace{5pt}$\blacksquare$\hspace{5pt}$\square$\hspace{5pt}$\square$\hspace{5pt}$\square$& 
        
        $\blacksquare$\hspace{5pt}$\square$& 
        
        $\blacksquare$\hspace{5pt}$\blacksquare$\hspace{5pt}$\square$  & 
        
        $\blacksquare$\hspace{5pt}$\blacksquare$\hspace{5pt}$\square$& 
        
        $\blacksquare$\hspace{5pt}$\square$\hspace{5pt}$\square$\hspace{5pt}$\square$\hspace{5pt}$\blacksquare$\hspace{5pt} & 
        
        $\square$\hspace{5pt}$\square$\hspace{5pt}$\square$\hspace{5pt}$\square$\hspace{5pt}$\square$\hspace{5pt}$\blacksquare$\hspace{5pt}$\square$\hspace{5pt}$\square$ & 
        
        $\square$\hspace{5pt}$\blacksquare$\hspace{5pt} & 
        
        $\blacksquare$\hspace{5pt}$\square$\hspace{5pt}$\square$\hspace{5pt}$\square$\hspace{5pt}$\blacksquare$\hspace{5pt}$\square$\hspace{5pt}$\square$\hspace{5pt}$\square$\hspace{5pt}$\blacksquare$\hspace{5pt}$\square$ &
        
        $\blacksquare$\hspace{5pt}$\square$\hspace{5pt}$\square$\hspace{5pt}$\square$\hspace{5pt}$\square$\hspace{5pt}$\square$         
        
        \\
        
        \begin{tabular}[c]{@{}l@{}} Shejwalkar and Houmansadr~\cite{shejwalkar2021manipulating} \end{tabular}
        &
        $\square$\hspace{5pt}$\blacksquare$\hspace{5pt}$\square$\hspace{5pt}$\square$\hspace{5pt}$\square$& 
        
        $\blacksquare$\hspace{5pt}$\square$& 
        
        $\blacksquare$\hspace{5pt}$\blacksquare$\hspace{5pt}$\square$  & 
        
        $\blacksquare$\hspace{5pt}$\blacksquare$\hspace{5pt}$\square$& 
        
        $\blacksquare$\hspace{5pt}$\square$\hspace{5pt}$\square$\hspace{5pt}$\square$\hspace{5pt}$\square$\hspace{5pt} & 
        
        $\square$\hspace{5pt}$\square$\hspace{5pt}$\square$\hspace{5pt}$\square$\hspace{5pt}$\square$\hspace{5pt}$\blacksquare$\hspace{5pt}$\square$\hspace{5pt}$\square$ & 
        
        $\square$\hspace{5pt}$\blacksquare$\hspace{5pt} & 
        
        $\blacksquare$\hspace{5pt}$\square$\hspace{5pt}$\square$\hspace{5pt}$\square$\hspace{5pt}$\square$\hspace{5pt}$\blacksquare$\hspace{5pt}$\square$\hspace{5pt}$\blacksquare$\hspace{5pt}$\blacksquare$\hspace{5pt}$\square$ &
        
        $\square$\hspace{5pt}$\square$\hspace{5pt}$\square$\hspace{5pt}$\square$\hspace{5pt}$\square$\hspace{5pt}$\blacksquare$         
        
        \\
        
        \begin{tabular}[c]{@{}l@{}} Tomsett \textit{et al.}~\cite{Tomsett}  \end{tabular}
        &
        $\square$\hspace{5pt}$\blacksquare$\hspace{5pt}$\square$\hspace{5pt}$\square$\hspace{5pt}$\square$& 
        
        $\blacksquare$\hspace{5pt}$\square$& 
        
        $\blacksquare$\hspace{5pt}$\square$\hspace{5pt}$\square$  & 
        
        $\blacksquare$\hspace{5pt}$\square$\hspace{5pt}$\square$& 
        
        $\square$\hspace{5pt}$\square$\hspace{5pt}$\square$\hspace{5pt}$\square$\hspace{5pt}$\blacksquare$\hspace{5pt} & 
        
        $\square$\hspace{5pt}$\blacksquare$\hspace{5pt}$\square$\hspace{5pt}$\square$\hspace{5pt}$\square$\hspace{5pt}$\square$\hspace{5pt}$\square$\hspace{5pt}$\square$ & 
        
        $\square$\hspace{5pt}$\blacksquare$ & 
        
        $\blacksquare$\hspace{5pt}$\square$\hspace{5pt}$\square$\hspace{5pt}$\square$\hspace{5pt}$\square$\hspace{5pt}$\square$\hspace{5pt}$\square$\hspace{5pt}$\square$\hspace{5pt}$\square$\hspace{5pt}$\square$ &
        
        $\blacksquare$\hspace{5pt}$\square$\hspace{5pt}$\square$\hspace{5pt}$\square$\hspace{5pt}$\square$\hspace{5pt}$\square$

        \\
        
        \begin{tabular}[c]{@{}l@{}} Chen \textit{et al.}~\cite{Chen2021} \end{tabular}
        &
        $\square$\hspace{5pt}$\blacksquare$\hspace{5pt}$\square$\hspace{5pt}$\square$\hspace{5pt}$\square$& 
        
        $\blacksquare$\hspace{5pt}$\square$& 
        
        $\blacksquare$\hspace{5pt}$\square$\hspace{5pt}$\square$  & 
        
        $\blacksquare$\hspace{5pt}$\square$\hspace{5pt}$\square$& 
        
        $\blacksquare$\hspace{5pt}$\square$\hspace{5pt}$\square$\hspace{5pt}$\blacksquare$\hspace{5pt}$\blacksquare$\hspace{5pt} & 
        
        $\square$\hspace{5pt}$\square$\hspace{5pt}$\square$\hspace{5pt}$\square$\hspace{5pt}$\square$\hspace{5pt}$\square$\hspace{5pt}$\square$\hspace{5pt}$\blacksquare$ & 
        
        $\square$\hspace{5pt}$\blacksquare$ & 
        
        $\blacksquare$\hspace{5pt}$\square$\hspace{5pt}$\square$\hspace{5pt}$\square$\hspace{5pt}$\square$\hspace{5pt}$\square$\hspace{5pt}$\square$\hspace{5pt}$\square$\hspace{5pt}$\square$\hspace{5pt}$\square$ &
        
        $\square$\hspace{5pt}$\square$\hspace{5pt}$\square$\hspace{5pt}$\square$\hspace{5pt}$\square$\hspace{5pt}$\blacksquare$

        \\
        
        \begin{tabular}[c]{@{}l@{}} Fang \textit{et al.}~\cite{Fang2020} \end{tabular}
        &
        $\square$\hspace{5pt}$\blacksquare$\hspace{5pt}$\square$\hspace{5pt}$\square$\hspace{5pt}$\square$& 
        
        $\blacksquare$\hspace{5pt}$\square$& 
        
        $\blacksquare$\hspace{5pt}$\blacksquare$\hspace{5pt}$\square$  & 
        
        $\blacksquare$\hspace{5pt}$\blacksquare$\hspace{5pt}$\square$& 
        
        $\blacksquare$\hspace{5pt}$\square$\hspace{5pt}$\square$\hspace{5pt}$\square$\hspace{5pt}$\square$\hspace{5pt} & 
        
        $\square$\hspace{5pt}$\square$\hspace{5pt}$\square$\hspace{5pt}$\square$\hspace{5pt}$\square$\hspace{5pt}$\blacksquare$\hspace{5pt}$\square$\hspace{5pt}$\square$ & 
        
        $\square$\hspace{5pt}$\blacksquare$ & 
        
        $\square$\hspace{5pt}$\blacksquare$\hspace{5pt}$\square$\hspace{5pt}$\square$\hspace{5pt}$\blacksquare$\hspace{5pt}$\square$\hspace{5pt}$\square$\hspace{5pt}$\square$\hspace{5pt}$\square$\hspace{5pt}$\square$ &
        
        $\square$\hspace{5pt}$\square$\hspace{5pt}$\square$\hspace{5pt}$\square$\hspace{5pt}$\square$\hspace{5pt}$\blacksquare$

        \\

        \begin{tabular}[c]{@{}l@{}} Nuding and Mayer \cite{Nuding}\dag \end{tabular}
        &
        $\square$\hspace{5pt}$\blacksquare$\hspace{5pt}$\square$\hspace{5pt}$\square$\hspace{5pt}$\square$& 
        
        $\blacksquare$\hspace{5pt}$\square$& 
        
        $\square$\hspace{5pt}$\square$\hspace{5pt}$\blacksquare$  & 
        
        $\blacksquare$\hspace{5pt}$\square$\hspace{5pt}$\square$& 
        
        $\blacksquare$\hspace{5pt}$\square$\hspace{5pt}$\square$\hspace{5pt}$\square$\hspace{5pt}$\square$\hspace{5pt} & 
        
        $\square$\hspace{5pt}$\square$\hspace{5pt}$\square$\hspace{5pt}$\square$\hspace{5pt}$\square$\hspace{5pt}$\square$\hspace{5pt}$\square$\hspace{5pt}$\blacksquare$ & 
        
        $\square$\hspace{5pt}$\blacksquare$ & 
        
        $\blacksquare$\hspace{5pt}$\square$\hspace{5pt}$\square$\hspace{5pt}$\square$\hspace{5pt}$\square$\hspace{5pt}$\square$\hspace{5pt}$\square$\hspace{5pt}$\square$\hspace{5pt}$\square$\hspace{5pt}$\square$ &
        
        $\square$\hspace{5pt}$\square$\hspace{5pt}$\square$\hspace{5pt}$\square$\hspace{5pt}$\square$\hspace{5pt}$\blacksquare$

        \\
        
        \begin{tabular}[c]{@{}l@{}} Sun \textit{et al.}~\cite{Sun}\dag \end{tabular}
        &
        $\square$\hspace{5pt}$\blacksquare$\hspace{5pt}$\square$\hspace{5pt}$\square$\hspace{5pt}$\square$& 
        
        $\blacksquare$\hspace{5pt}$\square$& 
        
        $\blacksquare$\hspace{5pt}$\blacksquare$\hspace{5pt}$\square$  & 
        
        $\blacksquare$\hspace{5pt}$\square$\hspace{5pt}$\square$& 
        
        $\blacksquare$\hspace{5pt}$\square$\hspace{5pt}$\square$\hspace{5pt}$\square$\hspace{5pt}$\square$\hspace{5pt} & 
        
        $\square$\hspace{5pt}$\blacksquare$\hspace{5pt}$\square$\hspace{5pt}$\square$\hspace{5pt}$\blacksquare$\hspace{5pt}$\square$\hspace{5pt}$\square$\hspace{5pt}$\square$ & 
        
        $\blacksquare$\hspace{5pt}$\square$ & 
        
        $\blacksquare$\hspace{5pt}$\square$\hspace{5pt}$\square$\hspace{5pt}$\square$\hspace{5pt}$\square$\hspace{5pt}$\square$\hspace{5pt}$\square$\hspace{5pt}$\square$\hspace{5pt}$\square$\hspace{5pt}$\square$ &
        
        $\square$\hspace{5pt}$\blacksquare$\hspace{5pt}$\square$\hspace{5pt}$\square$\hspace{5pt}$\square$\hspace{5pt}$\square$

        \\
        
        \begin{tabular}[c]{@{}l@{}} Bagdasaryan \textit{et al.}~\cite{Bagdasaryan2020}\dag \end{tabular}
        &
        $\square$\hspace{5pt}$\blacksquare$\hspace{5pt}$\square$\hspace{5pt}$\square$\hspace{5pt}$\square$& 
        
        $\blacksquare$\hspace{5pt}$\square$& 
        
        $\square$\hspace{5pt}$\blacksquare$\hspace{5pt}$\square$  & 
        
        $\blacksquare$\hspace{5pt}$\square$\hspace{5pt}$\blacksquare$& 
        
        $\blacksquare$\hspace{5pt}$\square$\hspace{5pt}$\square$\hspace{5pt}$\square$\hspace{5pt}$\square$\hspace{5pt} & 
        
        $\square$\hspace{5pt}$\square$\hspace{5pt}$\square$\hspace{5pt}$\square$\hspace{5pt}$\square$\hspace{5pt}$\square$\hspace{5pt}$\square$\hspace{5pt}$\blacksquare$ & 
        
        $\blacksquare$\hspace{5pt}$\square$ & 
        
        $\square$\hspace{5pt}$\blacksquare$\hspace{5pt}$\square$\hspace{5pt}$\blacksquare$\hspace{5pt}$\square$\hspace{5pt}$\square$\hspace{5pt}$\square$\hspace{5pt}$\square$\hspace{5pt}$\square$\hspace{5pt}$\square$ &
        
        $\blacksquare$\hspace{5pt}$\square$\hspace{5pt}$\square$\hspace{5pt}$\square$\hspace{5pt}$\square$\hspace{5pt}$\square$

        \\

        \begin{tabular}[c]{@{}l@{}} Xie \textit{et al.}~\cite{Xie}\dag \end{tabular}
        &
        $\square$\hspace{5pt}$\blacksquare$\hspace{5pt}$\square$\hspace{5pt}$\square$\hspace{5pt}$\square$& 
        
        $\blacksquare$\hspace{5pt}$\square$& 
        
        $\square$\hspace{5pt}$\blacksquare$\hspace{5pt}$\square$  & 
        
        $\blacksquare$\hspace{5pt}$\blacksquare$\hspace{5pt}$\square$& 
        
        $\square$\hspace{5pt}$\blacksquare$\hspace{5pt}$\square$\hspace{5pt}$\square$\hspace{5pt}$\square$\hspace{5pt} & 
        
        $\square$\hspace{5pt}$\square$\hspace{5pt}$\square$\hspace{5pt}$\square$\hspace{5pt}$\square$\hspace{5pt}$\blacksquare$\hspace{5pt}$\square$\hspace{5pt}$\square$ & 
        
        $\blacksquare$\hspace{5pt}$\square$ & 
        
        $\blacksquare$\hspace{5pt}$\blacksquare$\hspace{5pt}$\square$\hspace{5pt}$\square$\hspace{5pt}$\square$\hspace{5pt}$\square$\hspace{5pt}$\square$\hspace{5pt}$\square$\hspace{5pt}$\blacksquare$\hspace{5pt}$\square$ &
        
        $\blacksquare$\hspace{5pt}$\square$\hspace{5pt}$\square$\hspace{5pt}$\square$\hspace{5pt}$\square$\hspace{5pt}$\square$

        \\
        
        \begin{tabular}[c]{@{}l@{}} Zhou \textit{et al.}~\cite{Zhou2021}\dag \end{tabular}
        &
        $\square$\hspace{5pt}$\blacksquare$\hspace{5pt}$\square$\hspace{5pt}$\square$\hspace{5pt}$\square$& 
        
        $\blacksquare$\hspace{5pt}$\square$& 
        
        $\square$\hspace{5pt}$\blacksquare$\hspace{5pt}$\square$  & 
        
        $\blacksquare$\hspace{5pt}$\square$\hspace{5pt}$\square$& 
        
        $\square$\hspace{5pt}$\blacksquare$\hspace{5pt}$\square$\hspace{5pt}$\square$\hspace{5pt}$\square$\hspace{5pt} & 
        
        $\square$\hspace{5pt}$\square$\hspace{5pt}$\square$\hspace{5pt}$\square$\hspace{5pt}$\square$\hspace{5pt}$\blacksquare$\hspace{5pt}$\square$\hspace{5pt}$\square$ & 
        
        $\square$\hspace{5pt}$\blacksquare$ & 
        
        $\square$\hspace{5pt}$\blacksquare$\hspace{5pt}$\blacksquare$\hspace{5pt}$\square$\hspace{5pt}$\square$\hspace{5pt}$\square$\hspace{5pt}$\square$\hspace{5pt}$\square$\hspace{5pt}$\square$\hspace{5pt}$\square$ &
        
        $\blacksquare$\hspace{5pt}$\square$\hspace{5pt}$\square$\hspace{5pt}$\square$\hspace{5pt}$\square$\hspace{5pt}$\square$

        \\  
        
        \begin{tabular}[c]{@{}l@{}} Huang \cite{Huang}\dag \end{tabular}
        &
        $\square$\hspace{5pt}$\blacksquare$\hspace{5pt}$\square$\hspace{5pt}$\square$\hspace{5pt}$\square$& 
        
        $\blacksquare$\hspace{5pt}$\square$& 
        
        $\blacksquare$\hspace{5pt}$\square$\hspace{5pt}$\square$  & 
        
        $\blacksquare$\hspace{5pt}$\square$\hspace{5pt}$\square$& 
        
        $\blacksquare$\hspace{5pt}$\square$\hspace{5pt}$\square$\hspace{5pt}$\square$\hspace{5pt}$\square$\hspace{5pt} & 
        
        $\square$\hspace{5pt}$\square$\hspace{5pt}$\square$\hspace{5pt}$\square$\hspace{5pt}$\square$\hspace{5pt}$\square$\hspace{5pt}$\square$\hspace{5pt}$\blacksquare$ & 
        
        $\square$\hspace{5pt}$\blacksquare$ & 
        
        $\square$\hspace{5pt}$\blacksquare$\hspace{5pt}$\blacksquare$\hspace{5pt}$\square$\hspace{5pt}$\square$\hspace{5pt}$\square$\hspace{5pt}$\blacksquare$\hspace{5pt}$\square$\hspace{5pt}$\square$\hspace{5pt}$\square$ &
        
        $\blacksquare$\hspace{5pt}$\square$\hspace{5pt}$\square$\hspace{5pt}$\square$\hspace{5pt}$\square$\hspace{5pt}$\square$

        \\   
        
        \begin{tabular}[c]{@{}l@{}} Zhang \textit{et al.}~\cite{Zhang2020a}\dag \end{tabular}
        &
        $\square$\hspace{5pt}$\blacksquare$\hspace{5pt}$\square$\hspace{5pt}$\square$\hspace{5pt}$\square$& 
        
        $\blacksquare$\hspace{5pt}$\square$& 
        
        $\square$\hspace{5pt}$\blacksquare$\hspace{5pt}$\square$  & 
        
        $\blacksquare$\hspace{5pt}$\square$\hspace{5pt}$\square$& 
        
        $\blacksquare$\hspace{5pt}$\blacksquare$\hspace{5pt}$\square$\hspace{5pt}$\square$\hspace{5pt}$\square$\hspace{5pt} & 
        
        $\square$\hspace{5pt}$\square$\hspace{5pt}$\square$\hspace{5pt}$\square$\hspace{5pt}$\square$\hspace{5pt}$\square$\hspace{5pt}$\square$\hspace{5pt}$\blacksquare$ & 
        
        $\square$\hspace{5pt}$\blacksquare$ & 
        
        $\blacksquare$\hspace{5pt}$\square$\hspace{5pt}$\square$\hspace{5pt}$\square$\hspace{5pt}$\square$\hspace{5pt}$\square$\hspace{5pt}$\square$\hspace{5pt}$\square$\hspace{5pt}$\square$\hspace{5pt}$\square$ &
        
        $\blacksquare$\hspace{5pt}$\square$\hspace{5pt}$\square$\hspace{5pt}$\square$\hspace{5pt}$\square$\hspace{5pt}$\square$

        \\
        
        \begin{tabular}[c]{@{}l@{}} Chen \textit{et al.}~\cite{Chen} \end{tabular}
        &
        $\square$\hspace{5pt}$\square$\hspace{5pt}$\blacksquare$\hspace{5pt}$\square$\hspace{5pt}$\square$& 
        
        $\blacksquare$\hspace{5pt}$\square$& 
        
        $\square$\hspace{5pt}$\square$\hspace{5pt}$\blacksquare$  & 
        
        $\blacksquare$\hspace{5pt}$\square$\hspace{5pt}$\square$& 
        
        $\blacksquare$\hspace{5pt}$\square$\hspace{5pt}$\square$\hspace{5pt}$\square$\hspace{5pt}$\square$\hspace{5pt} & 
        
        $\square$\hspace{5pt}$\square$\hspace{5pt}$\square$\hspace{5pt}$\square$\hspace{5pt}$\square$\hspace{5pt}$\square$\hspace{5pt}$\square$\hspace{5pt}$\blacksquare$ & 
        
        $\square$\hspace{5pt}$\blacksquare$ & 
        
        $\blacksquare$\hspace{5pt}$\square$\hspace{5pt}$\square$\hspace{5pt}$\square$\hspace{5pt}$\square$\hspace{5pt}$\square$\hspace{5pt}$\square$\hspace{5pt}$\square$\hspace{5pt}$\square$\hspace{5pt}$\square$ &
        
        $\blacksquare$\hspace{5pt}$\blacksquare$\hspace{5pt}$\square$\hspace{5pt}$\square$\hspace{5pt}$\square$\hspace{5pt}$\square$

        \\        
        
        \begin{tabular}[c]{@{}l@{}} Nasr \textit{et al.}~\cite{nasr2019comprehensive} \end{tabular}
        &
        $\square$\hspace{5pt}$\square$\hspace{5pt}$\blacksquare$\hspace{5pt}$\square$\hspace{5pt}$\square$& 
        
        $\blacksquare$\hspace{5pt}$\square$& 
        
        $\blacksquare$\hspace{5pt}$\square$\hspace{5pt}$\square$  & 
        
        $\blacksquare$\hspace{5pt}$\blacksquare$\hspace{5pt}$\square$& 
        
        $\blacksquare$\hspace{5pt}$\square$\hspace{5pt}$\square$\hspace{5pt}$\square$\hspace{5pt}$\square$\hspace{5pt} & 
        
        $\square$\hspace{5pt}$\square$\hspace{5pt}$\square$\hspace{5pt}$\square$\hspace{5pt}$\square$\hspace{5pt}$\square$\hspace{5pt}$\square$\hspace{5pt}$\blacksquare$ & 
        
        $\square$\hspace{5pt}$\blacksquare$ & 
        
        $\square$\hspace{5pt}$\blacksquare$\hspace{5pt}$\square$\hspace{5pt}$\square$\hspace{5pt}$\square$\hspace{5pt}$\blacksquare$\hspace{5pt}$\blacksquare$\hspace{5pt}$\square$\hspace{5pt}$\square$\hspace{5pt}$\square$ &
        
        $\blacksquare$\hspace{5pt}$\square$\hspace{5pt}$\square$\hspace{5pt}$\square$\hspace{5pt}$\square$\hspace{5pt}$\square$

        \\   
        
        \begin{tabular}[c]{@{}l@{}} Luo \textit{et al.}~\cite{Luo} \end{tabular}
        &
        $\square$\hspace{5pt}$\square$\hspace{5pt}$\blacksquare$\hspace{5pt}$\square$\hspace{5pt}$\square$& 
        
        $\blacksquare$\hspace{5pt}$\square$& 
        
        $\blacksquare$\hspace{5pt}$\square$\hspace{5pt}$\square$  & 
        
        $\square$\hspace{5pt}$\blacksquare$\hspace{5pt}$\square$& 
        
        $\square$\hspace{5pt}$\square$\hspace{5pt}$\blacksquare$\hspace{5pt}$\square$\hspace{5pt}$\blacksquare$\hspace{5pt} & 
        
        $\square$\hspace{5pt}$\square$\hspace{5pt}$\square$\hspace{5pt}$\square$\hspace{5pt}$\square$\hspace{5pt}$\square$\hspace{5pt}$\square$\hspace{5pt}$\blacksquare$ & 
        
        $\square$\hspace{5pt}$\blacksquare$ & 
        
        $\square$\hspace{5pt}$\square$\hspace{5pt}$\square$\hspace{5pt}$\square$\hspace{5pt}$\blacksquare$\hspace{5pt}$\square$\hspace{5pt}$\square$\hspace{5pt}$\square$\hspace{5pt}$\blacksquare$\hspace{5pt}$\square$ &
        
        $\blacksquare$\hspace{5pt}$\square$\hspace{5pt}$\blacksquare$\hspace{5pt}$\square$\hspace{5pt}$\square$\hspace{5pt}$\square$

        \\       
        
        \begin{tabular}[c]{@{}l@{}} Zhang \textit{et al.}~\cite{Zhang2020} \end{tabular}
        &
        $\square$\hspace{5pt}$\square$\hspace{5pt}$\blacksquare$\hspace{5pt}$\square$\hspace{5pt}$\square$& 
        
        $\blacksquare$\hspace{5pt}$\square$& 
        
        $\blacksquare$\hspace{5pt}$\square$\hspace{5pt}$\square$  & 
        
        $\blacksquare$\hspace{5pt}$\square$\hspace{5pt}$\square$& 
        
        $\blacksquare$\hspace{5pt}$\square$\hspace{5pt}$\square$\hspace{5pt}$\square$\hspace{5pt}$\square$\hspace{5pt} & 
        
        $\square$\hspace{5pt}$\square$\hspace{5pt}$\square$\hspace{5pt}$\square$\hspace{5pt}$\square$\hspace{5pt}$\square$\hspace{5pt}$\square$\hspace{5pt}$\blacksquare$ & 
        
        $\square$\hspace{5pt}$\blacksquare$ & 
        
        $\blacksquare$\hspace{5pt}$\blacksquare$\hspace{5pt}$\square$\hspace{5pt}$\square$\hspace{5pt}$\square$\hspace{5pt}$\square$\hspace{5pt}$\square$\hspace{5pt}$\square$\hspace{5pt}$\square$\hspace{5pt}$\square$ &
        
        $\blacksquare$\hspace{5pt}$\square$\hspace{5pt}$\square$\hspace{5pt}$\square$\hspace{5pt}$\square$\hspace{5pt}$\square$

        \\     
        
        \begin{tabular}[c]{@{}l@{}} Li \textit{et al.}~\cite{Li2021a} \end{tabular}
        &
        $\square$\hspace{5pt}$\square$\hspace{5pt}$\blacksquare$\hspace{5pt}$\blacksquare$\hspace{5pt}$\square$& 
        
        $\blacksquare$\hspace{5pt}$\square$& 
        
        $\blacksquare$\hspace{5pt}$\blacksquare$\hspace{5pt}$\square$  & 
        
        $\blacksquare$\hspace{5pt}$\blacksquare$\hspace{5pt}$\square$& 
        
        $\square$\hspace{5pt}$\square$\hspace{5pt}$\square$\hspace{5pt}$\square$\hspace{5pt}$\blacksquare$\hspace{5pt} & 
        
        $\square$\hspace{5pt}$\square$\hspace{5pt}$\blacksquare$\hspace{5pt}$\square$\hspace{5pt}$\square$\hspace{5pt}$\square$\hspace{5pt}$\square$\hspace{5pt}$\square$ & 
        
        $\square$\hspace{5pt}$\blacksquare$ & 
        
        $\square$\hspace{5pt}$\square$\hspace{5pt}$\square$\hspace{5pt}$\square$\hspace{5pt}$\blacksquare$\hspace{5pt}$\square$\hspace{5pt}$\square$\hspace{5pt}$\square$\hspace{5pt}$\square$\hspace{5pt}$\square$ &
        
        $\square$\hspace{5pt}$\square$\hspace{5pt}$\square$\hspace{5pt}$\blacksquare$\hspace{5pt}$\blacksquare$\hspace{5pt}$\square$

        \\        
        
        \begin{tabular}[c]{@{}l@{}} Song \textit{et al.}~\cite{Song2020} \end{tabular}
        &
        $\square$\hspace{5pt}$\square$\hspace{5pt}$\square$\hspace{5pt}$\blacksquare$\hspace{5pt}$\square$& 
        
        $\blacksquare$\hspace{5pt}$\square$& 
        
        $\square$\hspace{5pt}$\blacksquare$\hspace{5pt}$\square$  & 
        
        $\blacksquare$\hspace{5pt}$\square$\hspace{5pt}$\square$& 
        
        $\blacksquare$\hspace{5pt}$\square$\hspace{5pt}$\square$\hspace{5pt}$\square$\hspace{5pt}$\square$\hspace{5pt} & 
        
        $\square$\hspace{5pt}$\square$\hspace{5pt}$\square$\hspace{5pt}$\square$\hspace{5pt}$\square$\hspace{5pt}$\square$\hspace{5pt}$\square$\hspace{5pt}$\blacksquare$ & 
        
        $\square$\hspace{5pt}$\blacksquare$ & 
        
        $\blacksquare$\hspace{5pt}$\square$\hspace{5pt}$\square$\hspace{5pt}$\square$\hspace{5pt}$\square$\hspace{5pt}$\square$\hspace{5pt}$\square$\hspace{5pt}$\square$\hspace{5pt}$\square$\hspace{5pt}$\square$ &
        
        $\square$\hspace{5pt}$\blacksquare$\hspace{5pt}$\square$\hspace{5pt}$\square$\hspace{5pt}$\square$\hspace{5pt}$\square$

        \\       
        
        \begin{tabular}[c]{@{}l@{}} Ren \textit{et al.}~\cite{Ren} \end{tabular}
        &
        $\square$\hspace{5pt}$\square$\hspace{5pt}$\square$\hspace{5pt}$\blacksquare$\hspace{5pt}$\square$& 
        
        $\blacksquare$\hspace{5pt}$\square$& 
        
        $\blacksquare$\hspace{5pt}$\square$\hspace{5pt}$\square$  & 
        
        $\blacksquare$\hspace{5pt}$\blacksquare$\hspace{5pt}$\square$& 
        
        $\blacksquare$\hspace{5pt}$\square$\hspace{5pt}$\blacksquare$\hspace{5pt}$\square$\hspace{5pt}$\blacksquare$\hspace{5pt} & 
        
        $\square$\hspace{5pt}$\square$\hspace{5pt}$\square$\hspace{5pt}$\square$\hspace{5pt}$\blacksquare$\hspace{5pt}$\square$\hspace{5pt}$\square$\hspace{5pt}$\square$ & 
        
        $\blacksquare$\hspace{5pt}$\square$ & 
        
        $\square$\hspace{5pt}$\blacksquare$\hspace{5pt}$\blacksquare$\hspace{5pt}$\square$\hspace{5pt}$\square$\hspace{5pt}$\square$\hspace{5pt}$\blacksquare$\hspace{5pt}$\square$\hspace{5pt}$\square$\hspace{5pt}$\square$ &
        
        $\blacksquare$\hspace{5pt}$\square$\hspace{5pt}$\square$\hspace{5pt}$\square$\hspace{5pt}$\square$\hspace{5pt}$\square$

        \\    
        
        \begin{tabular}[c]{@{}l@{}} Wu \cite{wu2020plfg} \end{tabular}
        &
        $\square$\hspace{5pt}$\square$\hspace{5pt}$\square$\hspace{5pt}$\blacksquare$\hspace{5pt}$\square$& 
        
        $\blacksquare$\hspace{5pt}$\square$& 
        
        $\square$\hspace{5pt}$\square$\hspace{5pt}$\blacksquare$  & 
        
        $\blacksquare$\hspace{5pt}$\square$\hspace{5pt}$\square$& 
        
        $\blacksquare$\hspace{5pt}$\square$\hspace{5pt}$\square$\hspace{5pt}$\square$\hspace{5pt}$\square$\hspace{5pt} & 
        
        $\square$\hspace{5pt}$\square$\hspace{5pt}$\square$\hspace{5pt}$\square$\hspace{5pt}$\square$\hspace{5pt}$\square$\hspace{5pt}$\square$\hspace{5pt}$\blacksquare$ & 
        
        $\square$\hspace{5pt}$\blacksquare$ & 
        
        $\blacksquare$\hspace{5pt}$\square$\hspace{5pt}$\square$\hspace{5pt}$\square$\hspace{5pt}$\square$\hspace{5pt}$\square$\hspace{5pt}$\square$\hspace{5pt}$\square$\hspace{5pt}$\square$\hspace{5pt}$\square$ &
        
        $\square$\hspace{5pt}$\blacksquare$\hspace{5pt}$\square$\hspace{5pt}$\square$\hspace{5pt}$\square$\hspace{5pt}$\square$

        \\        
        
        \begin{tabular}[c]{@{}l@{}} Fraboni \textit{et al.}~\cite{fraboni2021free} \end{tabular}
        &
        $\square$\hspace{5pt}$\square$\hspace{5pt}$\square$\hspace{5pt}$\square$\hspace{5pt}$\blacksquare$& 
        
        $\square$\hspace{5pt}$\blacksquare$& 
        
        $\blacksquare$\hspace{5pt}$\blacksquare$\hspace{5pt}$\square$  & 
        
        $\blacksquare$\hspace{5pt}$\square$\hspace{5pt}$\blacksquare$& 
        
        $\blacksquare$\hspace{5pt}$\square$\hspace{5pt}$\square$\hspace{5pt}$\square$\hspace{5pt}$\blacksquare$\hspace{5pt} & 
        
        $\square$\hspace{5pt}$\square$\hspace{5pt}$\square$\hspace{5pt}$\square$\hspace{5pt}$\square$\hspace{5pt}$\blacksquare$\hspace{5pt}$\square$\hspace{5pt}$\square$ & 
        
        $\blacksquare$\hspace{5pt}$\square$ & 
        
        $\blacksquare$\hspace{5pt}$\square$\hspace{5pt}$\square$\hspace{5pt}$\blacksquare$\hspace{5pt}$\blacksquare$\hspace{5pt}$\square$\hspace{5pt}$\square$\hspace{5pt}$\square$\hspace{5pt}$\square$\hspace{5pt}$\square$ &
        
        $\blacksquare$\hspace{5pt}$\square$\hspace{5pt}$\square$\hspace{5pt}$\square$\hspace{5pt}$\square$\hspace{5pt}$\square$        
        
        \\        
        
        \begin{tabular}[c]{@{}l@{}} Wan \textit{et al.}~\cite{Wan2021} \end{tabular}
        &
        $\square$\hspace{5pt}$\square$\hspace{5pt}$\square$\hspace{5pt}$\square$\hspace{5pt}$\blacksquare$& 
        
        $\blacksquare$\hspace{5pt}$\square$& 
        
        $\square$\hspace{5pt}$\square$\hspace{5pt}$\blacksquare$  & 
        
        $\blacksquare$\hspace{5pt}$\square$\hspace{5pt}$\square$& 
        
        $\blacksquare$\hspace{5pt}$\square$\hspace{5pt}$\square$\hspace{5pt}$\square$\hspace{5pt}$\square$\hspace{5pt} & 
        
        $\square$\hspace{5pt}$\blacksquare$\hspace{5pt}$\square$\hspace{5pt}$\square$\hspace{5pt}$\square$\hspace{5pt}$\square$\hspace{5pt}$\square$\hspace{5pt}$\square$ & 
        
        $\square$\hspace{5pt}$\blacksquare$ & 
        
        $\blacksquare$\hspace{5pt}$\square$\hspace{5pt}$\square$\hspace{5pt}$\square$\hspace{5pt}$\square$\hspace{5pt}$\square$\hspace{5pt}$\square$\hspace{5pt}$\square$\hspace{5pt}$\square$\hspace{5pt}$\square$ &
        
        $\square$\hspace{5pt}$\blacksquare$\hspace{5pt}$\square$\hspace{5pt}$\square$\hspace{5pt}$\square$\hspace{5pt}$\square$

        \\        
        
        \begin{tabular}[c]{@{}l@{}} Qureshi \textit{et. al}~\cite{qureshi2021performance} \end{tabular}
        &
        $\blacksquare$\hspace{5pt}$\square$\hspace{5pt}$\square$\hspace{5pt}$\square$\hspace{5pt}$\blacksquare$& 
        
        $\blacksquare$\hspace{5pt}$\blacksquare$& 
        
        $\square$\hspace{5pt}$\square$\hspace{5pt}$\blacksquare$  & 
        
        $\square$\hspace{5pt}$\square$\hspace{5pt}$\blacksquare$ & 
        
        $\blacksquare$\hspace{5pt}$\square$\hspace{5pt}$\blacksquare$\hspace{5pt}$\square$\hspace{5pt}$\square$\hspace{5pt} & 
        
        $\square$\hspace{5pt}$\square$\hspace{5pt}$\square$\hspace{5pt}$\square$\hspace{5pt}$\square$\hspace{5pt}$\square$\hspace{5pt}$\square$\hspace{5pt}$\blacksquare$ & 
        
        $\square$\hspace{5pt}$\blacksquare$ & 
        
        $\square$\hspace{5pt}$\square$\hspace{5pt}$\square$\hspace{5pt}$\blacksquare$\hspace{5pt}$\square$\hspace{5pt}$\square$\hspace{5pt}$\square$\hspace{5pt}$\square$\hspace{5pt}$\square$\hspace{5pt}$\square$ &
        
        $\square$\hspace{5pt}$\blacksquare$\hspace{5pt}$\square$\hspace{5pt}$\square$\hspace{5pt}$\square$\hspace{5pt}$\square$

        \\        
        
        \begin{tabular}[c]{@{}l@{}} Hossain \textit{et al.}t~\cite{hossain2021desmp} \end{tabular}
        &
        $\square$\hspace{5pt}$\blacksquare$\hspace{5pt}$\square$\hspace{5pt}$\square$\hspace{5pt}$\square$& 
        
        $\blacksquare$\hspace{5pt}$\square$& 
        
        $\square$\hspace{5pt}$\blacksquare$\hspace{5pt}$\square$  & 
        
        $\blacksquare$\hspace{5pt}$\square$\hspace{5pt}$\square$ & 
        
        $\blacksquare$\hspace{5pt}$\square$\hspace{5pt}$\square$\hspace{5pt}$\square$\hspace{5pt}$\blacksquare$\hspace{5pt} & 
        
        $\square$\hspace{5pt}$\square$\hspace{5pt}$\square$\hspace{5pt}$\square$\hspace{5pt}$\blacksquare$\hspace{5pt}$\square$\hspace{5pt}$\square$\hspace{5pt}$\square$ & 
        
        $\square$\hspace{5pt}$\blacksquare$ & 
        
        $\blacksquare$\hspace{5pt}$\square$\hspace{5pt}$\square$\hspace{5pt}$\square$\hspace{5pt}$\square$\hspace{5pt}$\square$\hspace{5pt}$\square$\hspace{5pt}$\square$\hspace{5pt}$\square$\hspace{5pt}$\square$ &
        
        $\blacksquare$\hspace{5pt}$\square$\hspace{5pt}$\square$\hspace{5pt}$\square$\hspace{5pt}$\square$\hspace{5pt}$\square$

        \\
        
        \begin{tabular}[c]{@{}l@{}} Lyu and Chen \cite{lyu2021novel} \end{tabular}
        &
        $\square$\hspace{5pt}$\blacksquare$\hspace{5pt}$\square$\hspace{5pt}$\square$\hspace{5pt}$\square$& 
        
        $\blacksquare$\hspace{5pt}$\square$& 
        
        $\blacksquare$\hspace{5pt}$\square$\hspace{5pt}$\square$  & 
        
        $\square$\hspace{5pt}$\blacksquare$\hspace{5pt}$\square$ & 
        
        $\blacksquare$\hspace{5pt}$\square$\hspace{5pt}$\square$\hspace{5pt}$\square$\hspace{5pt}$\square$\hspace{5pt} & 
        
        $\square$\hspace{5pt}$\square$\hspace{5pt}$\square$\hspace{5pt}$\square$\hspace{5pt}$\square$\hspace{5pt}$\square$\hspace{5pt}$\square$\hspace{5pt}$\blacksquare$ & 
        
        $\square$\hspace{5pt}$\blacksquare$ & 
        
        $\square$\hspace{5pt}$\square$\hspace{5pt}$\square$\hspace{5pt}$\square$\hspace{5pt}$\square$\hspace{5pt}$\square$\hspace{5pt}$\square$\hspace{5pt}$\square$\hspace{5pt}$\blacksquare$\hspace{5pt}$\square$ &
        
        $\square$\hspace{5pt}$\square$\hspace{5pt}$\square$\hspace{5pt}$\square$\hspace{5pt}$\square$\hspace{5pt}$\blacksquare$

        \\

        \begin{tabular}[c]{@{}l@{}} Hu \textit{et al.}~\cite{hu2021source} \end{tabular}
        &
        $\square$\hspace{5pt}$\square$\hspace{5pt}$\blacksquare$\hspace{5pt}$\blacksquare$\hspace{5pt}$\square$& 
        
        $\blacksquare$\hspace{5pt}$\square$& 
        
        $\blacksquare$\hspace{5pt}$\blacksquare$\hspace{5pt}$\square$  & 
        
        $\blacksquare$\hspace{5pt}$\blacksquare$\hspace{5pt}$\square$ & 
        
        $\square$\hspace{5pt}$\blacksquare$\hspace{5pt}$\square$\hspace{5pt}$\square$\hspace{5pt}$\square$\hspace{5pt} & 
        
        $\square$\hspace{5pt}$\square$\hspace{5pt}$\square$\hspace{5pt}$\square$\hspace{5pt}$\blacksquare$\hspace{5pt}$\square$\hspace{5pt}$\square$\hspace{5pt}$\square$ & 
        
        $\blacksquare$\hspace{5pt}$\square$ & 
        
        $\blacksquare$\hspace{5pt}$\square$\hspace{5pt}$\square$\hspace{5pt}$\square$\hspace{5pt}$\square$\hspace{5pt}$\square$\hspace{5pt}$\square$\hspace{5pt}$\square$\hspace{5pt}$\square$\hspace{5pt}$\blacksquare$ &
        
        $\blacksquare$\hspace{5pt}$\square$\hspace{5pt}$\square$\hspace{5pt}$\square$\hspace{5pt}$\square$\hspace{5pt}$\square$

        \\
        
        \begin{tabular}[c]{@{}l@{}} Sun \textit{et al.}~\cite{sun2021information} \end{tabular}
        &
        $\square$\hspace{5pt}$\square$\hspace{5pt}$\square$\hspace{5pt}$\blacksquare$\hspace{5pt}$\square$& 
        
        $\blacksquare$\hspace{5pt}$\square$& 
        
        $\square$\hspace{5pt}$\square$\hspace{5pt}$\blacksquare$  & 
        
        $\blacksquare$\hspace{5pt}$\square$\hspace{5pt}$\square$ & 
        
        $\blacksquare$\hspace{5pt}$\square$\hspace{5pt}$\square$\hspace{5pt}$\square$\hspace{5pt}$\square$\hspace{5pt} & 
        
        $\square$\hspace{5pt}$\square$\hspace{5pt}$\square$\hspace{5pt}$\square$\hspace{5pt}$\square$\hspace{5pt}$\square$\hspace{5pt}$\square$\hspace{5pt}$\blacksquare$ & 
        
        $\square$\hspace{5pt}$\blacksquare$ & 
        
        $\square$\hspace{5pt}$\square$\hspace{5pt}$\square$\hspace{5pt}$\square$\hspace{5pt}$\square$\hspace{5pt}$\square$\hspace{5pt}$\square$\hspace{5pt}$\square$\hspace{5pt}$\square$\hspace{5pt}$\blacksquare$ &
        
        $\square$\hspace{5pt}$\square$\hspace{5pt}$\square$\hspace{5pt}$\square$\hspace{5pt}$\square$\hspace{5pt}$\blacksquare$

        \\ 
        
        \begin{tabular}[c]{@{}l@{}} Gong \textit{et al.}~\cite{gong2022coordinated}~\dag \end{tabular}
        &
        $\square$\hspace{5pt}$\blacksquare$\hspace{5pt}$\square$\hspace{5pt}$\square$\hspace{5pt}$\square$& 
        
        $\blacksquare$\hspace{5pt}$\square$& 
        
        $\square$\hspace{5pt}$\blacksquare$\hspace{5pt}$\square$  & 
        
        $\blacksquare$\hspace{5pt}$\square$\hspace{5pt}$\square$ & 
        
        $\blacksquare$\hspace{5pt}$\blacksquare$\hspace{5pt}$\square$\hspace{5pt}$\square$\hspace{5pt}$\square$\hspace{5pt} & 
        
        $\square$\hspace{5pt}$\square$\hspace{5pt}$\square$\hspace{5pt}$\square$\hspace{5pt}$\square$\hspace{5pt}$\blacksquare$\hspace{5pt}$\square$\hspace{5pt}$\square$ & 
        
        $\square$\hspace{5pt}$\blacksquare$ & 
        
        $\square$\hspace{5pt}$\blacksquare$\hspace{5pt}$\blacksquare$\hspace{5pt}$\square$\hspace{5pt}$\square$\hspace{5pt}$\square$\hspace{5pt}$\square$\hspace{5pt}$\square$\hspace{5pt}$\square$\hspace{5pt}$\square$ &
        
        $\square$\hspace{5pt}$\square$\hspace{5pt}$\square$\hspace{5pt}$\square$\hspace{5pt}$\square$\hspace{5pt}$\blacksquare$

        \\
      
        \hline

    \end{tabular} 
}
\end{sidewaystable}

%% file: 07-TowardsSecureFL.tex
\section{Towards Secure FL} 
\label{sec:secureFL}

\begin{figure}[!htb]
    \centering
    \includegraphics[width=0.7\columnwidth]{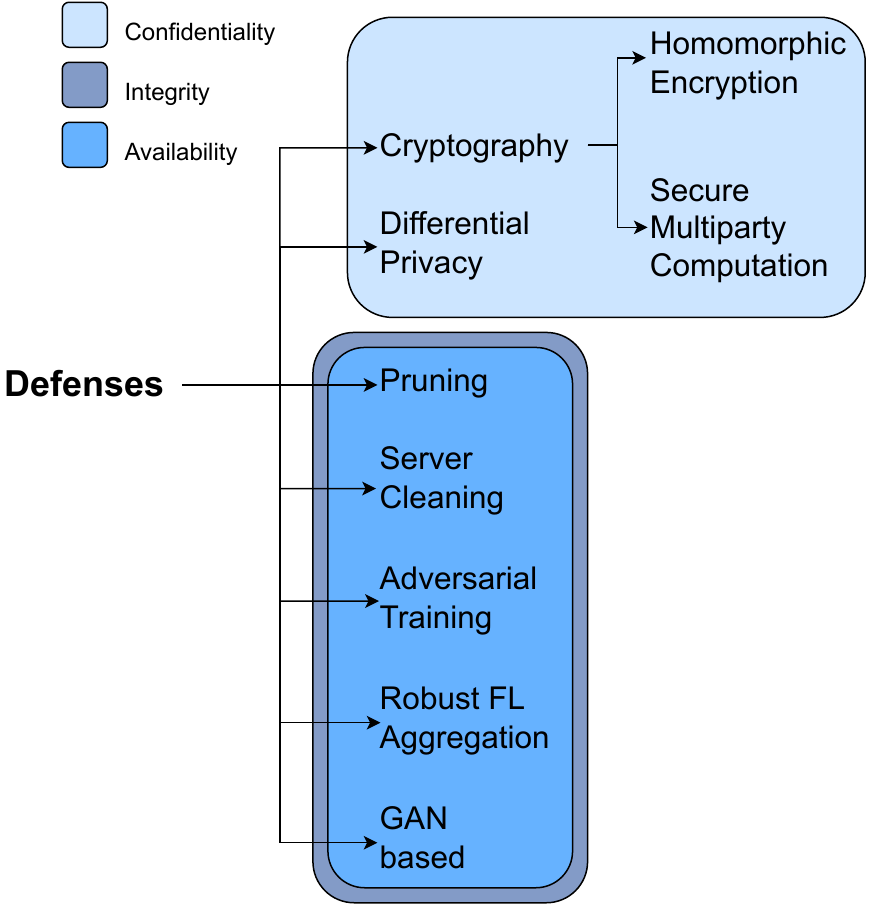}
    \caption{Proposed taxonomy of defenses.}
    \label{fig:taxonomyDef}
\end{figure}

This section explains the SoTA defensive mechanisms and groups them according to the CIA triad. We propose a taxonomy for each defense type by analyzing the SoTA defenses presented in Table~\ref{tab:DefencesTable} and considering the adversarial goal and capabilities introduced in Fig.~\ref{fig:taxonomyDef}.

\newtheorem{challenge}{Challenge}

\subsection{Defending Confidentiality}

\subsubsection{Cryptographic Methods}

Cryptographic Methods (CMs) prevent Inference attacks in FL. For a general use case, the clients or the aggregator perform local or in-server encryption to the updates. Thus, securing the communications between them. This defensive mechanism prevents an attacker from privacy leaks, but it is computationally expensive.

Secure Multi-party Computation (SMC) enabled secret sharing without a trusted third party~\cite{bonawitz2017practical, agrawal2019quotient}. Furthermore, SMC deals with cross-silo settings, where clients may be unreachable, dropouts are expected, and communications expensive. Other authors~\cite{boutet2021mixnn} proposed leveraging a proxy-like Trusted Execution Environment (TEE) that mixes layers from the updates of different clients before sending these to the aggregator. Similarly, Homomorphic Encryption (HE) relies on encrypting data, where mathematical operators over encrypted and non-encrypted data result in the same. Furthermore, HE theoretically does not carry a performance loss on the model convergence~\cite{aono2017privacy}, which enabled its use in decentralized settings~\cite{yang2019federated, kairouz2019advances, paillier1999public, liu2019secure}.

The authors in~\cite{zhang2020batchcrypt, hardy2017private} enabled HE for FL, where the server is \textit{honest-but-curious}. Before training, the server randomly selects a client as a leader. The leader generates and shares key pairs with the rest of the clients. The clients compute gradients locally and encrypt them. Then, they upload those to the server that aggregates the encrypted weights. The aggregator sends the encrypted model back to the clients for another FL round. However, CMs could not provide privacy guarantees since authors in~\cite{Li2021a} successfully performed Inference attacks over encrypted data.

\begin{challenge}
\textbf{\textit{CMs defend against Inference attacks by encrypting data. However, recent attacks evade such defenses by extracting information over encrypted data. Further research is needed to determine CMs viability in FL.}}
\end{challenge}


\subsubsection{Differential Privacy}

\textit{Differential Privacy} is a privacy-preserving mechanism that adds noise to the model for limiting a wide range of attacks~\cite{abadi2016deep, dwork2014algorithmic, mcmahan2017learning}. The core idea is that noise alters the weights of the model. Therefore, an Inference attack, such as Model Inversion, will contain noise, degrading attack efficiency. However, the model accuracy will also drop. Each client adds the noise locally during the training process or in the server by the aggregator. In general, the reviewed papers (see Table~\ref{tab:DefencesTable}) that applied Differential Privacy proved its versatility against a wide variety of Inference attacks. 

Another approach followed by~\cite{sun2020provable, sun2021defending, lee2021defensive} added noise to the updates, similar to~\cite{xiong2021privacy}, which applied Differential Privacy in the server and the clients.
Compressing the model before sharing it with the aggregator is another defensive mechanism that reduces the communications overloads apart from improving confidentiality~\cite{Kerkouche, wei2020framework, li2021adaptive, sun2021soteria}. Likewise, the authors in~\cite{ibitoye2021dipsen} proposed using a Self Normalizing Neural Network, which combined normalizing and Differential Privacy techniques, adding noise per NN layer. The authors in~\cite{luo2020exploiting} developed an anti-GAN mechanism to prevent GAN-based Inference attacks. They leveraged a Wasserstein GAN for distorting the datasets of clients. Thus, an attacker using a GAN-based attack would learn over a wrong data distribution, reducing the attack performance. 

\begin{challenge}
\textbf{\textit{Differential Privacy provides a defense against multiple attacks while degrading the model performance. Further comparisons are required for establishing if Differential Privacy is the most versatile defense mechanism.}}
\end{challenge}

\subsection{Defending Integrity \& Availability}

\subsubsection{Pruning}


Data-filtering is a common technique for preventing Poisoning attacks in ML~\cite{steinhardt2017certified} by removing the poisoned data from the dataset. However, clients' datasets are not accessible to the aggregator under FL settings. While the mentioned defensive mechanisms are effective during training, they do not defend once the model has been poisoned. As a solution, \textbf{post-training defenses} aim to remove the negative effect after training~\cite{goldblum2020dataset}. In classical ML, Neural Cleanse~\cite{wang2019neural} identifies backdoors in the model and reduces its impact. Similarly, fine-pruning~\cite{liu2018fine} works by disabling neurons that have a negative effect. Authors~\cite{Wu} successfully applied Pruning to prevent Poisoning attacks in the context of FL.

\begin{challenge}
\textbf{\textit{Post-training defenses are common in classical ML. Nevertheless, only one proposal uses Pruning against Model Poisoning attacks after our research. It is not clear whether similar post-training defenses can defend against other types of attacks in FL or how the aggregation affects these defensive mechanisms.}}
\end{challenge}


\subsubsection{Server Cleaning}

Server Cleaning, also named Sanitization~\cite{kairouz2019advances}, focuses on filtering malicious updates, similarly to Robust FL Aggregators, where the server discards them, i.e., does not modify the base aggregation algorithm. However, Server Cleaning requires a separated aggregator algorithm. For such purposes, the server might leverage different techniques based on the intrinsic properties of the updates~\cite{goldblum2020dataset}.

Typical approaches used statistical deviation or patterns that differentiate a malicious update. The authors in~\cite{barreno2010security} presented a defense based on RONI, where the server discarded updates that negatively affect the accuracy. Towards calculating such, they proposed training a classification model based on a clean-holdout dataset. Then they compared the classification accuracy between the clean model and the uploaded. If the uploaded model's accuracy was lower than the clean's, it got discarded. This approach required a heterogeneous dataset to generalize properly and not discard benign updates. 

Similarly, the authors in~\cite{rodriguez2020dynamic} dynamically discarded adversarial clients. They adapted the Induced Ordered Weighted Averaging (IOWA) operator for the FL environment. IOWA is a function for weighting the updates of the clients on a distributed system. Similarly, the authors in~\cite{singh2020fair} proposed a self adversarial-detecting system where the clients were clustered and shared relevant features. Other approaches, as in~\cite{Chen2020}, performed clustering on the server for detecting malicious updates. A similar procedure was followed by~\cite{Doku2021} applying Support Vector Machines, dimensionality reduction using Principal Component Analysis~\cite{Tolpegin}, or Digestive Neural Networks~\cite{lee2021digestive}. Similarly, the authors in~\cite{chen2020backdoor} applied Matching Networks for discarding updates based on similarity. Other authors~\cite{Liu2020} used Differential Privacy locally and tested the models' quality on edge, improving communications efficiency while providing privacy guarantees. The authors in~\cite{Sun2020BC} leveraged a Blockchained structure for ensuring data integrity and traceability, where a committee validates updates via Proof-of-Work. For further read in Blockchain-based FL, refer to~\cite{li2021blockchain}.

The authors in~\cite{lin2019free} defended Free-riders attacks by applying a Deep Autoencoding Gaussian Mixture Model (DAGMM). DAGMM is composed of a compression and estimation net, where the updates are passed through the networks, and their distances are measured between dimensions and their standard deviation. Similarly, the authors in~\cite{zong2018deep} applied Autoencoders for the defense. In particular, they proposed a DAGMM, where the encoder performed a dimensionality reduction. Afterward, the remaining dimensions were passed to an estimation network that predicted the likelihood density. The larger the density more likely it was to be a malicious update. 
Similarly, in~\cite{Wan2021}, the authors proposed a density-based anomaly detection. They assigned an anomaly score to each model, where the aggregator weighted them according to the score. Therefore, reducing the adversarial relevance on the global model. Other confidence or score-based anomaly detection have been proposed~\cite{Wan2021, mallah2021untargeted, wang2020model}.

Furthermore, the authors in~\cite{jagielski2018manipulating} suggested training a Linear Regression model to identify data pieces close to legitimate data pieces from a poisoned dataset. Differently, the authors in~\cite{DucNguyen} proposed using the K-means clustering algorithm, leveraging a clean-holdout dataset for removing outliners based on trimmed means.

Also, during the Boosting process in Model Poisoning attacks, the weights were adapted for increasing relevance in the aggregated model~\cite{Sun}, which carried an increase in the $l_n$ norm. Compared with benign updates, the server might define a threshold and discard the updates that do not satisfy it~\cite{varma2021legato}. Likewise, the authors in~\cite{Tomsett} proposed using the Kolmogorov-Smirnov (K-S) test for cleaning the model. The K-S test is used in non-parametric statistics for measuring the distance between two data distributions. They compared the distance between data distributions, removing those different from the majority. However, they only demonstrated its effectiveness under IID settings.

TEE enables a secure area that provides confidentiality and integrity guarantees. The authors in~\cite{mo2020darknetz} applied TEE to centralized ML to train a Deep Neural Network (DNN) and analyze its security. Similarly, the authors in~\cite{mo2021ppfl} used TEE locally for the clients for training and on the server for securing the aggregation. This scheme, as proved, successfully defended against inference-time attacks while also guaranteeing data integrity. TEE ensures that the aggregation procedure or the model's parameters are not haltered.

\begin{challenge}
\textbf{\textit{Server Cleaning methods can defend against the majority of attacks except for Adversarial Examples. However, this method has to be implemented explicitly for defending against a particular threat. Therefore, a further study focusing on generalizing Server Cleaning techniques is needed.}}
\end{challenge}

\subsubsection{Adversarial Training}

As previously introduced, Data Poisoning attacks alter the local dataset. Because of the decentralized setting of FL, Data Cleaning methods were not applicable. Adversarial Training inserts poisoned data on the dataset during the training phase~\cite{shafahi2019adversarial, Shah}. Once trained, the model can differentiate poisoned data. However, this defense requires a large dataset and is computationally challenging. As a solution, the authors in~\cite{Hong} proposed sharing resources, where the most prepared or powerful devices could share the secure model with other clients. Furthermore, for preventing Evasion attacks, the authors ~\cite{chen2021certifiably} smoothed the training data by applying Gaussian noise and including adversarial data in the training dataset.

\begin{challenge}
\textbf{\textit{Adversarial Training exclusively defends against Adversarial Examples in FL environments. It is crucial to understand how Adversarial Training could be used for protecting against other types of attacks.}}
\end{challenge}

\subsubsection{Robust FL Aggregation}

The server could discard malicious updates~\cite{Zhang2021, xie2021crfl, zhao2021federatedreverse, wan2021robust, zhang2021safelearning, fu2019attack, liu2021privacy} or remove the poisoning effect~\cite{sun2021fl}. As previously mentioned, FedAvg is the baseline aggregation algorithm. Still, the server could use a robust aggregator instead. For instance, the authors in~\cite{blanchard2017machine} described Krum. For each client's update, the server calculated the sum of the Euclidean distance towards the other models. Then, the server selected the model with a lower distance, i.e., the most similar. Therefore, the chosen model was benign for networks over 50\% is not an adversary. Krum notably reduced the adversarial accuracy but was not as effective in non-IID settings. The same authors presented a variation of Krum, named multi-Krum~\cite{blanchard2017machine}, which smartly altered between Krum and averaging for improving performance while providing resilience.

In~\cite{ozdayi2020defending}, they set the robust algorithm to adjust the learning rate per round and dimension to maximize the loss on adversarial updates instead of minimizing it. The authors in~\cite{shejwalkar2021manipulating} designed a divide-and-conquer dimensionally reduction defense. After reducing the dimensions, the search spectrum was smaller and thus more efficient, then the authors applied single value decomposition for detecting outliners. Another novel approach named Baffle was presented by~\cite{andreina2020baffle}. They designed an algorithm where the clients send the updates to some chosen clients evaluating the updates against their local dataset and sharing the results back to the aggregator. After benchmarking, the aggregator could detect inconsistencies in clients.

Analogously, FoolsGold~\cite{fung2018mitigating} and Sniper~\cite{cao2019understanding} are defense mechanisms against Sybil attacks. The authors suggested that since Sybils work together towards a common adversarial goal, their updates were cosine-similar. Therefore, the benign and malicious updates were not cosine-similar. Thus, the server could remove those. Even though this defense performed correctly against Sybil's targeted poisoning attacks over non-IID and IID, it did not function with a single attacker~\cite{fung2018mitigating, cao2019understanding}. Similarly, following the same intuition, the authors in~\cite{mao2021romoa} presented Romoa, which discarded malicious ones based on historical updates.

Moreover, coordinate-wise trimmed median~\cite{yin2018byzantine} aimed to remove Sybil updates of Poisoning attacks. The intuition was to eliminate updates that are dissimilar from harmless updates. Then the remaining updates were averaged. In addition, the authors in~\cite{pillutla2019robust} presented the Robust Federated Aggregation algorithm, which aggregated the updates by calculating the weighted geometric median using the smoothed Weiszfeld's algorithm. Similarly, authors in~\cite{yin2018byzantine} proposed the coordinate-wise median defense, which performed an arithmetic median for each update. Bulyan~\cite{guerraoui2018hidden} used an aggregator algorithm such as Krum for interactively selecting some updates, aggregated by trimmed-mean.

Other approaches focused on setting confidence values or scores for each update for corresponding participation on the joint model~\cite{awan2021contra, ge2021chain, sharma2021tesseract, manna2021moat, li2021lomar}. Similarly, the authors in~\cite{xi2021batfl} relied on the Shapley value for eliminating Backdoor attacks. The Shapley value yields how each client collaborates to the common objective, thus removing low collaborative updates.


\begin{challenge}
\textbf{\textit{Robust Aggregators try to defend training time attacks. However, recent adaptative attacks evade such defenses~\cite{Fang2020}. This observation sets the need for further research on how promising the Robust Aggregators are against such adaptative evasion attacks.}}
\end{challenge}

\subsubsection{GAN-based Defenses}

GANs~\cite{goodfellow2014generative} have demonstrated outstanding performance in the last years~\cite{creswell2018generative}. Designed for images, GANs can create pictures and video from noise. They are composed of two networks, a generator $\mathcal{G}$ and a discriminator $\mathcal{D}$. They both compete against each other simultaneously where $\mathcal{G}$ tries to create new, non-seen images similar to the real ones. $\mathcal{D}$ tries to differentiate actual inputs from generated ones. After achieving the Nash equilibrium, $\mathcal{G}$ can create new inputs that $\mathcal{D}$ cannot differentiate from real ones.
Therefore, GANs may be split and use $\mathcal{G}$ autonomously. The authors in~\cite{Zhang2020a, Zhang2019} used GANs for datasets augmentation, enhancing the performance of Poisoning attacks. Similarly, an adversary might perform Inference attacks by reconstructing data using $\mathcal{G}$~\cite{Zhang2020}.

As shown, Model Poisoning attacks are more powerful and more popular than Data Poisoning. Their growth in popularity influenced, along with GANs, the GAN-based defensive mechanisms development. The authors in~\cite{zhang2020defending, zhao2020detecting, Zhao2020a} used GANs for creating an auditing dataset, which, provided to a classifier, could differentiate adversarial updates. Another approach followed by~\cite{jiang2020mitigating} consisted in adding noise to the updates. They also used dimensionality reduction for improving training time in the cost of model accuracy.

GANs' versatility allowed their use in both attacks and defenses successfully. However, as FL rounds go by, the data distribution drifts, requiring continuous training of GANs. Furthermore, in non-IID environments, GANs did not achieve as exceptional quality as in IID.

\begin{challenge}\textbf{\textit{GANs were only applied for defending against Model Poisoning attacks. It is necessary to determine GANs' viability for other attacks and non-IID environments.}}\end{challenge}

\begin{sidewaystable}[!htbp]
    \caption{Comparison of defenses regarding different features.~\dag~denotes a backdoor attack.}
    \label{tab:DefencesTable}
    \resizebox{\textwidth}{!}{

    \begin{tabular}{lllllllll}

        \hline
        
        & \begin{tabular}[c]{@{}l@{}} Defences  \\ \fbox{\tiny{1}} \fbox{\tiny{2}} \fbox{\tiny{3}} \fbox{\tiny{4}} \fbox{\tiny{5}} \fbox{\tiny{6}} \fbox{\tiny{7}} \end{tabular}
        
        & \begin{tabular}[c]{@{}l@{}} Data distribution   \\ \fbox{\tiny{1}} \fbox{\tiny{2}} \fbox{\tiny{3}} \end{tabular}
        
        & \begin{tabular}[c]{@{}l@{}} Dataset \\ \fbox{\tiny{1}} \fbox{\tiny{2}} \fbox{\tiny{3}}\end{tabular} 
    
        & \begin{tabular}[c]{@{}l@{}} Evaluation metrics   \\ \fbox{\tiny{1}} \fbox{\tiny{2}} \fbox{\tiny{3}} \fbox{\tiny{4}} \fbox{\tiny{5}}\end{tabular}
        
        & \begin{tabular}[c]{@{}l@{}} Attack   \\ \fbox{\tiny{1}} \fbox{\tiny{2}} \fbox{\tiny{3}} \fbox{\tiny{4}} \fbox{\tiny{5}} \fbox{\tiny{6}} \fbox{\tiny{7}} \end{tabular}        
        
        & \begin{tabular}[c]{@{}l@{}} Source code \\ \fbox{\tiny{1}} \fbox{\tiny{2}} \end{tabular}  
        
        & \begin{tabular}[c]{@{}l@{}} FL algorithm  \\ \fbox{\tiny{1}} \fbox{\tiny{2}} \fbox{\tiny{3}} \fbox{\tiny{4}} \fbox{\tiny{5}} \fbox{\tiny{6}} \fbox{\tiny{7}} \fbox{\tiny{8}} \fbox{\tiny{9}} \end{tabular}
        
        & \begin{tabular}[c]{@{}l@{}} Framework  \\ \fbox{\tiny{1}} \fbox{\tiny{2}} \fbox{\tiny{3}} \fbox{\tiny{4}} \fbox{\tiny{5}} \fbox{\tiny{6}} \fbox{\tiny{7}} \fbox{\tiny{8}} \fbox{\tiny{9}} \end{tabular} \\

        \hline
            
        & \begin{tabular}[c]{@{}l@{}}1. Pruning \\ 2. Server Cleaning \\ 3. Cryptographic Methods \\ 4. Adversarial Training \\ 5. Differential Privacy  \\ 6. Robust FL Aggregation \\ 7. GAN based \end{tabular}
        
        & \begin{tabular}[c]{@{}l@{}}1. IID \\ 2. Non-IID \\3. Not defined\end{tabular}
        
        & \begin{tabular}[c]{@{}l@{}} 1. Image\\ 2. Multivariate\\ 3. Text\end{tabular} 
        
        & \begin{tabular}[c]{@{}l@{}} 1. Confusion matrix\\ 2. ASR\\ 3. MSE\\ 4. SSIM \\ 5. Others\end{tabular}
        
        & \begin{tabular}[c]{@{}l@{}} 1. Data Poisoning\\ 2. Model Poisoning\\ 
        3. Inference\\ 4. Model Inversion\\ 5. Free-rider\\ 6. Adversarial Examples\\ 7. None \end{tabular}        
        
        & \begin{tabular}[c]{@{}l@{}} 1. Yes \\ 2. No \end{tabular}
        
        & \begin{tabular}[c]{@{}l@{}} 1. CNN \\ 2. ResNet \\ 3. LeNet \\ 4. LSTM \\ 5. DNN \\ 6. AlexNet \\ 7. VGG \\ 8. Others \\ 9. Not defined \end{tabular} 

        & \begin{tabular}[c]{@{}l@{}} 1. PyTorch \\ 2. TensorFlow \\ 3. Scikit-Learn \\ 4. PySyft \\ 5. ART \\ 6. Sherpa \\ 7. FATE \\ 8. IBM FL \\ 9. Not defined \end{tabular} \\

        \hline
        
        \begin{tabular}[c]{@{}l@{}} Shah \textit{et al.}~\cite{Shah} \end{tabular}
        &
        
        $\square$\hspace{5pt}$\square$\hspace{5pt}$\square$\hspace{5pt}$\blacksquare$\hspace{5pt}$\square$\hspace{5pt}$\square$\hspace{5pt}$\square$ & 
        
        $\blacksquare$\hspace{5pt}$\blacksquare$\hspace{5pt}$\square$  & 
        
        $\blacksquare$\hspace{5pt}$\square$\hspace{5pt}$\square$& 
        
        $\blacksquare$\hspace{5pt}$\square$\hspace{5pt}$\square$\hspace{5pt}$\square$\hspace{5pt}$\blacksquare$ & 
        
        $\square$\hspace{5pt}$\square$\hspace{5pt}$\square$\hspace{5pt}$\square$\hspace{5pt}$\square$\hspace{5pt}$\square$\hspace{5pt}$\blacksquare$ & 
        
        $\square$\hspace{5pt}$\blacksquare$ & 
        
        $\blacksquare$\hspace{5pt}$\square$\hspace{5pt}$\square$\hspace{5pt}$\square$\hspace{5pt}$\square$\hspace{5pt}$\square$\hspace{5pt}$\blacksquare$\hspace{5pt}$\blacksquare$\hspace{5pt}$\square$ &
        
        $\square$\hspace{5pt}$\blacksquare$\hspace{5pt}$\square$\hspace{5pt}$\square$\hspace{5pt}$\square$\hspace{5pt}$\square$\hspace{5pt}$\square$\hspace{5pt}$\square$\hspace{5pt}$\square$

        \\
                
        \begin{tabular}[c]{@{}l@{}} Sun \textit{et al.}~\cite{Sun2020BC} \end{tabular}
        &
        
        $\square$\hspace{5pt}$\blacksquare$\hspace{5pt}$\square$\hspace{5pt}$\square$\hspace{5pt}$\square$\hspace{5pt}$\square$\hspace{5pt}$\square$ & 
        
        $\square$\hspace{5pt}$\square$\hspace{5pt}$\blacksquare$  & 
        
        $\blacksquare$\hspace{5pt}$\square$\hspace{5pt}$\square$& 
        
        $\blacksquare$\hspace{5pt}$\square$\hspace{5pt}$\square$\hspace{5pt}$\square$\hspace{5pt}$\square$\hspace{5pt} & 
        
        $\blacksquare$\hspace{5pt}$\square$\hspace{5pt}$\square$\hspace{5pt}$\square$\hspace{5pt}$\square$\hspace{5pt}$\square$\hspace{5pt}$\square$ & 
        
        $\square$\hspace{5pt}$\blacksquare$ & 
        
        $\blacksquare$\hspace{5pt}$\square$\hspace{5pt}$\square$\hspace{5pt}$\square$\hspace{5pt}$\square$\hspace{5pt}$\square$\hspace{5pt}$\square$\hspace{5pt}$\square$\hspace{5pt}$\square$ &
        
        $\square$\hspace{5pt}$\square$\hspace{5pt}$\square$\hspace{5pt}$\square$\hspace{5pt}$\square$\hspace{5pt}$\square$\hspace{5pt}$\square$\hspace{5pt}$\square$\hspace{5pt}$\blacksquare$

        \\
        
        \begin{tabular}[c]{@{}l@{}} Chen \textit{et al.}~\cite{chen2021certifiably} \end{tabular}
        &
        
        $\square$\hspace{5pt}$\square$\hspace{5pt}$\square$\hspace{5pt}$\blacksquare$\hspace{5pt}$\square$\hspace{5pt}$\square$\hspace{5pt}$\square$ & 
        
        $\square$\hspace{5pt}$\blacksquare$\hspace{5pt}$\square$  & 
        
        $\blacksquare$\hspace{5pt}$\square$\hspace{5pt}$\square$& 
        
        $\blacksquare$\hspace{5pt}$\square$\hspace{5pt}$\square$\hspace{5pt}$\square$\hspace{5pt}$\square$\hspace{5pt} & 
        
        $\square$\hspace{5pt}$\square$\hspace{5pt}$\square$\hspace{5pt}$\square$\hspace{5pt}$\square$\hspace{5pt}$\blacksquare$\hspace{5pt}$\square$ & 
        
        $\square$\hspace{5pt}$\blacksquare$ & 
        
        $\square$\hspace{5pt}$\square$\hspace{5pt}$\square$\hspace{5pt}$\square$\hspace{5pt}$\square$\hspace{5pt}$\blacksquare$\hspace{5pt}$\square$\hspace{5pt}$\square$\hspace{5pt}$\square$ &
        
        $\square$\hspace{5pt}$\square$\hspace{5pt}$\square$\hspace{5pt}$\square$\hspace{5pt}$\square$\hspace{5pt}$\square$\hspace{5pt}$\square$\hspace{5pt}$\square$\hspace{5pt}$\blacksquare$

        \\  
        
        \begin{tabular}[c]{@{}l@{}} Zhang \textit{et al.}~\cite{zhang2020defending} \end{tabular}
        &
        
        $\square$\hspace{5pt}$\square$\hspace{5pt}$\square$\hspace{5pt}$\blacksquare$\hspace{5pt}$\square$\hspace{5pt}$\square$\hspace{5pt}$\blacksquare$ & 
        
        $\blacksquare$\hspace{5pt}$\square$\hspace{5pt}$\square$  & 
        
        $\blacksquare$\hspace{5pt}$\square$\hspace{5pt}$\square$& 
        
        $\blacksquare$\hspace{5pt}$\square$\hspace{5pt}$\square$\hspace{5pt}$\square$\hspace{5pt}$\square$\hspace{5pt} & 
        
        $\square$\hspace{5pt}$\blacksquare$\hspace{5pt}$\square$\hspace{5pt}$\square$\hspace{5pt}$\square$\hspace{5pt}$\square$\hspace{5pt}$\square$ & 
        
        $\square$\hspace{5pt}$\blacksquare$ & 
        
        $\blacksquare$\hspace{5pt}$\square$\hspace{5pt}$\square$\hspace{5pt}$\square$\hspace{5pt}$\blacksquare$\hspace{5pt}$\square$\hspace{5pt}$\square$\hspace{5pt}$\square$\hspace{5pt}$\square$ &
        
        $\square$\hspace{5pt}$\square$\hspace{5pt}$\square$\hspace{5pt}$\square$\hspace{5pt}$\square$\hspace{5pt}$\square$\hspace{5pt}$\square$\hspace{5pt}$\square$\hspace{5pt}$\blacksquare$

        \\

        \begin{tabular}[c]{@{}l@{}} Zhao \textit{et al.}~\cite{zhao2020detecting} \end{tabular}
        &
        
        $\square$\hspace{5pt}$\square$\hspace{5pt}$\square$\hspace{5pt}$\square$\hspace{5pt}$\square$\hspace{5pt}$\square$\hspace{5pt}$\blacksquare$ & 
        
        $\square$\hspace{5pt}$\blacksquare$\hspace{5pt}$\square$  & 
        
        $\blacksquare$\hspace{5pt}$\square$\hspace{5pt}$\square$& 
        
        $\blacksquare$\hspace{5pt}$\square$\hspace{5pt}$\square$\hspace{5pt}$\square$\hspace{5pt}$\square$\hspace{5pt} & 
        
        $\square$\hspace{5pt}$\blacksquare$\hspace{5pt}$\square$\hspace{5pt}$\square$\hspace{5pt}$\square$\hspace{5pt}$\square$\hspace{5pt}$\square$ & 
        
        $\square$\hspace{5pt}$\blacksquare$ & 
        
        $\blacksquare$\hspace{5pt}$\square$\hspace{5pt}$\square$\hspace{5pt}$\square$\hspace{5pt}$\square$\hspace{5pt}$\square$\hspace{5pt}$\square$\hspace{5pt}$\square$\hspace{5pt}$\square$ &
        
        $\blacksquare$\hspace{5pt}$\square$\hspace{5pt}$\square$\hspace{5pt}$\square$\hspace{5pt}$\square$\hspace{5pt}$\square$\hspace{5pt}$\square$\hspace{5pt}$\square$\hspace{5pt}$\square$

        \\      
        
        \begin{tabular}[c]{@{}l@{}} Ibitoye \textit{et al.}~\cite{ibitoye2021dipsen} \end{tabular}
        &
        
        $\square$\hspace{5pt}$\square$\hspace{5pt}$\square$\hspace{5pt}$\square$\hspace{5pt}$\blacksquare$\hspace{5pt}$\square$\hspace{5pt}$\square$ & 
        
        $\square$\hspace{5pt}$\blacksquare$\hspace{5pt}$\square$  & 
        
        $\blacksquare$\hspace{5pt}$\blacksquare$\hspace{5pt}$\square$& 
        
        $\blacksquare$\hspace{5pt}$\square$\hspace{5pt}$\square$\hspace{5pt}$\square$\hspace{5pt}$\blacksquare$\hspace{5pt} & 
        
        $\square$\hspace{5pt}$\square$\hspace{5pt}$\square$\hspace{5pt}$\square$\hspace{5pt}$\square$\hspace{5pt}$\blacksquare$\hspace{5pt}$\square$ & 
        
        $\square$\hspace{5pt}$\blacksquare$ & 
        
        $\blacksquare$\hspace{5pt}$\square$\hspace{5pt}$\square$\hspace{5pt}$\square$\hspace{5pt}$\square$\hspace{5pt}$\square$\hspace{5pt}$\square$\hspace{5pt}$\square$\hspace{5pt}$\square$ &
        
        $\square$\hspace{5pt}$\blacksquare$\hspace{5pt}$\square$\hspace{5pt}$\square$\hspace{5pt}$\blacksquare$\hspace{5pt}$\square$\hspace{5pt}$\square$\hspace{5pt}$\square$\hspace{5pt}$\square$

        \\        
        
        \begin{tabular}[c]{@{}l@{}} Rodriguez \textit{et al.}~\cite{rodriguez2020dynamic} \end{tabular}
        &
        
        $\square$\hspace{5pt}$\blacksquare$\hspace{5pt}$\square$\hspace{5pt}$\square$\hspace{5pt}$\square$\hspace{5pt}$\square$\hspace{5pt}$\square$ & 
        
        $\square$\hspace{5pt}$\blacksquare$\hspace{5pt}$\square$  & 
        
        $\blacksquare$\hspace{5pt}$\square$\hspace{5pt}$\square$& 
        
        $\blacksquare$\hspace{5pt}$\square$\hspace{5pt}$\square$\hspace{5pt}$\square$\hspace{5pt}$\square$\hspace{5pt} & 
        
        $\blacksquare$\hspace{5pt}$\square$\hspace{5pt}$\square$\hspace{5pt}$\square$\hspace{5pt}$\square$\hspace{5pt}$\square$\hspace{5pt}$\square$ & 
        
        $\square$\hspace{5pt}$\blacksquare$ & 
        
        $\blacksquare$\hspace{5pt}$\square$\hspace{5pt}$\square$\hspace{5pt}$\square$\hspace{5pt}$\square$\hspace{5pt}$\square$\hspace{5pt}$\square$\hspace{5pt}$\square$\hspace{5pt}$\square$ &
        
        $\square$\hspace{5pt}$\square$\hspace{5pt}$\square$\hspace{5pt}$\square$\hspace{5pt}$\square$\hspace{5pt}$\blacksquare$\hspace{5pt}$\square$\hspace{5pt}$\square$\hspace{5pt}$\square$

        \\    
        
        \begin{tabular}[c]{@{}l@{}} Kerkouche \textit{et al.}~\cite{Kerkouche}~\dag \end{tabular}
        &
        
        $\square$\hspace{5pt}$\square$\hspace{5pt}$\square$\hspace{5pt}$\square$\hspace{5pt}$\blacksquare$\hspace{5pt}$\square$\hspace{5pt}$\square$ & 
        
        $\square$\hspace{5pt}$\blacksquare$\hspace{5pt}$\square$  & 
        
        $\blacksquare$\hspace{5pt}$\square$\hspace{5pt}$\blacksquare$& 
        
        $\blacksquare$\hspace{5pt}$\square$\hspace{5pt}$\square$\hspace{5pt}$\square$\hspace{5pt}$\square$\hspace{5pt} & 
        
        $\square$\hspace{5pt}$\blacksquare$\hspace{5pt}$\square$\hspace{5pt}$\square$\hspace{5pt}$\square$\hspace{5pt}$\square$\hspace{5pt}$\square$ & 
        
        $\square$\hspace{5pt}$\blacksquare$ & 
        
        $\blacksquare$\hspace{5pt}$\square$\hspace{5pt}$\square$\hspace{5pt}$\square$\hspace{5pt}$\square$\hspace{5pt}$\square$\hspace{5pt}$\square$\hspace{5pt}$\square$\hspace{5pt}$\square$ &
        
        $\square$\hspace{5pt}$\blacksquare$\hspace{5pt}$\square$\hspace{5pt}$\square$\hspace{5pt}$\square$\hspace{5pt}$\square$\hspace{5pt}$\square$\hspace{5pt}$\square$\hspace{5pt}$\square$

        \\        
        
        \begin{tabular}[c]{@{}l@{}} Zhao \textit{et al.}~\cite{Zhao2020a} \end{tabular}
        &
        
        $\square$\hspace{5pt}$\blacksquare$\hspace{5pt}$\square$\hspace{5pt}$\square$\hspace{5pt}$\square$\hspace{5pt}$\square$\hspace{5pt}$\blacksquare$ & 
        
        $\square$\hspace{5pt}$\blacksquare$\hspace{5pt}$\square$  & 
        
        $\blacksquare$\hspace{5pt}$\square$\hspace{5pt}$\square$& 
        
        $\blacksquare$\hspace{5pt}$\square$\hspace{5pt}$\square$\hspace{5pt}$\square$\hspace{5pt}$\square$\hspace{5pt} & 
        
        $\square$\hspace{5pt}$\blacksquare$\hspace{5pt}$\square$\hspace{5pt}$\square$\hspace{5pt}$\square$\hspace{5pt}$\square$\hspace{5pt}$\square$ & 
        
        $\square$\hspace{5pt}$\blacksquare$ & 
        
        $\square$\hspace{5pt}$\square$\hspace{5pt}$\square$\hspace{5pt}$\square$\hspace{5pt}$\square$\hspace{5pt}$\square$\hspace{5pt}$\square$\hspace{5pt}$\square$\hspace{5pt}$\blacksquare$ &
        
        $\blacksquare$\hspace{5pt}$\square$\hspace{5pt}$\square$\hspace{5pt}$\square$\hspace{5pt}$\square$\hspace{5pt}$\square$\hspace{5pt}$\square$\hspace{5pt}$\square$\hspace{5pt}$\square$

        \\        
        
        \begin{tabular}[c]{@{}l@{}} Chen \textit{et al.}~\cite{Chen2020} \end{tabular}
        &
        
        $\square$\hspace{5pt}$\blacksquare$\hspace{5pt}$\square$\hspace{5pt}$\square$\hspace{5pt}$\square$\hspace{5pt}$\square$\hspace{5pt}$\square$ & 
        
        $\square$\hspace{5pt}$\blacksquare$\hspace{5pt}$\square$  & 
        
        $\blacksquare$\hspace{5pt}$\square$\hspace{5pt}$\square$& 
        
        $\blacksquare$\hspace{5pt}$\square$\hspace{5pt}$\square$\hspace{5pt}$\square$\hspace{5pt}$\blacksquare$\hspace{5pt} & 
        
        $\blacksquare$\hspace{5pt}$\blacksquare$\hspace{5pt}$\square$\hspace{5pt}$\square$\hspace{5pt}$\square$\hspace{5pt}$\square$\hspace{5pt}$\square$ & 
        
        $\square$\hspace{5pt}$\blacksquare$ & 
        
        $\blacksquare$\hspace{5pt}$\square$\hspace{5pt}$\square$\hspace{5pt}$\square$\hspace{5pt}$\square$\hspace{5pt}$\square$\hspace{5pt}$\square$\hspace{5pt}$\square$\hspace{5pt}$\square$ &
        
        $\square$\hspace{5pt}$\square$\hspace{5pt}$\square$\hspace{5pt}$\square$\hspace{5pt}$\square$\hspace{5pt}$\square$\hspace{5pt}$\square$\hspace{5pt}$\square$\hspace{5pt}$\blacksquare$

        \\ 
        
        \begin{tabular}[c]{@{}l@{}} Hong \textit{et al.}~\cite{Hong} \end{tabular}
        &
        
        $\square$\hspace{5pt}$\square$\hspace{5pt}$\square$\hspace{5pt}$\blacksquare$\hspace{5pt}$\square$\hspace{5pt}$\square$\hspace{5pt}$\square$ & 
        
        $\square$\hspace{5pt}$\blacksquare$\hspace{5pt}$\square$  & 
        
        $\blacksquare$\hspace{5pt}$\square$\hspace{5pt}$\square$& 
        
        $\blacksquare$\hspace{5pt}$\square$\hspace{5pt}$\square$\hspace{5pt}$\square$\hspace{5pt}$\square$\hspace{5pt} & 
        
        $\square$\hspace{5pt}$\square$\hspace{5pt}$\square$\hspace{5pt}$\square$\hspace{5pt}$\square$\hspace{5pt}$\square$\hspace{5pt}$\blacksquare$ & 
        
        $\square$\hspace{5pt}$\blacksquare$ & 
        
        $\square$\hspace{5pt}$\square$\hspace{5pt}$\square$\hspace{5pt}$\square$\hspace{5pt}$\square$\hspace{5pt}$\blacksquare$\hspace{5pt}$\square$\hspace{5pt}$\blacksquare$\hspace{5pt}$\square$ &
        
        $\blacksquare$\hspace{5pt}$\square$\hspace{5pt}$\square$\hspace{5pt}$\square$\hspace{5pt}$\square$\hspace{5pt}$\square$\hspace{5pt}$\square$\hspace{5pt}$\square$\hspace{5pt}$\square$

        \\        
        
        \begin{tabular}[c]{@{}l@{}} Singh \textit{et al.}~\cite{singh2020fair} \end{tabular}
        &
        
        $\square$\hspace{5pt}$\blacksquare$\hspace{5pt}$\square$\hspace{5pt}$\square$\hspace{5pt}$\square$\hspace{5pt}$\square$\hspace{5pt}$\square$ & 
        
        $\blacksquare$\hspace{5pt}$\square$\hspace{5pt}$\square$  & 
        
        $\square$\hspace{5pt}$\blacksquare$\hspace{5pt}$\square$& 
        
        $\blacksquare$\hspace{5pt}$\square$\hspace{5pt}$\square$\hspace{5pt}$\square$\hspace{5pt}$\square$\hspace{5pt} & 
        
        $\blacksquare$\hspace{5pt}$\square$\hspace{5pt}$\square$\hspace{5pt}$\square$\hspace{5pt}$\square$\hspace{5pt}$\square$\hspace{5pt}$\square$ & 
        
        $\square$\hspace{5pt}$\blacksquare$ & 
        
        $\square$\hspace{5pt}$\square$\hspace{5pt}$\square$\hspace{5pt}$\square$\hspace{5pt}$\square$\hspace{5pt}$\square$\hspace{5pt}$\square$\hspace{5pt}$\square$\hspace{5pt}$\blacksquare$ &
        
        $\square$\hspace{5pt}$\square$\hspace{5pt}$\blacksquare$\hspace{5pt}$\square$\hspace{5pt}$\square$\hspace{5pt}$\square$\hspace{5pt}$\square$\hspace{5pt}$\square$\hspace{5pt}$\square$

        \\     
        
        \begin{tabular}[c]{@{}l@{}} Shejwalkar and Houmansadr \cite{shejwalkar2021manipulating} \end{tabular}
        &
        
        $\square$\hspace{5pt}$\square$\hspace{5pt}$\square$\hspace{5pt}$\square$\hspace{5pt}$\square$\hspace{5pt}$\blacksquare$\hspace{5pt}$\square$ & 
        
        $\blacksquare$\hspace{5pt}$\blacksquare$\hspace{5pt}$\square$  & 
        
        $\blacksquare$\hspace{5pt}$\blacksquare$\hspace{5pt}$\square$& 
        
        $\blacksquare$\hspace{5pt}$\square$\hspace{5pt}$\square$\hspace{5pt}$\square$\hspace{5pt}$\square$\hspace{5pt} & 
        
        $\square$\hspace{5pt}$\blacksquare$\hspace{5pt}$\square$\hspace{5pt}$\square$\hspace{5pt}$\square$\hspace{5pt}$\square$\hspace{5pt}$\square$ & 
        
        $\square$\hspace{5pt}$\blacksquare$ & 
        
        $\square$\hspace{5pt}$\square$\hspace{5pt}$\square$\hspace{5pt}$\square$\hspace{5pt}$\square$\hspace{5pt}$\blacksquare$\hspace{5pt}$\blacksquare$\hspace{5pt}$\blacksquare$\hspace{5pt}$\square$ &
        
        $\square$\hspace{5pt}$\square$\hspace{5pt}$\square$\hspace{5pt}$\square$\hspace{5pt}$\square$\hspace{5pt}$\square$\hspace{5pt}$\square$\hspace{5pt}$\square$\hspace{5pt}$\blacksquare$

        \\        
        
        \begin{tabular}[c]{@{}l@{}} Doku and Rawat \cite{Doku2021} \end{tabular}
        &
        
        $\square$\hspace{5pt}$\blacksquare$\hspace{5pt}$\square$\hspace{5pt}$\square$\hspace{5pt}$\square$\hspace{5pt}$\square$\hspace{5pt}$\square$ & 
        
        $\square$\hspace{5pt}$\square$\hspace{5pt}$\blacksquare$  & 
        
        $\square$\hspace{5pt}$\square$\hspace{5pt}$\blacksquare$& 
        
        $\blacksquare$\hspace{5pt}$\square$\hspace{5pt}$\square$\hspace{5pt}$\square$\hspace{5pt}$\blacksquare$\hspace{5pt} & 
        
        $\blacksquare$\hspace{5pt}$\square$\hspace{5pt}$\square$\hspace{5pt}$\square$\hspace{5pt}$\square$\hspace{5pt}$\square$\hspace{5pt}$\square$ & 
        
        $\square$\hspace{5pt}$\blacksquare$ & 
        
        $\square$\hspace{5pt}$\square$\hspace{5pt}$\square$\hspace{5pt}$\square$\hspace{5pt}$\square$\hspace{5pt}$\square$\hspace{5pt}$\square$\hspace{5pt}$\blacksquare$\hspace{5pt}$\square$ &
        
        $\square$\hspace{5pt}$\square$\hspace{5pt}$\square$\hspace{5pt}$\square$\hspace{5pt}$\square$\hspace{5pt}$\square$\hspace{5pt}$\square$\hspace{5pt}$\square$\hspace{5pt}$\blacksquare$

        \\        
        
        \begin{tabular}[c]{@{}l@{}} Fung \textit{et al.}~\cite{fung2018mitigating}~\dag \end{tabular}
        &
        
        $\square$\hspace{5pt}$\square$\hspace{5pt}$\square$\hspace{5pt}$\square$\hspace{5pt}$\square$\hspace{5pt}$\blacksquare$\hspace{5pt}$\square$ & 
        
        $\blacksquare$\hspace{5pt}$\blacksquare$\hspace{5pt}$\square$  & 
        
        $\blacksquare$\hspace{5pt}$\square$\hspace{5pt}$\square$& 
        
        $\blacksquare$\hspace{5pt}$\blacksquare$\hspace{5pt}$\square$\hspace{5pt}$\square$\hspace{5pt}$\square$\hspace{5pt} & 
        
        $\blacksquare$\hspace{5pt}$\square$\hspace{5pt}$\square$\hspace{5pt}$\square$\hspace{5pt}$\square$\hspace{5pt}$\square$\hspace{5pt}$\square$ & 
        
        $\blacksquare$\hspace{5pt}$\square$ & 
        
        $\square$\hspace{5pt}$\square$\hspace{5pt}$\square$\hspace{5pt}$\square$\hspace{5pt}$\square$\hspace{5pt}$\square$\hspace{5pt}$\blacksquare$\hspace{5pt}$\blacksquare$\hspace{5pt}$\square$ &
        
        $\square$\hspace{5pt}$\square$\hspace{5pt}$\blacksquare$\hspace{5pt}$\square$\hspace{5pt}$\square$\hspace{5pt}$\square$\hspace{5pt}$\square$\hspace{5pt}$\square$\hspace{5pt}$\square$

        \\        
        
        \begin{tabular}[c]{@{}l@{}} Andreina \textit{et al.}~\cite{andreina2020baffle}~\dag \end{tabular}
        &
        
        $\square$\hspace{5pt}$\square$\hspace{5pt}$\square$\hspace{5pt}$\square$\hspace{5pt}$\square$\hspace{5pt}$\blacksquare$\hspace{5pt}$\square$ & 
        
        $\square$\hspace{5pt}$\blacksquare$\hspace{5pt}$\square$  & 
        
        $\blacksquare$\hspace{5pt}$\square$\hspace{5pt}$\square$& 
        
        $\blacksquare$\hspace{5pt}$\square$\hspace{5pt}$\square$\hspace{5pt}$\square$\hspace{5pt}$\square$\hspace{5pt} & 
        
        $\square$\hspace{5pt}$\blacksquare$\hspace{5pt}$\square$\hspace{5pt}$\square$\hspace{5pt}$\square$\hspace{5pt}$\square$\hspace{5pt}$\square$ & 
        
        $\square$\hspace{5pt}$\blacksquare$ & 
        
        $\square$\hspace{5pt}$\blacksquare$\hspace{5pt}$\square$\hspace{5pt}$\square$\hspace{5pt}$\square$\hspace{5pt}$\square$\hspace{5pt}$\square$\hspace{5pt}$\square$\hspace{5pt}$\square$ &
        
        $\blacksquare$\hspace{5pt}$\square$\hspace{5pt}$\square$\hspace{5pt}$\square$\hspace{5pt}$\square$\hspace{5pt}$\square$\hspace{5pt}$\square$\hspace{5pt}$\square$\hspace{5pt}$\square$

        \\        
        
        \begin{tabular}[c]{@{}l@{}} Ozdayi \textit{et al.}~\cite{ozdayi2020defending}~\dag \end{tabular}
        &
        
        $\square$\hspace{5pt}$\square$\hspace{5pt}$\square$\hspace{5pt}$\square$\hspace{5pt}$\square$\hspace{5pt}$\blacksquare$\hspace{5pt}$\square$ & 
        
        $\blacksquare$\hspace{5pt}$\blacksquare$\hspace{5pt}$\square$  & 
        
        $\blacksquare$\hspace{5pt}$\square$\hspace{5pt}$\square$& 
        
        $\blacksquare$\hspace{5pt}$\square$\hspace{5pt}$\square$\hspace{5pt}$\square$\hspace{5pt}$\square$\hspace{5pt} & 
        
        $\square$\hspace{5pt}$\blacksquare$\hspace{5pt}$\square$\hspace{5pt}$\square$\hspace{5pt}$\square$\hspace{5pt}$\square$\hspace{5pt}$\square$ & 
        
        $\square$\hspace{5pt}$\blacksquare$ & 
        
        $\blacksquare$\hspace{5pt}$\square$\hspace{5pt}$\square$\hspace{5pt}$\square$\hspace{5pt}$\square$\hspace{5pt}$\square$\hspace{5pt}$\square$\hspace{5pt}$\square$\hspace{5pt}$\square$ &
        
        $\square$\hspace{5pt}$\square$\hspace{5pt}$\square$\hspace{5pt}$\square$\hspace{5pt}$\square$\hspace{5pt}$\square$\hspace{5pt}$\square$\hspace{5pt}$\square$\hspace{5pt}$\blacksquare$

        \\        
        
        \begin{tabular}[c]{@{}l@{}} Wu \textit{et al.}~\cite{Wu}~\dag \end{tabular}
        &
        
        $\blacksquare$\hspace{5pt}$\square$\hspace{5pt}$\square$\hspace{5pt}$\square$\hspace{5pt}$\square$\hspace{5pt}$\square$\hspace{5pt}$\square$ & 
        
        $\square$\hspace{5pt}$\blacksquare$\hspace{5pt}$\square$  & 
        
        $\blacksquare$\hspace{5pt}$\square$\hspace{5pt}$\square$& 
        
        $\blacksquare$\hspace{5pt}$\square$\hspace{5pt}$\square$\hspace{5pt}$\square$\hspace{5pt}$\square$\hspace{5pt} & 
        
        $\square$\hspace{5pt}$\blacksquare$\hspace{5pt}$\square$\hspace{5pt}$\square$\hspace{5pt}$\square$\hspace{5pt}$\square$\hspace{5pt}$\square$ & 
        
        $\square$\hspace{5pt}$\blacksquare$ & 
        
        $\blacksquare$\hspace{5pt}$\square$\hspace{5pt}$\square$\hspace{5pt}$\square$\hspace{5pt}$\square$\hspace{5pt}$\square$\hspace{5pt}$\square$\hspace{5pt}$\square$\hspace{5pt}$\square$ &
        
        $\square$\hspace{5pt}$\square$\hspace{5pt}$\square$\hspace{5pt}$\square$\hspace{5pt}$\square$\hspace{5pt}$\square$\hspace{5pt}$\square$\hspace{5pt}$\square$\hspace{5pt}$\blacksquare$

        \\   
        
        \begin{tabular}[c]{@{}l@{}} Zhang \textit{et al.}~\cite{zhang2021safelearning}~\dag \end{tabular}
        &
        
        $\square$\hspace{5pt}$\square$\hspace{5pt}$\square$\hspace{5pt}$\square$\hspace{5pt}$\square$\hspace{5pt}$\blacksquare$\hspace{5pt}$\square$ & 
        
        $\square$\hspace{5pt}$\square$\hspace{5pt}$\blacksquare$  & 
        
        $\blacksquare$\hspace{5pt}$\square$\hspace{5pt}$\square$& 
        
        $\blacksquare$\hspace{5pt}$\blacksquare$\hspace{5pt}$\square$\hspace{5pt}$\square$\hspace{5pt}$\square$\hspace{5pt} & 
        
        $\square$\hspace{5pt}$\blacksquare$\hspace{5pt}$\square$\hspace{5pt}$\square$\hspace{5pt}$\square$\hspace{5pt}$\square$\hspace{5pt}$\square$ & 
        
        $\square$\hspace{5pt}$\blacksquare$ & 
        
        $\square$\hspace{5pt}$\blacksquare$\hspace{5pt}$\square$\hspace{5pt}$\square$\hspace{5pt}$\square$\hspace{5pt}$\square$\hspace{5pt}$\square$\hspace{5pt}$\square$\hspace{5pt}$\square$ &
        
        $\square$\hspace{5pt}$\square$\hspace{5pt}$\square$\hspace{5pt}$\square$\hspace{5pt}$\square$\hspace{5pt}$\square$\hspace{5pt}$\square$\hspace{5pt}$\square$\hspace{5pt}$\blacksquare$

        \\     
        
        \begin{tabular}[c]{@{}l@{}} Xie \textit{et al.}~\cite{xie2021crfl}~\dag \end{tabular}
        &
        
        $\square$\hspace{5pt}$\square$\hspace{5pt}$\square$\hspace{5pt}$\square$\hspace{5pt}$\square$\hspace{5pt}$\blacksquare$\hspace{5pt}$\square$ & 
        
        $\blacksquare$\hspace{5pt}$\blacksquare$\hspace{5pt}$\square$  & 
        
        $\blacksquare$\hspace{5pt}$\blacksquare$\hspace{5pt}$\square$& 
        
        $\square$\hspace{5pt}$\square$\hspace{5pt}$\square$\hspace{5pt}$\square$\hspace{5pt}$\blacksquare$\hspace{5pt} & 
        
        $\square$\hspace{5pt}$\blacksquare$\hspace{5pt}$\square$\hspace{5pt}$\square$\hspace{5pt}$\square$\hspace{5pt}$\square$\hspace{5pt}$\square$ & 
        
        $\blacksquare$\hspace{5pt}$\square$ & 
        
        $\square$\hspace{5pt}$\square$\hspace{5pt}$\square$\hspace{5pt}$\square$\hspace{5pt}$\square$\hspace{5pt}$\square$\hspace{5pt}$\square$\hspace{5pt}$\blacksquare$\hspace{5pt}$\square$ &
        
        $\blacksquare$\hspace{5pt}$\square$\hspace{5pt}$\square$\hspace{5pt}$\square$\hspace{5pt}$\square$\hspace{5pt}$\square$\hspace{5pt}$\square$\hspace{5pt}$\square$\hspace{5pt}$\square$

        \\        
        
        \begin{tabular}[c]{@{}l@{}} Zhao \textit{et al.}~\cite{zhao2021federatedreverse}~\dag \end{tabular}
        &
        
        $\square$\hspace{5pt}$\square$\hspace{5pt}$\square$\hspace{5pt}$\square$\hspace{5pt}$\square$\hspace{5pt}$\blacksquare$\hspace{5pt}$\square$ & 
        
        $\square$\hspace{5pt}$\blacksquare$\hspace{5pt}$\square$  & 
        
        $\blacksquare$\hspace{5pt}$\square$\hspace{5pt}$\square$& 
        
        $\blacksquare$\hspace{5pt}$\square$\hspace{5pt}$\square$\hspace{5pt}$\square$\hspace{5pt}$\square$\hspace{5pt} & 
        
        $\square$\hspace{5pt}$\blacksquare$\hspace{5pt}$\square$\hspace{5pt}$\square$\hspace{5pt}$\square$\hspace{5pt}$\square$\hspace{5pt}$\square$ & 
        
        $\square$\hspace{5pt}$\blacksquare$ & 
        
        $\blacksquare$\hspace{5pt}$\blacksquare$\hspace{5pt}$\square$\hspace{5pt}$\square$\hspace{5pt}$\square$\hspace{5pt}$\square$\hspace{5pt}$\square$\hspace{5pt}$\square$\hspace{5pt}$\square$ &
        
        $\blacksquare$\hspace{5pt}$\square$\hspace{5pt}$\square$\hspace{5pt}$\square$\hspace{5pt}$\square$\hspace{5pt}$\square$\hspace{5pt}$\square$\hspace{5pt}$\square$\hspace{5pt}$\square$

        \\        
        
        
        
        
        
        
        
        
        

        
        \begin{tabular}[c]{@{}l@{}} Wei \textit{et al.}~\cite{wei2020framework} \end{tabular}
        &
        
        $\square$\hspace{5pt}$\square$\hspace{5pt}$\square$\hspace{5pt}$\square$\hspace{5pt}$\blacksquare$\hspace{5pt}$\square$\hspace{5pt}$\square$ & 
        
        $\square$\hspace{5pt}$\blacksquare$\hspace{5pt}$\square$  & 
        
        $\blacksquare$\hspace{5pt}$\blacksquare$\hspace{5pt}$\square$& 
        
        $\square$\hspace{5pt}$\blacksquare$\hspace{5pt}$\blacksquare$\hspace{5pt}$\blacksquare$\hspace{5pt}$\square$\hspace{5pt} & 
        
        $\square$\hspace{5pt}$\square$\hspace{5pt}$\square$\hspace{5pt}$\blacksquare$\hspace{5pt}$\square$\hspace{5pt}$\square$\hspace{5pt}$\square$ & 
        
        $\blacksquare$\hspace{5pt}$\square$ & 
        
        $\square$\hspace{5pt}$\blacksquare$\hspace{5pt}$\blacksquare$\hspace{5pt}$\square$\hspace{5pt}$\square$\hspace{5pt}$\square$\hspace{5pt}$\square$\hspace{5pt}$\square$\hspace{5pt}$\square$ &
        
        $\blacksquare$\hspace{5pt}$\square$\hspace{5pt}$\square$\hspace{5pt}$\square$\hspace{5pt}$\square$\hspace{5pt}$\square$\hspace{5pt}$\square$\hspace{5pt}$\square$\hspace{5pt}$\square$

        \\

        \begin{tabular}[c]{@{}l@{}} Li \textit{et al.}~\cite{li2021adaptive} \end{tabular}
        &
        
        $\square$\hspace{5pt}$\square$\hspace{5pt}$\square$\hspace{5pt}$\square$\hspace{5pt}$\square$\hspace{5pt}$\blacksquare$\hspace{5pt}$\square$ & 
        
        $\square$\hspace{5pt}$\blacksquare$\hspace{5pt}$\square$  & 
        
        $\blacksquare$\hspace{5pt}$\square$\hspace{5pt}$\square$& 
        
        $\blacksquare$\hspace{5pt}$\square$\hspace{5pt}$\blacksquare$\hspace{5pt}$\blacksquare$\hspace{5pt}$\square$ & 
        
        $\square$\hspace{5pt}$\square$\hspace{5pt}$\square$\hspace{5pt}$\blacksquare$\hspace{5pt}$\square$\hspace{5pt}$\square$\hspace{5pt}$\square$ & 
        
        $\square$\hspace{5pt}$\blacksquare$ & 
        
        $\square$\hspace{5pt}$\square$\hspace{5pt}$\blacksquare$\hspace{5pt}$\square$\hspace{5pt}$\square$\hspace{5pt}$\square$\hspace{5pt}$\square$\hspace{5pt}$\square$\hspace{5pt}$\square$ &
        
        $\square$\hspace{5pt}$\blacksquare$\hspace{5pt}$\square$\hspace{5pt}$\square$\hspace{5pt}$\square$\hspace{5pt}$\square$\hspace{5pt}$\square$\hspace{5pt}$\square$\hspace{5pt}$\square$

        \\     
        
        \begin{tabular}[c]{@{}l@{}} Fu \textit{et al.}~\cite{fu2019attack}~\dag \end{tabular}
        &
        
        $\square$\hspace{5pt}$\square$\hspace{5pt}$\square$\hspace{5pt}$\square$\hspace{5pt}$\square$\hspace{5pt}$\blacksquare$\hspace{5pt}$\square$ & 
        
        $\square$\hspace{5pt}$\blacksquare$\hspace{5pt}$\square$  & 
        
        $\blacksquare$\hspace{5pt}$\blacksquare$\hspace{5pt}$\blacksquare$& 
        
        $\blacksquare$\hspace{5pt}$\blacksquare$\hspace{5pt}$\square$\hspace{5pt}$\square$\hspace{5pt}$\blacksquare$ & 
        
        $\square$\hspace{5pt}$\blacksquare$\hspace{5pt}$\square$\hspace{5pt}$\square$\hspace{5pt}$\square$\hspace{5pt}$\square$\hspace{5pt}$\square$ & 
        
        $\blacksquare$\hspace{5pt}$\square$ & 
        
        $\blacksquare$\hspace{5pt}$\blacksquare$\hspace{5pt}$\square$\hspace{5pt}$\square$\hspace{5pt}$\blacksquare$\hspace{5pt}$\square$\hspace{5pt}$\square$\hspace{5pt}$\blacksquare$\hspace{5pt}$\square$ &
        
        $\blacksquare$\hspace{5pt}$\square$\hspace{5pt}$\square$\hspace{5pt}$\square$\hspace{5pt}$\square$\hspace{5pt}$\square$\hspace{5pt}$\square$\hspace{5pt}$\square$\hspace{5pt}$\square$

        \\        
        
        \begin{tabular}[c]{@{}l@{}} Chen \textit{et al.}~\cite{chen2020backdoor}~\dag \end{tabular}
        &
        
        $\square$\hspace{5pt}$\blacksquare$\hspace{5pt}$\square$\hspace{5pt}$\square$\hspace{5pt}$\square$\hspace{5pt}$\square$\hspace{5pt}$\square$ & 
        
        $\square$\hspace{5pt}$\square$\hspace{5pt}$\blacksquare$  & 
        
        $\blacksquare$\hspace{5pt}$\square$\hspace{5pt}$\square$& 
        
        $\blacksquare$\hspace{5pt}$\square$\hspace{5pt}$\square$\hspace{5pt}$\square$\hspace{5pt}$\square$ & 
        
        $\blacksquare$\hspace{5pt}$\blacksquare$\hspace{5pt}$\square$\hspace{5pt}$\square$\hspace{5pt}$\square$\hspace{5pt}$\square$\hspace{5pt}$\square$ & 
        
        $\square$\hspace{5pt}$\blacksquare$ & 
        
        $\square$\hspace{5pt}$\square$\hspace{5pt}$\square$\hspace{5pt}$\square$\hspace{5pt}$\square$\hspace{5pt}$\square$\hspace{5pt}$\square$\hspace{5pt}$\blacksquare$\hspace{5pt}$\square$ &
        
        $\square$\hspace{5pt}$\square$\hspace{5pt}$\square$\hspace{5pt}$\square$\hspace{5pt}$\square$\hspace{5pt}$\square$\hspace{5pt}$\square$\hspace{5pt}$\square$\hspace{5pt}$\blacksquare$

        \\   
        
        \begin{tabular}[c]{@{}l@{}} Tolpegin \textit{et al.}~\cite{Tolpegin}~\dag \end{tabular}
        &
        
        $\square$\hspace{5pt}$\blacksquare$\hspace{5pt}$\square$\hspace{5pt}$\square$\hspace{5pt}$\square$\hspace{5pt}$\square$\hspace{5pt}$\square$ & 
        
        $\blacksquare$\hspace{5pt}$\square$\hspace{5pt}$\square$  & 
        
        $\blacksquare$\hspace{5pt}$\square$\hspace{5pt}$\square$& 
        
        $\blacksquare$\hspace{5pt}$\square$\hspace{5pt}$\square$\hspace{5pt}$\square$\hspace{5pt}$\square$ & 
        
        $\blacksquare$\hspace{5pt}$\square$\hspace{5pt}$\square$\hspace{5pt}$\square$\hspace{5pt}$\square$\hspace{5pt}$\square$\hspace{5pt}$\square$ & 
        
        $\blacksquare$\hspace{5pt}$\square$ & 
        
        $\blacksquare$\hspace{5pt}$\square$\hspace{5pt}$\square$\hspace{5pt}$\square$\hspace{5pt}$\square$\hspace{5pt}$\square$\hspace{5pt}$\square$\hspace{5pt}$\blacksquare$\hspace{5pt}$\square$ &
        
        $\blacksquare$\hspace{5pt}$\square$\hspace{5pt}$\square$\hspace{5pt}$\square$\hspace{5pt}$\square$\hspace{5pt}$\square$\hspace{5pt}$\square$\hspace{5pt}$\square$\hspace{5pt}$\square$

        \\    
        
        \begin{tabular}[c]{@{}l@{}} Jiang \textit{et al.}~\cite{jiang2020mitigating} \end{tabular}
        &
        
        $\square$\hspace{5pt}$\square$\hspace{5pt}$\square$\hspace{5pt}$\square$\hspace{5pt}$\square$\hspace{5pt}$\blacksquare$\hspace{5pt}$\square$ & 
        
        $\blacksquare$\hspace{5pt}$\blacksquare$\hspace{5pt}$\square$  & 
        
        $\blacksquare$\hspace{5pt}$\square$\hspace{5pt}$\square$& 
        
        $\square$\hspace{5pt}$\square$\hspace{5pt}$\blacksquare$\hspace{5pt}$\square$\hspace{5pt}$\square$ & 
        
        $\square$\hspace{5pt}$\blacksquare$\hspace{5pt}$\square$\hspace{5pt}$\square$\hspace{5pt}$\square$\hspace{5pt}$\square$\hspace{5pt}$\square$ & 
        
        $\square$\hspace{5pt}$\blacksquare$ & 
        
        $\blacksquare$\hspace{5pt}$\square$\hspace{5pt}$\square$\hspace{5pt}$\square$\hspace{5pt}$\square$\hspace{5pt}$\square$\hspace{5pt}$\square$\hspace{5pt}$\blacksquare$\hspace{5pt}$\square$ &
        
        $\blacksquare$\hspace{5pt}$\square$\hspace{5pt}$\square$\hspace{5pt}$\square$\hspace{5pt}$\square$\hspace{5pt}$\square$\hspace{5pt}$\square$\hspace{5pt}$\square$\hspace{5pt}$\square$

        \\        
        
        \begin{tabular}[c]{@{}l@{}} Wan and Chen \cite{wan2021robust}~\dag \end{tabular}
        &
        
        $\square$\hspace{5pt}$\square$\hspace{5pt}$\square$\hspace{5pt}$\square$\hspace{5pt}$\square$\hspace{5pt}$\blacksquare$\hspace{5pt}$\square$ & 
        
        $\blacksquare$\hspace{5pt}$\square$\hspace{5pt}$\square$  & 
        
        $\blacksquare$\hspace{5pt}$\square$\hspace{5pt}$\blacksquare$& 
        
        $\blacksquare$\hspace{5pt}$\blacksquare$\hspace{5pt}$\square$\hspace{5pt}$\square$\hspace{5pt}$\square$ & 
        
        $\square$\hspace{5pt}$\blacksquare$\hspace{5pt}$\square$\hspace{5pt}$\square$\hspace{5pt}$\square$\hspace{5pt}$\square$\hspace{5pt}$\square$ & 
        
        $\blacksquare$\hspace{5pt}$\square$ & 
        
        $\square$\hspace{5pt}$\blacksquare$\hspace{5pt}$\blacksquare$\hspace{5pt}$\square$\hspace{5pt}$\square$\hspace{5pt}$\square$\hspace{5pt}$\square$\hspace{5pt}$\blacksquare$\hspace{5pt}$\square$ &
        
        $\blacksquare$\hspace{5pt}$\square$\hspace{5pt}$\square$\hspace{5pt}$\square$\hspace{5pt}$\square$\hspace{5pt}$\square$\hspace{5pt}$\square$\hspace{5pt}$\square$\hspace{5pt}$\square$

        \\
        
        \begin{tabular}[c]{@{}l@{}} Liu \textit{et al.}~\cite{Liu2020} \end{tabular}
        &
        
        $\square$\hspace{5pt}$\blacksquare$\hspace{5pt}$\square$\hspace{5pt}$\square$\hspace{5pt}$\square$\hspace{5pt}$\square$\hspace{5pt}$\square$ & 
        
        $\square$\hspace{5pt}$\square$\hspace{5pt}$\blacksquare$  & 
        
        $\blacksquare$\hspace{5pt}$\square$\hspace{5pt}$\square$& 
        
        $\blacksquare$\hspace{5pt}$\blacksquare$\hspace{5pt}$\square$\hspace{5pt}$\square$\hspace{5pt}$\square$ & 
        
        $\blacksquare$\hspace{5pt}$\square$\hspace{5pt}$\blacksquare$\hspace{5pt}$\square$\hspace{5pt}$\square$\hspace{5pt}$\square$\hspace{5pt}$\square$ & 
        
        $\square$\hspace{5pt}$\blacksquare$ & 
        
        $\blacksquare$\hspace{5pt}$\square$\hspace{5pt}$\square$\hspace{5pt}$\square$\hspace{5pt}$\square$\hspace{5pt}$\square$\hspace{5pt}$\square$\hspace{5pt}$\square$\hspace{5pt}$\square$ &
        
        $\blacksquare$\hspace{5pt}$\square$\hspace{5pt}$\square$\hspace{5pt}$\blacksquare$\hspace{5pt}$\square$\hspace{5pt}$\square$\hspace{5pt}$\square$\hspace{5pt}$\square$\hspace{5pt}$\square$

        \\      
        
        \begin{tabular}[c]{@{}l@{}} Cao \textit{et al.}~\cite{cao2019understanding} \end{tabular}
        &
        
        $\square$\hspace{5pt}$\square$\hspace{5pt}$\square$\hspace{5pt}$\square$\hspace{5pt}$\square$\hspace{5pt}$\blacksquare$\hspace{5pt}$\square$ & 
        
        $\square$\hspace{5pt}$\square$\hspace{5pt}$\blacksquare$  & 
        
        $\blacksquare$\hspace{5pt}$\square$\hspace{5pt}$\square$& 
        
        $\blacksquare$\hspace{5pt}$\square$\hspace{5pt}$\square$\hspace{5pt}$\square$\hspace{5pt}$\square$ & 
        
        $\blacksquare$\hspace{5pt}$\square$\hspace{5pt}$\square$\hspace{5pt}$\square$\hspace{5pt}$\square$\hspace{5pt}$\square$\hspace{5pt}$\square$ & 
        
        $\square$\hspace{5pt}$\blacksquare$ & 
        
        $\blacksquare$\hspace{5pt}$\square$\hspace{5pt}$\square$\hspace{5pt}$\square$\hspace{5pt}$\square$\hspace{5pt}$\square$\hspace{5pt}$\square$\hspace{5pt}$\square$\hspace{5pt}$\square$ &
        
        $\square$\hspace{5pt}$\square$\hspace{5pt}$\square$\hspace{5pt}$\square$\hspace{5pt}$\square$\hspace{5pt}$\square$\hspace{5pt}$\square$\hspace{5pt}$\square$\hspace{5pt}$\blacksquare$

        \\        
        
        \begin{tabular}[c]{@{}l@{}} Mallah \textit{et al.}~\cite{mallah2021untargeted} \end{tabular}
        &
        
        $\square$\hspace{5pt}$\blacksquare$\hspace{5pt}$\square$\hspace{5pt}$\square$\hspace{5pt}$\square$\hspace{5pt}$\square$\hspace{5pt}$\square$ & 
        
        $\square$\hspace{5pt}$\blacksquare$\hspace{5pt}$\square$  & 
        
        $\square$\hspace{5pt}$\blacksquare$\hspace{5pt}$\square$& 
        
        $\square$\hspace{5pt}$\square$\hspace{5pt}$\square$\hspace{5pt}$\square$\hspace{5pt}$\blacksquare$ & 
        
        $\square$\hspace{5pt}$\blacksquare$\hspace{5pt}$\square$\hspace{5pt}$\square$\hspace{5pt}$\square$\hspace{5pt}$\square$\hspace{5pt}$\square$ & 
        
        $\blacksquare$\hspace{5pt}$\square$ & 
        
        $\blacksquare$\hspace{5pt}$\square$\hspace{5pt}$\square$\hspace{5pt}$\square$\hspace{5pt}$\square$\hspace{5pt}$\square$\hspace{5pt}$\square$\hspace{5pt}$\square$\hspace{5pt}$\square$ &
        
        $\square$\hspace{5pt}$\blacksquare$\hspace{5pt}$\square$\hspace{5pt}$\square$\hspace{5pt}$\square$\hspace{5pt}$\square$\hspace{5pt}$\square$\hspace{5pt}$\square$\hspace{5pt}$\square$

        \\       
        
        \begin{tabular}[c]{@{}l@{}} Lee \textit{et al.}~\cite{lee2021digestive} \end{tabular}
        &
        
        $\square$\hspace{5pt}$\blacksquare$\hspace{5pt}$\square$\hspace{5pt}$\square$\hspace{5pt}$\square$\hspace{5pt}$\square$\hspace{5pt}$\square$ & 
        
        $\square$\hspace{5pt}$\square$\hspace{5pt}$\blacksquare$  & 
        
        $\blacksquare$\hspace{5pt}$\square$\hspace{5pt}$\square$& 
        
        $\blacksquare$\hspace{5pt}$\blacksquare$\hspace{5pt}$\square$\hspace{5pt}$\square$\hspace{5pt}$\square$ & 
        
        $\square$\hspace{5pt}$\square$\hspace{5pt}$\blacksquare$\hspace{5pt}$\blacksquare$\hspace{5pt}$\square$\hspace{5pt}$\square$\hspace{5pt}$\square$ & 
        
        $\square$\hspace{5pt}$\blacksquare$ & 
        
        $\square$\hspace{5pt}$\square$\hspace{5pt}$\square$\hspace{5pt}$\square$\hspace{5pt}$\blacksquare$\hspace{5pt}$\blacksquare$\hspace{5pt}$\square$\hspace{5pt}$\square$\hspace{5pt}$\square$ &
        
        $\blacksquare$\hspace{5pt}$\square$\hspace{5pt}$\square$\hspace{5pt}$\square$\hspace{5pt}$\square$\hspace{5pt}$\square$\hspace{5pt}$\square$\hspace{5pt}$\square$\hspace{5pt}$\square$

        \\      
        
        \begin{tabular}[c]{@{}l@{}} Sun \textit{et al.}~\cite{sun2021defending} \end{tabular}
        &
        
        $\square$\hspace{5pt}$\square$\hspace{5pt}$\square$\hspace{5pt}$\square$\hspace{5pt}$\blacksquare$\hspace{5pt}$\square$\hspace{5pt}$\square$ & 
        
        $\blacksquare$\hspace{5pt}$\square$\hspace{5pt}$\square$  & 
        
        $\square$\hspace{5pt}$\blacksquare$\hspace{5pt}$\square$& 
        
        $\square$\hspace{5pt}$\blacksquare$\hspace{5pt}$\blacksquare$\hspace{5pt}$\square$\hspace{5pt}$\square$ & 
        
        $\square$\hspace{5pt}$\square$\hspace{5pt}$\blacksquare$\hspace{5pt}$\square$\hspace{5pt}$\square$\hspace{5pt}$\square$\hspace{5pt}$\square$ & 
        
        $\square$\hspace{5pt}$\blacksquare$ & 
        
        $\square$\hspace{5pt}$\square$\hspace{5pt}$\square$\hspace{5pt}$\square$\hspace{5pt}$\square$\hspace{5pt}$\square$\hspace{5pt}$\square$\hspace{5pt}$\blacksquare$\hspace{5pt}$\square$ &
        
        $\square$\hspace{5pt}$\square$\hspace{5pt}$\square$\hspace{5pt}$\square$\hspace{5pt}$\square$\hspace{5pt}$\square$\hspace{5pt}$\square$\hspace{5pt}$\square$\hspace{5pt}$\blacksquare$

        \\   
        
        \begin{tabular}[c]{@{}l@{}} Xiong \textit{et al.}~\cite{xiong2021privacy} \end{tabular}
        &
        
        $\square$\hspace{5pt}$\square$\hspace{5pt}$\square$\hspace{5pt}$\square$\hspace{5pt}$\blacksquare$\hspace{5pt}$\square$\hspace{5pt}$\square$ & 
        
        $\square$\hspace{5pt}$\blacksquare$\hspace{5pt}$\square$  & 
        
        $\blacksquare$\hspace{5pt}$\square$\hspace{5pt}$\square$& 
        
        $\blacksquare$\hspace{5pt}$\square$\hspace{5pt}$\square$\hspace{5pt}$\blacksquare$\hspace{5pt}$\blacksquare$ & 
        
        $\square$\hspace{5pt}$\square$\hspace{5pt}$\square$\hspace{5pt}$\blacksquare$\hspace{5pt}$\square$\hspace{5pt}$\square$\hspace{5pt}$\square$ & 
        
        $\square$\hspace{5pt}$\blacksquare$ & 
        
        $\square$\hspace{5pt}$\square$\hspace{5pt}$\square$\hspace{5pt}$\square$\hspace{5pt}$\square$\hspace{5pt}$\square$\hspace{5pt}$\square$\hspace{5pt}$\square$\hspace{5pt}$\blacksquare$ &
        
        $\blacksquare$\hspace{5pt}$\square$\hspace{5pt}$\square$\hspace{5pt}$\blacksquare$\hspace{5pt}$\square$\hspace{5pt}$\square$\hspace{5pt}$\square$\hspace{5pt}$\square$\hspace{5pt}$\square$

        \\   
        
        \begin{tabular}[c]{@{}l@{}} Luo and Zhu \cite{luo2020exploiting} \end{tabular}
        &
        
        $\square$\hspace{5pt}$\square$\hspace{5pt}$\square$\hspace{5pt}$\square$\hspace{5pt}$\blacksquare$\hspace{5pt}$\square$\hspace{5pt}$\square$ & 
        
        $\square$\hspace{5pt}$\blacksquare$\hspace{5pt}$\square$  & 
        
        $\blacksquare$\hspace{5pt}$\square$\hspace{5pt}$\square$& 
        
        $\square$\hspace{5pt}$\square$\hspace{5pt}$\square$\hspace{5pt}$\square$\hspace{5pt}$\blacksquare$ & 
        
        $\square$\hspace{5pt}$\square$\hspace{5pt}$\blacksquare$\hspace{5pt}$\square$\hspace{5pt}$\square$\hspace{5pt}$\square$\hspace{5pt}$\square$ & 
        
        $\square$\hspace{5pt}$\blacksquare$ & 
        
        $\square$\hspace{5pt}$\square$\hspace{5pt}$\square$\hspace{5pt}$\square$\hspace{5pt}$\blacksquare$\hspace{5pt}$\square$\hspace{5pt}$\square$\hspace{5pt}$\square$\hspace{5pt}$\square$ &
        
        $\square$\hspace{5pt}$\square$\hspace{5pt}$\square$\hspace{5pt}$\square$\hspace{5pt}$\square$\hspace{5pt}$\square$\hspace{5pt}$\square$\hspace{5pt}$\square$\hspace{5pt}$\blacksquare$

        \\    
        
        \begin{tabular}[c]{@{}l@{}} Lin \textit{et al.}~\cite{lin2019free} \end{tabular}
        &
        
        $\square$\hspace{5pt}$\blacksquare$\hspace{5pt}$\square$\hspace{5pt}$\square$\hspace{5pt}$\square$\hspace{5pt}$\square$\hspace{5pt}$\square$ & 
        
        $\square$\hspace{5pt}$\square$\hspace{5pt}$\blacksquare$  & 
        
        $\blacksquare$\hspace{5pt}$\square$\hspace{5pt}$\square$& 
        
        $\blacksquare$\hspace{5pt}$\square$\hspace{5pt}$\square$\hspace{5pt}$\square$\hspace{5pt}$\blacksquare$ & 
        
        $\square$\hspace{5pt}$\square$\hspace{5pt}$\square$\hspace{5pt}$\square$\hspace{5pt}$\blacksquare$\hspace{5pt}$\square$\hspace{5pt}$\square$ & 
        
        $\square$\hspace{5pt}$\blacksquare$ & 
        
        $\square$\hspace{5pt}$\square$\hspace{5pt}$\square$\hspace{5pt}$\square$\hspace{5pt}$\blacksquare$\hspace{5pt}$\square$\hspace{5pt}$\square$\hspace{5pt}$\square$\hspace{5pt}$\square$ &
        
        $\square$\hspace{5pt}$\square$\hspace{5pt}$\square$\hspace{5pt}$\square$\hspace{5pt}$\square$\hspace{5pt}$\square$\hspace{5pt}$\square$\hspace{5pt}$\square$\hspace{5pt}$\blacksquare$

        \\  
        
        \begin{tabular}[c]{@{}l@{}} Sun \textit{et al.}~\cite{sun2020provable} \end{tabular}
        &
        
        $\square$\hspace{5pt}$\square$\hspace{5pt}$\square$\hspace{5pt}$\square$\hspace{5pt}$\blacksquare$\hspace{5pt}$\square$\hspace{5pt}$\square$ & 
        
        $\square$\hspace{5pt}$\blacksquare$\hspace{5pt}$\square$  & 
        
        $\blacksquare$\hspace{5pt}$\square$\hspace{5pt}$\square$& 
        
        $\blacksquare$\hspace{5pt}$\square$\hspace{5pt}$\blacksquare$\hspace{5pt}$\square$\hspace{5pt}$\square$ & 
        
        $\square$\hspace{5pt}$\square$\hspace{5pt}$\blacksquare$\hspace{5pt}$\blacksquare$\hspace{5pt}$\square$\hspace{5pt}$\square$\hspace{5pt}$\square$ & 
        
        $\square$\hspace{5pt}$\blacksquare$ & 
        
        $\blacksquare$\hspace{5pt}$\square$\hspace{5pt}$\blacksquare$\hspace{5pt}$\square$\hspace{5pt}$\square$\hspace{5pt}$\square$\hspace{5pt}$\square$\hspace{5pt}$\blacksquare$\hspace{5pt}$\square$ &
        
        $\square$\hspace{5pt}$\square$\hspace{5pt}$\square$\hspace{5pt}$\square$\hspace{5pt}$\square$\hspace{5pt}$\square$\hspace{5pt}$\square$\hspace{5pt}$\square$\hspace{5pt}$\blacksquare$

        \\        
        
        \begin{tabular}[c]{@{}l@{}} Wang \textit{et al.}~\cite{wang2020model} \end{tabular}
        &
        
        $\square$\hspace{5pt}$\blacksquare$\hspace{5pt}$\square$\hspace{5pt}$\square$\hspace{5pt}$\square$\hspace{5pt}$\square$\hspace{5pt}$\square$ & 
        
        $\blacksquare$\hspace{5pt}$\blacksquare$\hspace{5pt}$\square$  & 
        
        $\blacksquare$\hspace{5pt}$\square$\hspace{5pt}$\square$& 
        
        $\blacksquare$\hspace{5pt}$\square$\hspace{5pt}$\square$\hspace{5pt}$\square$\hspace{5pt}$\square$ & 
        
        $\square$\hspace{5pt}$\blacksquare$\hspace{5pt}$\square$\hspace{5pt}$\square$\hspace{5pt}$\square$\hspace{5pt}$\square$\hspace{5pt}$\square$ & 
        
        $\square$\hspace{5pt}$\blacksquare$ & 
        
        $\square$\hspace{5pt}$\square$\hspace{5pt}$\square$\hspace{5pt}$\square$\hspace{5pt}$\blacksquare$\hspace{5pt}$\square$\hspace{5pt}$\square$\hspace{5pt}$\square$\hspace{5pt}$\square$ &
        
        $\square$\hspace{5pt}$\square$\hspace{5pt}$\square$\hspace{5pt}$\square$\hspace{5pt}$\square$\hspace{5pt}$\square$\hspace{5pt}$\square$\hspace{5pt}$\square$\hspace{5pt}$\blacksquare$

        \\        
        
        \begin{tabular}[c]{@{}l@{}} Wan \textit{et al.}~\cite{Wan2021} \end{tabular}
        &
        
        $\square$\hspace{5pt}$\blacksquare$\hspace{5pt}$\square$\hspace{5pt}$\square$\hspace{5pt}$\square$\hspace{5pt}$\square$\hspace{5pt}$\square$ & 
        
        $\square$\hspace{5pt}$\square$\hspace{5pt}$\blacksquare$  & 
        
        $\blacksquare$\hspace{5pt}$\square$\hspace{5pt}$\square$& 
        
        $\blacksquare$\hspace{5pt}$\square$\hspace{5pt}$\square$\hspace{5pt}$\square$\hspace{5pt}$\square$ & 
        
        $\square$\hspace{5pt}$\square$\hspace{5pt}$\square$\hspace{5pt}$\square$\hspace{5pt}$\blacksquare$\hspace{5pt}$\square$\hspace{5pt}$\square$ & 
        
        $\square$\hspace{5pt}$\blacksquare$ & 
        
        $\blacksquare$\hspace{5pt}$\square$\hspace{5pt}$\square$\hspace{5pt}$\square$\hspace{5pt}$\square$\hspace{5pt}$\square$\hspace{5pt}$\square$\hspace{5pt}$\square$\hspace{5pt}$\square$ &
        
        $\square$\hspace{5pt}$\blacksquare$\hspace{5pt}$\square$\hspace{5pt}$\square$\hspace{5pt}$\square$\hspace{5pt}$\square$\hspace{5pt}$\square$\hspace{5pt}$\square$\hspace{5pt}$\square$

        \\        
        
        \begin{tabular}[c]{@{}l@{}} Zhang \textit{et al.}~\cite{zhang2020batchcrypt} \end{tabular}
        &
        
        $\square$\hspace{5pt}$\square$\hspace{5pt}$\blacksquare$\hspace{5pt}$\square$\hspace{5pt}$\square$\hspace{5pt}$\square$\hspace{5pt}$\square$ & 
        
        $\square$\hspace{5pt}$\blacksquare$\hspace{5pt}$\square$  & 
        
        $\blacksquare$\hspace{5pt}$\square$\hspace{5pt}$\blacksquare$& 
        
        $\blacksquare$\hspace{5pt}$\square$\hspace{5pt}$\square$\hspace{5pt}$\square$\hspace{5pt}$\blacksquare$ & 
        
        $\square$\hspace{5pt}$\square$\hspace{5pt}$\square$\hspace{5pt}$\square$\hspace{5pt}$\square$\hspace{5pt}$\square$\hspace{5pt}$\blacksquare$ & 
        
        $\blacksquare$\hspace{5pt}$\square$ & 
        
        $\square$\hspace{5pt}$\square$\hspace{5pt}$\square$\hspace{5pt}$\blacksquare$\hspace{5pt}$\blacksquare$\hspace{5pt}$\blacksquare$\hspace{5pt}$\square$\hspace{5pt}$\square$\hspace{5pt}$\square$ &
        
        $\square$\hspace{5pt}$\square$\hspace{5pt}$\square$\hspace{5pt}$\square$\hspace{5pt}$\square$\hspace{5pt}$\square$\hspace{5pt}$\blacksquare$\hspace{5pt}$\square$\hspace{5pt}$\square$

        \\
        
        \begin{tabular}[c]{@{}l@{}} Hardy \textit{et al.}~\cite{hardy2017private} \end{tabular}
        &
        
        $\square$\hspace{5pt}$\square$\hspace{5pt}$\blacksquare$\hspace{5pt}$\square$\hspace{5pt}$\square$\hspace{5pt}$\square$\hspace{5pt}$\square$ & 
        
        $\blacksquare$\hspace{5pt}$\square$\hspace{5pt}$\square$  & 
        
        $\blacksquare$\hspace{5pt}$\square$\hspace{5pt}$\square$& 
        
        $\blacksquare$\hspace{5pt}$\square$\hspace{5pt}$\square$\hspace{5pt}$\square$\hspace{5pt}$\blacksquare$ & 
        
        $\square$\hspace{5pt}$\square$\hspace{5pt}$\blacksquare$\hspace{5pt}$\square$\hspace{5pt}$\square$\hspace{5pt}$\square$\hspace{5pt}$\square$ & 
        
        $\square$\hspace{5pt}$\blacksquare$ & 
        
        $\square$\hspace{5pt}$\square$\hspace{5pt}$\square$\hspace{5pt}$\square$\hspace{5pt}$\square$\hspace{5pt}$\square$\hspace{5pt}$\square$\hspace{5pt}$\blacksquare$\hspace{5pt}$\square$ &
        
        $\square$\hspace{5pt}$\square$\hspace{5pt}$\square$\hspace{5pt}$\square$\hspace{5pt}$\square$\hspace{5pt}$\square$\hspace{5pt}$\square$\hspace{5pt}$\square$\hspace{5pt}$\blacksquare$

        \\

        \begin{tabular}[c]{@{}l@{}} Ge \textit{et al.}~\cite{ge2021chain} \end{tabular}
        &
        
        $\square$\hspace{5pt}$\square$\hspace{5pt}$\square$\hspace{5pt}$\square$\hspace{5pt}$\square$\hspace{5pt}$\blacksquare$\hspace{5pt}$\square$ & 
        
        $\square$\hspace{5pt}$\square$\hspace{5pt}$\blacksquare$  & 
        
        $\blacksquare$\hspace{5pt}$\square$\hspace{5pt}$\square$& 
        
        $\blacksquare$\hspace{5pt}$\square$\hspace{5pt}$\square$\hspace{5pt}$\square$\hspace{5pt}$\square$ & 
        
        $\square$\hspace{5pt}$\blacksquare$\hspace{5pt}$\square$\hspace{5pt}$\blacksquare$\hspace{5pt}$\square$\hspace{5pt}$\square$\hspace{5pt}$\square$ & 
        
        $\square$\hspace{5pt}$\blacksquare$ & 
        
        $\blacksquare$\hspace{5pt}$\square$\hspace{5pt}$\square$\hspace{5pt}$\square$\hspace{5pt}$\square$\hspace{5pt}$\square$\hspace{5pt}$\square$\hspace{5pt}$\square$\hspace{5pt}$\square$ &
        
        $\blacksquare$\hspace{5pt}$\square$\hspace{5pt}$\square$\hspace{5pt}$\square$\hspace{5pt}$\square$\hspace{5pt}$\square$\hspace{5pt}$\square$\hspace{5pt}$\square$\hspace{5pt}$\square$

        \\
  
        \begin{tabular}[c]{@{}l@{}} Liu \textit{et al.}~\cite{liu2021privacy} \dag \end{tabular}
        &
        
        $\square$\hspace{5pt}$\square$\hspace{5pt}$\square$\hspace{5pt}$\square$\hspace{5pt}$\square$\hspace{5pt}$\blacksquare$\hspace{5pt}$\square$ & 
        
        $\blacksquare$\hspace{5pt}$\square$\hspace{5pt}$\square$  & 
        
        $\blacksquare$\hspace{5pt}$\square$\hspace{5pt}$\square$& 
        
        $\blacksquare$\hspace{5pt}$\square$\hspace{5pt}$\square$\hspace{5pt}$\square$\hspace{5pt}$\blacksquare$ & 
        
        $\blacksquare$\hspace{5pt}$\square$\hspace{5pt}$\square$\hspace{5pt}$\square$\hspace{5pt}$\square$\hspace{5pt}$\square$\hspace{5pt}$\square$ & 
        
        $\square$\hspace{5pt}$\blacksquare$ & 
        
        $\blacksquare$\hspace{5pt}$\square$\hspace{5pt}$\square$\hspace{5pt}$\square$\hspace{5pt}$\blacksquare$\hspace{5pt}$\square$\hspace{5pt}$\square$\hspace{5pt}$\square$\hspace{5pt}$\square$ &
        
        $\square$\hspace{5pt}$\square$\hspace{5pt}$\square$\hspace{5pt}$\square$\hspace{5pt}$\square$\hspace{5pt}$\square$\hspace{5pt}$\square$\hspace{5pt}$\square$\hspace{5pt}$\blacksquare$

        \\      
        
        \begin{tabular}[c]{@{}l@{}} Awan \textit{et al.}~\cite{awan2021contra} \dag \end{tabular}
        &
        
        $\square$\hspace{5pt}$\square$\hspace{5pt}$\square$\hspace{5pt}$\square$\hspace{5pt}$\square$\hspace{5pt}$\blacksquare$\hspace{5pt}$\square$ & 
        
        $\blacksquare$\hspace{5pt}$\blacksquare$\hspace{5pt}$\square$  & 
        
        $\blacksquare$\hspace{5pt}$\blacksquare$\hspace{5pt}$\square$& 
        
        $\blacksquare$\hspace{5pt}$\blacksquare$\hspace{5pt}$\square$\hspace{5pt}$\square$\hspace{5pt}$\square$ & 
        
        $\blacksquare$\hspace{5pt}$\square$\hspace{5pt}$\square$\hspace{5pt}$\square$\hspace{5pt}$\square$\hspace{5pt}$\square$\hspace{5pt}$\square$ & 
        
        $\square$\hspace{5pt}$\blacksquare$ & 
        
        $\square$\hspace{5pt}$\blacksquare$\hspace{5pt}$\square$\hspace{5pt}$\square$\hspace{5pt}$\blacksquare$\hspace{5pt}$\square$\hspace{5pt}$\square$\hspace{5pt}$\square$\hspace{5pt}$\square$ &
        
        $\square$\hspace{5pt}$\square$\hspace{5pt}$\square$\hspace{5pt}$\square$\hspace{5pt}$\square$\hspace{5pt}$\square$\hspace{5pt}$\square$\hspace{5pt}$\square$\hspace{5pt}$\blacksquare$

        \\   

        \begin{tabular}[c]{@{}l@{}} Mao \textit{et al.}~\cite{mao2021romoa} \end{tabular}
        &
        
        $\square$\hspace{5pt}$\square$\hspace{5pt}$\square$\hspace{5pt}$\square$\hspace{5pt}$\square$\hspace{5pt}$\blacksquare$\hspace{5pt}$\square$ & 
        
        $\blacksquare$\hspace{5pt}$\blacksquare$\hspace{5pt}$\square$  & 
        
        $\blacksquare$\hspace{5pt}$\square$\hspace{5pt}$\square$& 
        
        $\blacksquare$\hspace{5pt}$\square$\hspace{5pt}$\blacksquare$\hspace{5pt}$\square$\hspace{5pt}$\blacksquare$ & 
        
        $\square$\hspace{5pt}$\blacksquare$\hspace{5pt}$\square$\hspace{5pt}$\square$\hspace{5pt}$\square$\hspace{5pt}$\square$\hspace{5pt}$\square$ & 
        
        $\square$\hspace{5pt}$\blacksquare$ & 
        
        $\square$\hspace{5pt}$\square$\hspace{5pt}$\square$\hspace{5pt}$\square$\hspace{5pt}$\blacksquare$\hspace{5pt}$\square$\hspace{5pt}$\square$\hspace{5pt}$\square$\hspace{5pt}$\square$ &
        
        $\square$\hspace{5pt}$\square$\hspace{5pt}$\square$\hspace{5pt}$\square$\hspace{5pt}$\square$\hspace{5pt}$\square$\hspace{5pt}$\square$\hspace{5pt}$\square$\hspace{5pt}$\blacksquare$

        \\  
        
        \begin{tabular}[c]{@{}l@{}} Manna ~\cite{manna2021moat} \dag \end{tabular}
        &
        
        $\square$\hspace{5pt}$\square$\hspace{5pt}$\square$\hspace{5pt}$\square$\hspace{5pt}$\square$\hspace{5pt}$\blacksquare$\hspace{5pt}$\square$ & 
        
        $\blacksquare$\hspace{5pt}$\blacksquare$\hspace{5pt}$\square$  & 
        
        $\blacksquare$\hspace{5pt}$\square$\hspace{5pt}$\square$& 
        
        $\square$\hspace{5pt}$\blacksquare$\hspace{5pt}$\square$\hspace{5pt}$\square$\hspace{5pt}$\blacksquare$ & 
        
        $\blacksquare$\hspace{5pt}$\blacksquare$\hspace{5pt}$\square$\hspace{5pt}$\square$\hspace{5pt}$\square$\hspace{5pt}$\square$\hspace{5pt}$\square$ & 
        
        $\square$\hspace{5pt}$\blacksquare$ & 
        
        $\blacksquare$\hspace{5pt}$\square$\hspace{5pt}$\square$\hspace{5pt}$\square$\hspace{5pt}$\square$\hspace{5pt}$\square$\hspace{5pt}$\square$\hspace{5pt}$\square$\hspace{5pt}$\square$ &
        
        $\square$\hspace{5pt}$\square$\hspace{5pt}$\square$\hspace{5pt}$\square$\hspace{5pt}$\square$\hspace{5pt}$\square$\hspace{5pt}$\square$\hspace{5pt}$\square$\hspace{5pt}$\blacksquare$

        \\   
        
        \begin{tabular}[c]{@{}l@{}} Xi \textit{et al.}~\cite{xi2021batfl} \dag \end{tabular}
        &
        
        $\square$\hspace{5pt}$\square$\hspace{5pt}$\square$\hspace{5pt}$\square$\hspace{5pt}$\square$\hspace{5pt}$\blacksquare$\hspace{5pt}$\square$ & 
        
        $\square$\hspace{5pt}$\square$\hspace{5pt}$\blacksquare$  & 
        
        $\blacksquare$\hspace{5pt}$\blacksquare$\hspace{5pt}$\square$& 
        
        $\square$\hspace{5pt}$\blacksquare$\hspace{5pt}$\square$\hspace{5pt}$\square$\hspace{5pt}$\blacksquare$ & 
        
        $\square$\hspace{5pt}$\blacksquare$\hspace{5pt}$\square$\hspace{5pt}$\square$\hspace{5pt}$\square$\hspace{5pt}$\square$\hspace{5pt}$\square$ & 
        
        $\square$\hspace{5pt}$\blacksquare$ & 
        
        $\blacksquare$\hspace{5pt}$\square$\hspace{5pt}$\square$\hspace{5pt}$\square$\hspace{5pt}$\square$\hspace{5pt}$\square$\hspace{5pt}$\square$\hspace{5pt}$\square$\hspace{5pt}$\square$ &
        
        $\square$\hspace{5pt}$\square$\hspace{5pt}$\square$\hspace{5pt}$\square$\hspace{5pt}$\square$\hspace{5pt}$\square$\hspace{5pt}$\square$\hspace{5pt}$\square$\hspace{5pt}$\blacksquare$

        \\           
        
        \begin{tabular}[c]{@{}l@{}} Lee \textit{et al.}~\cite{lee2021defensive} \end{tabular}
        &
        
        $\square$\hspace{5pt}$\square$\hspace{5pt}$\square$\hspace{5pt}$\square$\hspace{5pt}$\blacksquare$\hspace{5pt}$\square$\hspace{5pt}$\square$ & 
        
        $\square$\hspace{5pt}$\square$\hspace{5pt}$\blacksquare$  & 
        
        $\blacksquare$\hspace{5pt}$\square$\hspace{5pt}$\square$& 
        
        $\blacksquare$\hspace{5pt}$\square$\hspace{5pt}$\square$\hspace{5pt}$\square$\hspace{5pt}$\square$ & 
        
        $\square$\hspace{5pt}$\square$\hspace{5pt}$\square$\hspace{5pt}$\blacksquare$\hspace{5pt}$\square$\hspace{5pt}$\square$\hspace{5pt}$\square$ & 
        
        $\square$\hspace{5pt}$\blacksquare$ & 
        
        $\square$\hspace{5pt}$\blacksquare$\hspace{5pt}$\square$\hspace{5pt}$\square$\hspace{5pt}$\square$\hspace{5pt}$\square$\hspace{5pt}$\square$\hspace{5pt}$\square$\hspace{5pt}$\square$ &
        
        $\blacksquare$\hspace{5pt}$\square$\hspace{5pt}$\square$\hspace{5pt}$\square$\hspace{5pt}$\square$\hspace{5pt}$\square$\hspace{5pt}$\square$\hspace{5pt}$\square$\hspace{5pt}$\square$

        \\
        
        \begin{tabular}[c]{@{}l@{}} Varma \textit{et al.}~\cite{varma2021legato} \end{tabular}
        &
        
        $\square$\hspace{5pt}$\square$\hspace{5pt}$\square$\hspace{5pt}$\square$\hspace{5pt}$\square$\hspace{5pt}$\blacksquare$\hspace{5pt}$\square$ & 
        
        $\blacksquare$\hspace{5pt}$\blacksquare$\hspace{5pt}$\square$  & 
        
        $\blacksquare$\hspace{5pt}$\square$\hspace{5pt}$\square$& 
        
        $\blacksquare$\hspace{5pt}$\square$\hspace{5pt}$\square$\hspace{5pt}$\square$\hspace{5pt}$\square$ & 
        
        $\square$\hspace{5pt}$\square$\hspace{5pt}$\square$\hspace{5pt}$\square$\hspace{5pt}$\blacksquare$\hspace{5pt}$\square$\hspace{5pt}$\square$ & 
        
        $\square$\hspace{5pt}$\blacksquare$ & 
        
        $\square$\hspace{5pt}$\square$\hspace{5pt}$\square$\hspace{5pt}$\square$\hspace{5pt}$\blacksquare$\hspace{5pt}$\square$\hspace{5pt}$\square$\hspace{5pt}$\square$\hspace{5pt}$\square$ &
        
        $\square$\hspace{5pt}$\square$\hspace{5pt}$\square$\hspace{5pt}$\square$\hspace{5pt}$\square$\hspace{5pt}$\square$\hspace{5pt}$\square$\hspace{5pt}$\blacksquare$\hspace{5pt}$\square$

        \\
        
        \begin{tabular}[c]{@{}l@{}} Boutet \textit{et al.}~\cite{boutet2021mixnn} \end{tabular}
        &
        
        $\square$\hspace{5pt}$\square$\hspace{5pt}$\blacksquare$\hspace{5pt}$\square$\hspace{5pt}$\square$\hspace{5pt}$\square$\hspace{5pt}$\square$ & 
        
        $\blacksquare$\hspace{5pt}$\square$\hspace{5pt}$\square$  & 
        
        $\blacksquare$\hspace{5pt}$\blacksquare$\hspace{5pt}$\square$& 
        
        $\blacksquare$\hspace{5pt}$\square$\hspace{5pt}$\square$\hspace{5pt}$\square$\hspace{5pt}$\blacksquare$ & 
        
        $\square$\hspace{5pt}$\square$\hspace{5pt}$\blacksquare$\hspace{5pt}$\square$\hspace{5pt}$\square$\hspace{5pt}$\square$\hspace{5pt}$\square$ & 
        
        $\square$\hspace{5pt}$\blacksquare$ & 
        
        $\blacksquare$\hspace{5pt}$\square$\hspace{5pt}$\square$\hspace{5pt}$\square$\hspace{5pt}$\square$\hspace{5pt}$\square$\hspace{5pt}$\square$\hspace{5pt}$\blacksquare$\hspace{5pt}$\square$ &
        
        $\square$\hspace{5pt}$\square$\hspace{5pt}$\square$\hspace{5pt}$\square$\hspace{5pt}$\square$\hspace{5pt}$\square$\hspace{5pt}$\square$\hspace{5pt}$\square$\hspace{5pt}$\blacksquare$  

        \\
        
        \begin{tabular}[c]{@{}l@{}} Sharma \textit{et. al}~\cite{sharma2021tesseract} \end{tabular}
        &
        
        $\square$\hspace{5pt}$\square$\hspace{5pt}$\square$\hspace{5pt}$\square$\hspace{5pt}$\square$\hspace{5pt}$\blacksquare$\hspace{5pt}$\square$ & 
        
        $\square$\hspace{5pt}$\blacksquare$\hspace{5pt}$\square$  & 
        
        $\blacksquare$\hspace{5pt}$\square$\hspace{5pt}$\blacksquare$& 
        
        $\blacksquare$\hspace{5pt}$\square$\hspace{5pt}$\square$\hspace{5pt}$\square$\hspace{5pt}$\blacksquare$ & 
        
        $\square$\hspace{5pt}$\blacksquare$\hspace{5pt}$\square$\hspace{5pt}$\square$\hspace{5pt}$\square$\hspace{5pt}$\square$\hspace{5pt}$\square$ & 
        
        $\blacksquare$\hspace{5pt}$\square$ & 
        
        $\square$\hspace{5pt}$\blacksquare$\hspace{5pt}$\square$\hspace{5pt}$\square$\hspace{5pt}$\blacksquare$\hspace{5pt}$\square$\hspace{5pt}$\square$\hspace{5pt}$\blacksquare$\hspace{5pt}$\square$ &
        
        $\blacksquare$\hspace{5pt}$\square$\hspace{5pt}$\square$\hspace{5pt}$\square$\hspace{5pt}$\square$\hspace{5pt}$\square$\hspace{5pt}$\square$\hspace{5pt}$\square$\hspace{5pt}$\square$

        \\
        
        \begin{tabular}[c]{@{}l@{}} Sun \textit{et. al}~\cite{sun2021fl} \end{tabular}
        &
        
        $\square$\hspace{5pt}$\square$\hspace{5pt}$\square$\hspace{5pt}$\square$\hspace{5pt}$\square$\hspace{5pt}$\blacksquare$\hspace{5pt}$\square$ & 
        
        $\blacksquare$\hspace{5pt}$\blacksquare$\hspace{5pt}$\square$  & 
        
        $\blacksquare$\hspace{5pt}$\square$\hspace{5pt}$\square$& 
        
        $\blacksquare$\hspace{5pt}$\square$\hspace{5pt}$\square$\hspace{5pt}$\square$\hspace{5pt}$\blacksquare$ & 
        
        $\blacksquare$\hspace{5pt}$\square$\hspace{5pt}$\square$\hspace{5pt}$\square$\hspace{5pt}$\square$\hspace{5pt}$\square$\hspace{5pt}$\square$ & 
        
        $\blacksquare$\hspace{5pt}$\square$ & 
        
        $\blacksquare$\hspace{5pt}$\square$\hspace{5pt}$\blacksquare$\hspace{5pt}$\square$\hspace{5pt}$\blacksquare$\hspace{5pt}$\square$\hspace{5pt}$\square$\hspace{5pt}$\square$\hspace{5pt}$\square$ &
        
        $\blacksquare$\hspace{5pt}$\square$\hspace{5pt}$\square$\hspace{5pt}$\square$\hspace{5pt}$\square$\hspace{5pt}$\square$\hspace{5pt}$\square$\hspace{5pt}$\square$\hspace{5pt}$\square$

        \\
        
        \begin{tabular}[c]{@{}l@{}} Sun \textit{et. al}~\cite{sun2021soteria} \end{tabular}
        &
        
        $\square$\hspace{5pt}$\square$\hspace{5pt}$\square$\hspace{5pt}$\square$\hspace{5pt}$\blacksquare$\hspace{5pt}$\square$\hspace{5pt}$\square$ & 
        
        $\square$\hspace{5pt}$\blacksquare$\hspace{5pt}$\square$  & 
        
        $\blacksquare$\hspace{5pt}$\square$\hspace{5pt}$\square$& 
        
        $\blacksquare$\hspace{5pt}$\square$\hspace{5pt}$\blacksquare$\hspace{5pt}$\square$\hspace{5pt}$\square$ & 
        
        $\square$\hspace{5pt}$\square$\hspace{5pt}$\square$\hspace{5pt}$\blacksquare$\hspace{5pt}$\square$\hspace{5pt}$\square$\hspace{5pt}$\square$ & 
        
        $\blacksquare$\hspace{5pt}$\square$ & 
        
        $\blacksquare$\hspace{5pt}$\square$\hspace{5pt}$\blacksquare$\hspace{5pt}$\square$\hspace{5pt}$\square$\hspace{5pt}$\square$\hspace{5pt}$\square$\hspace{5pt}$\square$\hspace{5pt}$\square$ &
        
        $\blacksquare$\hspace{5pt}$\square$\hspace{5pt}$\square$\hspace{5pt}$\square$\hspace{5pt}$\square$\hspace{5pt}$\square$\hspace{5pt}$\square$\hspace{5pt}$\square$\hspace{5pt}$\square$

        \\             
        
        \begin{tabular}[c]{@{}l@{}} Li \textit{et al.}~\cite{li2021lomar} \end{tabular}
        &
        
        $\square$\hspace{5pt}$\square$\hspace{5pt}$\square$\hspace{5pt}$\square$\hspace{5pt}$\square$\hspace{5pt}$\blacksquare$\hspace{5pt}$\square$ & 
        
        $\square$\hspace{5pt}$\blacksquare$\hspace{5pt}$\square$  & 
        
        $\blacksquare$\hspace{5pt}$\blacksquare$\hspace{5pt}$\blacksquare$& 
        
        $\blacksquare$\hspace{5pt}$\square$\hspace{5pt}$\square$\hspace{5pt}$\square$\hspace{5pt}$\square$ & 
        
        $\square$\hspace{5pt}$\blacksquare$\hspace{5pt}$\square$\hspace{5pt}$\square$\hspace{5pt}$\square$\hspace{5pt}$\square$\hspace{5pt}$\square$ & 
        
        $\square$\hspace{5pt}$\blacksquare$ & 
        
        $\square$\hspace{5pt}$\square$\hspace{5pt}$\square$\hspace{5pt}$\square$\hspace{5pt}$\square$\hspace{5pt}$\blacksquare$\hspace{5pt}$\square$\hspace{5pt}$\square$\hspace{5pt}$\square$ &
        
        $\blacksquare$\hspace{5pt}$\square$\hspace{5pt}$\square$\hspace{5pt}$\square$\hspace{5pt}$\square$\hspace{5pt}$\square$\hspace{5pt}$\square$\hspace{5pt}$\square$\hspace{5pt}$\square$

        \\

        \hline

    \end{tabular} 
}
\end{sidewaystable}
        
        
        
        
        
        
        
        
        


%% file: 08-Challenges.tex
\section{Discussion}
\label{sec:challenges}

\subsection{Recognizing the FL Threats}

\begin{figure}[!htb]
\centering
\includegraphics[width = 0.7\columnwidth]{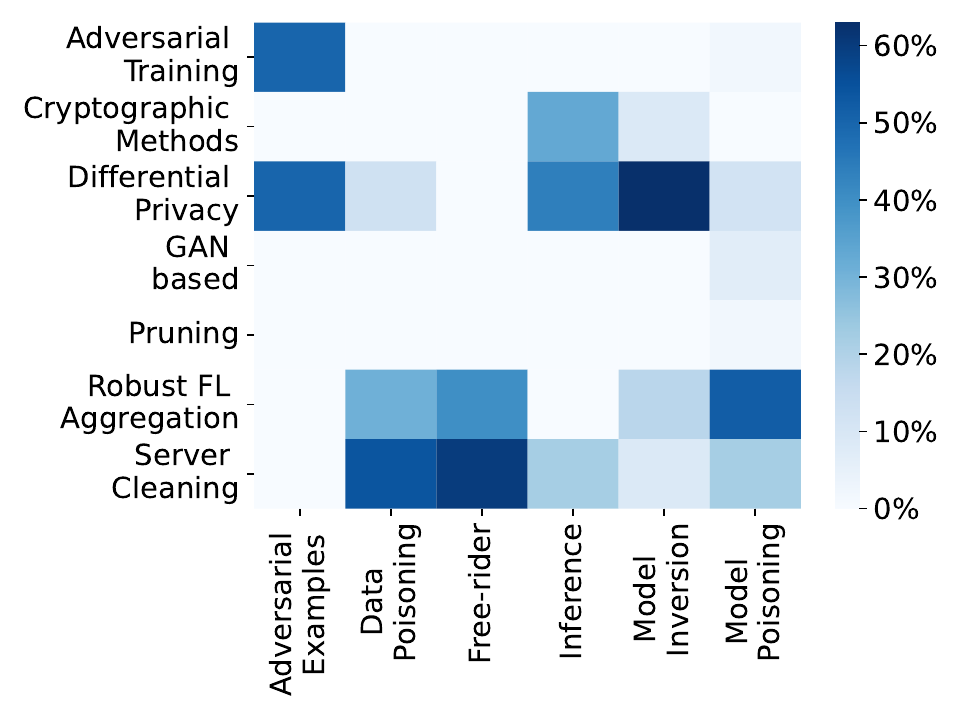}
\caption{Attacks and defenses relation.}
\label{fig:heatmapAttacksDefenses}
\end{figure}

After extensive analysis of the SoTA, we mapped each attack type regarding the defensive mechanisms used in the experiments and vice versa (see Fig.~\ref{fig:heatmapAttacksDefenses}). Each column represents the attacks, while defenses are shown horizontally. The color scheme plots the percentage of used defensive mechanisms for an attack. By analyzing the figure, we extract that for training-time attacks, the most common countermeasures are applied during the training phase, i.e., Server Cleaning, Robust FL Aggregation, Differential Privacy, and GAN-based. Contrary, for preventing Inference attacks by an outsider, CMs are the most common choice. To stop Adversarial Examples, Adversarial Training successfully prevents them by inserting adversarial data during training.

Regarding Pruning defenses, as seen in Table~\ref{tab:attacks2} and the heat map in Fig.~\ref{fig:heatmapAttacksDefenses}, none of the evaluated attacks tested its performance against Pruning defenses.

\begin{challenge}
\textbf{\textit{Since Pruning has not been yet bypassed in FL, we propose to develop attacks that embed the backdoor effect inside non-Poisoned neurons so that the adversarial effect remains present.}}
\end{challenge}

In other than FL fields, i.e., adversarial machine learning, the authors~\cite{che2020new} developed ensemble pre-trained adversarial models for transferring the adverse effect to a target model. This approach has not been applied to FL yet.

\begin{challenge}
\textbf{\textit{Since ensemble attacks are applied in other settings like centralized ML, we suggest adapting them for FL and exploring their effectiveness.}}
\end{challenge}


\subsection{Attacks \& Defenses Performance}

\begin{figure}[ht]
     \centering
     \begin{subfigure}[b]{0.6\textwidth}
         \centering
        \includegraphics[width = \textwidth]{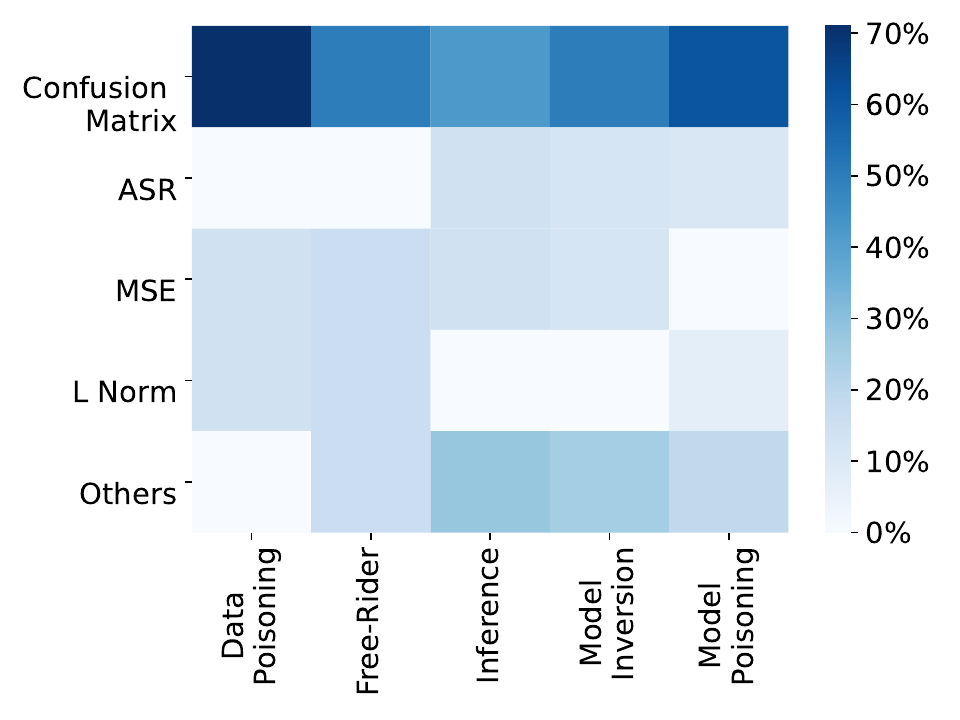}
        \caption{Attacks and evaluation metrics relation.}
     \end{subfigure}
     \hfill
     \begin{subfigure}[b]{0.6\textwidth}
         \centering
        \includegraphics[width =\textwidth]{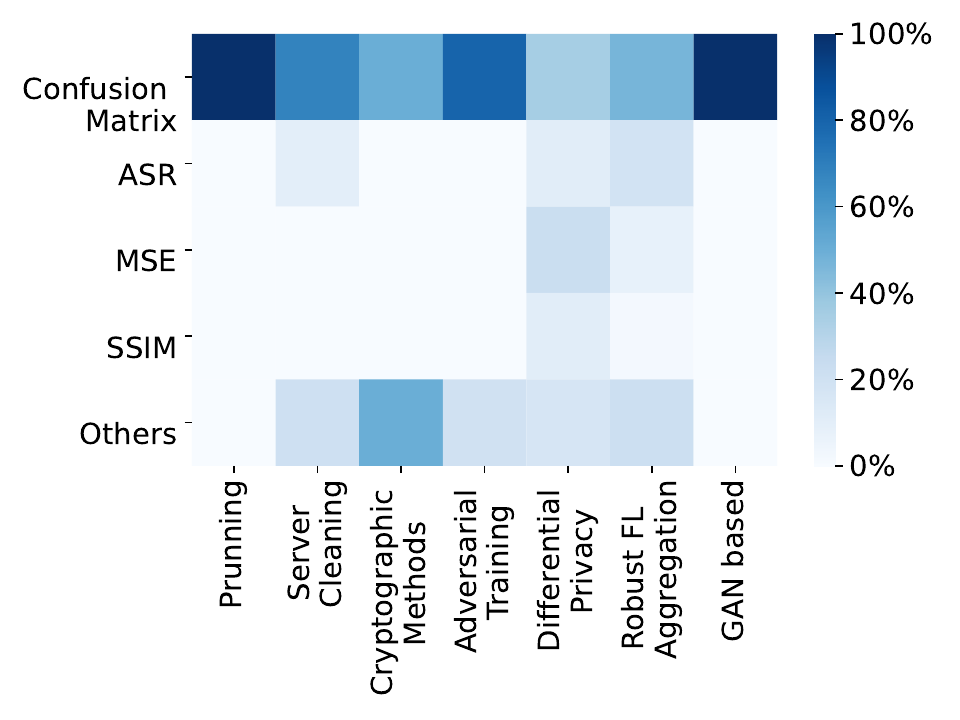}
        \caption{Defenses and evaluation metrics relation.}
     \end{subfigure}
    \caption{Defenses and Attacks relation with evaluation metrics.}
    \label{fig:heatmapMetricsDefenseAttack}
\end{figure}

The quality of an attack or a defense is influenced by different means: the dataset type, data distribution, model type, training procedure, number of clients, number of adversarial clients, and selected clients per round. We investigate this effect, represented in Fig.~\ref{fig:heatmapMetricsDefenseAttack} for attacks and defenses.

In Model Poisoning attacks, how many attacks have been successful, the Attack Success Rate (ASR) is relevant for adjusting the boosting factor. However, augmenting the boosting factor entails increasing the $l_n$-norm of the weights vector. Defense mechanisms also use this information to prevent Model Poisoning attacks. Furthermore, an attacker could leverage this information for increasing its stealthiness by minimizing the $l_n$-norm while maximizing the ASR. For Inference and Model Inversion attacks, since it is complicated to quantify the similarity between ground truth data and a reconstructed one, mostly in images, instead of using the ASR, it is measured by the Mean Squared Error (MSE). Likewise, an attacker may minimize the MSE to achieve better attack performance. For the rest of the cases, i.e., Free-Riders, Adversarial Samples, and Data Poisoning attacks, the attacker may leverage the information provided by the confusion matrix, e.g., accuracy, True Positive Rate, Area Under the ROC Curve, and Structural Similarity (SSIM).

Evaluating the defenses for preventing privacy degradation in Inference attacks, the countermeasures based on CMs measure its effectiveness with Fake Attack Rate (FAR). FAR measures the number of times the attacks failed. Similarly, cryptography-based defenses also use Known Data Ratio as the ratio between the amount of known data by the attacker used for Model Inversion and the amount of data of the victim. For Differential Privacy, Robust FL Aggregator, or Server Cleaning, the evaluation metrics vary, caused by the wide threat surface. In contrast, the confusion matrix is the most common metric for evaluating GAN-based defenses. 

\begin{challenge} \textbf{\textit{The evaluation metrics used to define a satisfactory attack or defense vary in the literature. For instance, as seen in Table~\ref{tab:attacks2}, the authors~\cite{Xie} proposed using the ASR, which describes the ability of the attack to stay during rounds, while other authors suggested sticking with the accuracy~\cite{DucNguyen}. Due to the lack of standard metrics that define the attack or defense success, we find it relevant to research and conveniently propose the best usable metrics.}}\end{challenge}

Furthermore, we encounter an inconsistency when evaluating the effect of the number of clients on the attacks and defenses. Few papers provide the number of clients, the number of adversarial clients, and the number of clients selected per round, complicating experiments' evaluation and repeatability.

\begin{challenge} \textbf{\textit{We propose to perform a further study to evaluate the effect of the number of clients on attacks and defenses performance.}}\end{challenge}

As previously demonstrated, any defensive mechanism will somewhat reduce the model's quality. Further research should be performed to investigate the trade-off between security, privacy, and accuracy.


\begin{challenge}
\textbf{\textit{There is no research on finding the appropriate level of security against an attack. Defining security and privacy boundaries for guaranteeing a safe model is therefore relevant.}}
\end{challenge}



\subsection{Fields of Application}

In our research, we find that image-based tasks are the most common. Thus, attacks and defenses are primarily developed in this field. However, FL architectures, such as HFL, are well suited for many low-power devices, such as IoT~\cite{savazzi2020federated}. FL lacks benchmark testing in this real-world case scenario. Therefore, extending the application field of attacks and defenses to a broader range, specifically IoT environments, is a natural extension.

\begin{sidewaystable}[!htbp]
	\caption{Overview of related works. C = Considered. NC = Not Considered.}
	   
	\label{tab:SuveysComparison}
	   
	\resizebox{\textwidth}{!}{

		\begin{tabular}{lllllllllll}
			
			\hline
			        
			& \begin{tabular}[c]{@{}l@{}} Data \\partitioning  
			\\ \fbox{\tiny{1}} \fbox{\tiny{2}} \fbox{\tiny{3}} \fbox{\tiny{4}}  \end{tabular}
			  
			& \begin{tabular}[c]{@{}l@{}} Architecture \\ \fbox{\tiny{1}} \fbox{\tiny{2}} \fbox{\tiny{3}}\end{tabular}
			        
			& \begin{tabular}[c]{@{}l@{}} Data \\distribution  \\  \fbox{\tiny{1}} \fbox{\tiny{2}} \end{tabular}
			    
			& \begin{tabular}[c]{@{}l@{}}  Attacks \\ \fbox{\tiny{1}} \fbox{\tiny{2}} \fbox{\tiny{3}} \fbox{\tiny{4}} \fbox{\tiny{5}} \fbox{\tiny{6}}\end{tabular}
			        
			& \begin{tabular}[c]{@{}l@{}} Defences  \\ \fbox{\tiny{1}} \fbox{\tiny{2}} \fbox{\tiny{3}} \fbox{\tiny{4}} \fbox{\tiny{5}} \fbox{\tiny{6}} \fbox{\tiny{7}} \fbox{\tiny{8}} \end{tabular}
			        
			& \begin{tabular}[c]{@{}l@{}} Evaluation \\metrics  
			\\ \fbox{\tiny{1}} \fbox{\tiny{2}} \end{tabular}
			        
			& \begin{tabular}[c]{@{}l@{}} Number of \\Clients \\ \fbox{\tiny{1}} \fbox{\tiny{2}} \end{tabular}  
			        
			& \begin{tabular}[c]{@{}l@{}} Frameworks \\ \fbox{\tiny{1}} \fbox{\tiny{2}} \end{tabular}
			
			& \begin{tabular}[c]{@{}l@{}} Datasets \\ \fbox{\tiny{1}} \fbox{\tiny{2}} \end{tabular}
			        
			& \begin{tabular}[c]{@{}l@{}} Taxonomy \\ \fbox{\tiny{1}} \fbox{\tiny{2}} \fbox{\tiny{3}}\end{tabular}  \\      
			
			\hline
			        
			& \begin{tabular}[c]{@{}l@{}} 1. VFL\\ 2. HFL\\ 3. Hybrid\\ 4. NC\end{tabular}
			        
			& \begin{tabular}[c]{@{}l@{}} 1. Centralized \\ 2. P2P \\3. NC\end{tabular}
			        
			& \begin{tabular}[c]{@{}l@{}} 1. C\\ 2. NC\end{tabular}
			        
			& \begin{tabular}[c]{@{}l@{}} 1. Poisoning\\ 2. Evasion\\
			3. Inference\\ 4. Free-rider\\ 5. Model Extraction \\ 6. NC\end{tabular}
			        
			& \begin{tabular}[c]{@{}l@{}}1. Pruning \\ 2. Server Cleaning \\ 3. Cryptographic Methods \\ 4. Adversarial Training \\ 5. Differential Privacy  \\ 6. Robust FL Aggregation \\ 7. GAN-based \\ 8. NC \end{tabular}
			        
			& \begin{tabular}[c]{@{}l@{}} 1. C \\ 2. NC \end{tabular}
			        
			& \begin{tabular}[c]{@{}l@{}} 1. C \\ 2. NC \end{tabular}
			        
			& \begin{tabular}[c]{@{}l@{}} 1. C \\ 2. NC \end{tabular}
			        
			& \begin{tabular}[c]{@{}l@{}} 1. C \\ 2. NC \end{tabular}
			        
			& \begin{tabular}[c]{@{}l@{}} 1. Proposed \\ 2. Partially \\ 3. Not Proposed \end{tabular}
			
			\\
			
			\hline

			\begin{tabular}[c]{@{}l@{}} Yin \textit{et al.}~\cite{yin2021comprehensive} \end{tabular}

			 & $\blacksquare$\hspace{5pt}$\blacksquare$\hspace{5pt}$\blacksquare$\hspace{5pt}$\square$ & 
			        
			$\blacksquare$\hspace{5pt}$\blacksquare$\hspace{5pt}$\square$  & 

			$\square$\hspace{5pt}$\blacksquare$ & 

			$\square$\hspace{5pt}$\square$\hspace{5pt}$\blacksquare$\hspace{5pt}$\square$\hspace{5pt}$\square$\hspace{5pt}$\square$ & 

			$\square$\hspace{5pt}$\square$\hspace{5pt}$\blacksquare$\hspace{5pt}$\square$\hspace{5pt}$\blacksquare$\hspace{5pt}$\square$\hspace{5pt}$\square$\hspace{5pt}$\square$ & 

			$\square$\hspace{5pt}$\blacksquare$ & 

			$\square$\hspace{5pt}$\blacksquare$ & 

			$\square$\hspace{5pt}$\blacksquare$ & 
			
			$\square$\hspace{5pt}$\blacksquare$ & 
			
			$\square$\hspace{5pt}$\blacksquare$\hspace{5pt}$\square$ 
			        
			\\

			\begin{tabular}[c]{@{}l@{}} Xia \textit{et al.}~\cite{xia2021survey} \end{tabular}

			  & $\blacksquare$\hspace{5pt}$\blacksquare$\hspace{5pt}$\blacksquare$\hspace{5pt}$\square$ & 

			$\square$\hspace{5pt}$\square$\hspace{5pt}$\blacksquare$  & 

			$\blacksquare$\hspace{5pt}$\square$ & 

			$\blacksquare$\hspace{5pt}$\square$\hspace{5pt}$\blacksquare$\hspace{5pt}$\square$\hspace{5pt}$\square$\hspace{5pt}$\square$ & 

			$\square$\hspace{5pt}$\blacksquare$\hspace{5pt}$\blacksquare$\hspace{5pt}$\square$\hspace{5pt}$\blacksquare$\hspace{5pt}$\square$\hspace{5pt}$\square$\hspace{5pt}$\square$ & 

			$\square$\hspace{5pt}$\blacksquare$ & 

			$\square$\hspace{5pt}$\blacksquare$ & 

			$\blacksquare$\hspace{5pt}$\square$ & 
			
			$\square$\hspace{5pt}$\blacksquare$ & 
			
			$\square$\hspace{5pt}$\square$\hspace{5pt}$\blacksquare$ 

			\\

			\begin{tabular}[c]{@{}l@{}} Zhang \textit{et al.}~\cite{zhang2018survey} \end{tabular}

			  & $\square$\hspace{5pt}$\square$\hspace{5pt}$\square$\hspace{5pt}$\blacksquare$           & 

			$\blacksquare$\hspace{5pt}$\square$\hspace{5pt}$\square$  & 

			$\square$\hspace{5pt}$\square$ & 

			$\blacksquare$\hspace{5pt}$\square$\hspace{5pt}$\blacksquare$\hspace{5pt}$\square$\hspace{5pt}$\square$\hspace{5pt}$\square$ & 

			$\square$\hspace{5pt}$\blacksquare$\hspace{5pt}$\blacksquare$\hspace{5pt}$\square$\hspace{5pt}$\blacksquare$\hspace{5pt}$\square$\hspace{5pt}$\square$\hspace{5pt}$\square$ & 

			$\square$\hspace{5pt}$\blacksquare$ & 

			$\square$\hspace{5pt}$\blacksquare$ & 

			$\square$\hspace{5pt}$\blacksquare$ & 
			
			$\square$\hspace{5pt}$\blacksquare$ & 
			
			$\square$\hspace{5pt}$\square$\hspace{5pt}$\blacksquare$ 
			        
			\\

			\begin{tabular}[c]{@{}l@{}} Li \textit{et al.}~\cite{li2019survey} \end{tabular}

			  & $\blacksquare$\hspace{5pt}$\blacksquare$\hspace{5pt}$\blacksquare$\hspace{5pt}$\square$ & 

			$\blacksquare$\hspace{5pt}$\blacksquare$\hspace{5pt}$\square$  & 

			$\blacksquare$\hspace{5pt}$\square$ & 

			$\square$\hspace{5pt}$\square$\hspace{5pt}$\square$\hspace{5pt}$\square$\hspace{5pt}$\square$\hspace{5pt}$\blacksquare$ & 

			$\square$\hspace{5pt}$\blacksquare$\hspace{5pt}$\blacksquare$\hspace{5pt}$\square$\hspace{5pt}$\blacksquare$\hspace{5pt}$\blacksquare$\hspace{5pt}$\square$\hspace{5pt}$\square$ & 

			$\square$\hspace{5pt}$\blacksquare$ & 

			$\square$\hspace{5pt}$\blacksquare$ & 

			$\blacksquare$\hspace{5pt}$\square$ & 
			
			$\square$\hspace{5pt}$\blacksquare$ & 
			
			$\square$\hspace{5pt}$\square$\hspace{5pt}$\blacksquare$ 
			        
			\\

			\begin{tabular}[c]{@{}l@{}} Mothukuri \textit{et al.}~\cite{mothukuri2021survey} \end{tabular}

			  & $\blacksquare$\hspace{5pt}$\blacksquare$\hspace{5pt}$\blacksquare$\hspace{5pt}$\square$ & 
			        
			$\blacksquare$\hspace{5pt}$\blacksquare$\hspace{5pt}$\square$  & 

			$\square$\hspace{5pt}$\blacksquare$ & 

			$\blacksquare$\hspace{5pt}$\blacksquare$\hspace{5pt}$\blacksquare$\hspace{5pt}$\blacksquare$\hspace{5pt}$\square$\hspace{5pt}$\square$ & 

			$\blacksquare$\hspace{5pt}$\blacksquare$\hspace{5pt}$\blacksquare$\hspace{5pt}$\blacksquare$\hspace{5pt}$\blacksquare$\hspace{5pt}$\blacksquare$\hspace{5pt}$\blacksquare$\hspace{5pt}$\square$ & 

			$\square$\hspace{5pt}$\blacksquare$ & 

			$\square$\hspace{5pt}$\blacksquare$ & 

			$\blacksquare$\hspace{5pt}$\square$ & 
			
			$\square$\hspace{5pt}$\blacksquare$ & 
			
			$\square$\hspace{5pt}$\square$\hspace{5pt}$\blacksquare$ 
			        
			\\

			\begin{tabular}[c]{@{}l@{}} Kairouz \textit{et al.}~\cite{kairouz2019advances} \end{tabular}

			  & $\square$\hspace{5pt}$\square$\hspace{5pt}$\square$\hspace{5pt}$\blacksquare$           & 

			$\blacksquare$\hspace{5pt}$\blacksquare$\hspace{5pt}$\square$  & 
			        
			$\blacksquare$\hspace{5pt}$\square$ & 

			$\blacksquare$\hspace{5pt}$\blacksquare$\hspace{5pt}$\blacksquare$\hspace{5pt}$\blacksquare$\hspace{5pt}$\square$\hspace{5pt}$\square$ & 

			$\blacksquare$\hspace{5pt}$\blacksquare$\hspace{5pt}$\blacksquare$\hspace{5pt}$\blacksquare$\hspace{5pt}$\blacksquare$\hspace{5pt}$\blacksquare$\hspace{5pt}$\square$\hspace{5pt}$\square$ & 

			$\square$\hspace{5pt}$\blacksquare$ & 

			$\square$\hspace{5pt}$\blacksquare$ & 

			$\blacksquare$\hspace{5pt}$\square$ & 
			
			$\square$\hspace{5pt}$\blacksquare$ & 
			
			$\square$\hspace{5pt}$\square$\hspace{5pt}$\blacksquare$ 

			\\

			\begin{tabular}[c]{@{}l@{}} Goldblum \textit{et al.}~\cite{goldblum2020dataset} \end{tabular}

			  & $\square$\hspace{5pt}$\square$\hspace{5pt}$\square$\hspace{5pt}$\blacksquare$           & 

			$\square$\hspace{5pt}$\square$\hspace{5pt}$\blacksquare$  & 

			$\blacksquare$\hspace{5pt}$\square$ & 

			$\blacksquare$\hspace{5pt}$\square$\hspace{5pt}$\square$\hspace{5pt}$\square$\hspace{5pt}$\square$\hspace{5pt}$\square$ & 

			$\square$\hspace{5pt}$\blacksquare$\hspace{5pt}$\square$\hspace{5pt}$\square$\hspace{5pt}$\blacksquare$\hspace{5pt}$\blacksquare$\hspace{5pt}$\square$\hspace{5pt}$\square$ & 
			        
			$\square$\hspace{5pt}$\blacksquare$ & 

			$\square$\hspace{5pt}$\blacksquare$ & 

			$\square$\hspace{5pt}$\blacksquare$ & 
			
			$\square$\hspace{5pt}$\blacksquare$ & 
			
			$\square$\hspace{5pt}$\square$\hspace{5pt}$\blacksquare$ 

			\\

			\begin{tabular}[c]{@{}l@{}} Lim \textit{et al.}~\cite{lim2020federated} \end{tabular}

			  & $\square$\hspace{5pt}$\square$\hspace{5pt}$\square$\hspace{5pt}$\blacksquare$           & 

			$\blacksquare$\hspace{5pt}$\square$\hspace{5pt}$\square$  & 

			$\blacksquare$\hspace{5pt}$\square$ & 

			$\square$\hspace{5pt}$\square$\hspace{5pt}$\blacksquare$\hspace{5pt}$\square$\hspace{5pt}$\square$\hspace{5pt}$\square$ & 

			$\square$\hspace{5pt}$\square$\hspace{5pt}$\square$\hspace{5pt}$\square$\hspace{5pt}$\blacksquare$\hspace{5pt}$\blacksquare$\hspace{5pt}$\square$\hspace{5pt}$\square$ & 

			$\square$\hspace{5pt}$\blacksquare$ & 
			        
			$\square$\hspace{5pt}$\blacksquare$ & 

			$\blacksquare$\hspace{5pt}$\square$ & 
			
			$\square$\hspace{5pt}$\blacksquare$ & 
			
			$\square$\hspace{5pt}$\blacksquare$\hspace{5pt}$\square$ 

			\\

			\begin{tabular}[c]{@{}l@{}} Li \textit{et al.}~\cite{li2020federated} \end{tabular}
			        
			  & $\square$\hspace{5pt}$\square$\hspace{5pt}$\square$\hspace{5pt}$\blacksquare$           & 

			$\blacksquare$\hspace{5pt}$\blacksquare$\hspace{5pt}$\square$  & 

			$\blacksquare$\hspace{5pt}$\square$ & 

			$\square$\hspace{5pt}$\square$\hspace{5pt}$\blacksquare$\hspace{5pt}$\square$\hspace{5pt}$\square$\hspace{5pt}$\square$ & 

			$\square$\hspace{5pt}$\square$\hspace{5pt}$\blacksquare$\hspace{5pt}$\blacksquare$\hspace{5pt}$\square$\hspace{5pt}$\square$\hspace{5pt}$\square$\hspace{5pt}$\square$ & 

			$\square$\hspace{5pt}$\blacksquare$ & 

			$\square$\hspace{5pt}$\blacksquare$ & 

			$\blacksquare$\hspace{5pt}$\square$ & 
			
			$\square$\hspace{5pt}$\blacksquare$ & 
			
			$\square$\hspace{5pt}$\square$\hspace{5pt}$\blacksquare$ 
			        
			\\

			\begin{tabular}[c]{@{}l@{}} Yang \textit{et al.}~\cite{yang2019federated} \end{tabular}

			  & $\blacksquare$\hspace{5pt}$\blacksquare$\hspace{5pt}$\blacksquare$\hspace{5pt}$\square$ & 

			$\blacksquare$\hspace{5pt}$\blacksquare$\hspace{5pt}$\square$  & 

			$\blacksquare$\hspace{5pt}$\square$ & 

			$\square$\hspace{5pt}$\square$\hspace{5pt}$\blacksquare$\hspace{5pt}$\square$\hspace{5pt}$\square$\hspace{5pt}$\square$ & 

			$\square$\hspace{5pt}$\blacksquare$\hspace{5pt}$\blacksquare$\hspace{5pt}$\square$\hspace{5pt}$\blacksquare$\hspace{5pt}$\blacksquare$\hspace{5pt}$\square$\hspace{5pt}$\square$ & 

			$\square$\hspace{5pt}$\blacksquare$ & 

			$\square$\hspace{5pt}$\blacksquare$ & 

			$\square$\hspace{5pt}$\blacksquare$ & 
			
			$\square$\hspace{5pt}$\blacksquare$ & 
			
			$\square$\hspace{5pt}$\square$\hspace{5pt}$\blacksquare$ 

			\\

			\begin{tabular}[c]{@{}l@{}} Li \textit{et al.}~\cite{li2020preserving} \end{tabular}

			  & $\square$\hspace{5pt}$\square$\hspace{5pt}$\square$\hspace{5pt}$\blacksquare$           & 

			$\blacksquare$\hspace{5pt}$\square$\hspace{5pt}$\square$  & 

			$\blacksquare$\hspace{5pt}$\square$ & 

			$\blacksquare$\hspace{5pt}$\square$\hspace{5pt}$\blacksquare$\hspace{5pt}$\square$\hspace{5pt}$\square$\hspace{5pt}$\square$ & 

			$\square$\hspace{5pt}$\square$\hspace{5pt}$\blacksquare$\hspace{5pt}$\square$\hspace{5pt}$\blacksquare$\hspace{5pt}$\square$\hspace{5pt}$\square$\hspace{5pt}$\square$ & 

			$\square$\hspace{5pt}$\blacksquare$ & 

			$\square$\hspace{5pt}$\blacksquare$ & 

			$\square$\hspace{5pt}$\blacksquare$ & 
			
			$\square$\hspace{5pt}$\blacksquare$ & 
			
			$\square$\hspace{5pt}$\square$\hspace{5pt}$\blacksquare$ 

			\\

			\begin{tabular}[c]{@{}l@{}} Kaissis \textit{et al.}~\cite{kaissis2020secure} \end{tabular}

			  & $\square$\hspace{5pt}$\square$\hspace{5pt}$\square$\hspace{5pt}$\blacksquare$           & 

			$\square$\hspace{5pt}$\square$\hspace{5pt}$\blacksquare$  & 
			        
			$\square$\hspace{5pt}$\blacksquare$ & 

			$\square$\hspace{5pt}$\square$\hspace{5pt}$\blacksquare$\hspace{5pt}$\square$\hspace{5pt}$\square$\hspace{5pt}$\square$ & 

			$\square$\hspace{5pt}$\blacksquare$\hspace{5pt}$\blacksquare$\hspace{5pt}$\square$\hspace{5pt}$\blacksquare$\hspace{5pt}$\square$\hspace{5pt}$\square$\hspace{5pt}$\square$ & 

			$\square$\hspace{5pt}$\blacksquare$ & 

			$\square$\hspace{5pt}$\blacksquare$ & 

			$\square$\hspace{5pt}$\blacksquare$ & 
			
			$\square$\hspace{5pt}$\blacksquare$ & 
			
			$\square$\hspace{5pt}$\square$\hspace{5pt}$\blacksquare$ 

			\\

			\begin{tabular}[c]{@{}l@{}} Lyu \textit{et al.}~\cite{lyu2020threats} \end{tabular}

			  & $\blacksquare$\hspace{5pt}$\blacksquare$\hspace{5pt}$\blacksquare$\hspace{5pt}$\square$ & 

			$\square$\hspace{5pt}$\square$\hspace{5pt}$\blacksquare$  & 

			$\square$\hspace{5pt}$\square$ & 

			$\blacksquare$\hspace{5pt}$\square$\hspace{5pt}$\blacksquare$\hspace{5pt}$\square$\hspace{5pt}$\square$\hspace{5pt}$\square$ & 

			$\square$\hspace{5pt}$\square$\hspace{5pt}$\square$\hspace{5pt}$\square$\hspace{5pt}$\square$\hspace{5pt}$\square$\hspace{5pt}$\square$\hspace{5pt}$\blacksquare$ & 

			$\square$\hspace{5pt}$\blacksquare$ & 

			$\square$\hspace{5pt}$\blacksquare$ & 

			$\square$\hspace{5pt}$\blacksquare$ & 
			
			$\square$\hspace{5pt}$\blacksquare$ & 
			
			$\square$\hspace{5pt}$\square$\hspace{5pt}$\blacksquare$ 

			\\

		\begin{tabular}[c]{@{}l@{}} Liu \textit{et al.}~\cite{liu2022threats} \end{tabular}

		 & $\square$\hspace{5pt}$\square$\hspace{5pt}$\square$\hspace{5pt}$\blacksquare$                & 

		$\square$\hspace{5pt}$\square$\hspace{5pt}$\blacksquare$  & 

		$\blacksquare$\hspace{5pt}$\square$ & 

		$\blacksquare$\hspace{5pt}$\blacksquare$\hspace{5pt}$\blacksquare$\hspace{5pt}$\square$\hspace{5pt}$\square$\hspace{5pt}$\square$ & 

		$\blacksquare$\hspace{5pt}$\square$\hspace{5pt}$\blacksquare$\hspace{5pt}$\square$\hspace{5pt}$\blacksquare$\hspace{5pt}$\square$\hspace{5pt}$\square$\hspace{5pt}$\square$ & 

		$\square$\hspace{5pt}$\blacksquare$ & 

		$\square$\hspace{5pt}$\blacksquare$ & 

		$\square$\hspace{5pt}$\blacksquare$ & 
		
		$\square$\hspace{5pt}$\blacksquare$ & 
	
	    $\square$\hspace{5pt}$\blacksquare$\hspace{5pt}$\square$ 
        
		\\
		
		\begin{tabular}[c]{@{}l@{}} Jere \textit{et al.}~\cite{jere2020taxonomy} \end{tabular}

		 & $\square$\hspace{5pt}$\square$\hspace{5pt}$\square$\hspace{5pt}$\blacksquare$                & 

		$\square$\hspace{5pt}$\square$\hspace{5pt}$\blacksquare$  & 

		$\square$\hspace{5pt}$\blacksquare$ & 

		$\blacksquare$\hspace{5pt}$\square$\hspace{5pt}$\blacksquare$\hspace{5pt}$\square$\hspace{5pt}$\square$\hspace{5pt}$\square$ & 

		$\square$\hspace{5pt}$\blacksquare$\hspace{5pt}$\square$\hspace{5pt}$\square$\hspace{5pt}$\blacksquare$\hspace{5pt}$\blacksquare$\hspace{5pt}$\square$\hspace{5pt}$\square$ & 

		$\square$\hspace{5pt}$\blacksquare$ & 

		$\blacksquare$\hspace{5pt}$\square$ & 

		$\square$\hspace{5pt}$\blacksquare$ & 
		
		$\blacksquare$\hspace{5pt}$\square$ & 
	
	    $\square$\hspace{5pt}$\blacksquare$\hspace{5pt}$\square$ 
        
		\\
		
		\begin{tabular}[c]{@{}l@{}} Ours \end{tabular}

    & $\blacksquare$\hspace{5pt}$\blacksquare$\hspace{5pt}$\blacksquare$\hspace{5pt}$\square$ & 

    $\blacksquare$\hspace{5pt}$\blacksquare$\hspace{5pt}$\square$  & 

    $\blacksquare$\hspace{5pt}$\square$ & 

    $\blacksquare$\hspace{5pt}$\blacksquare$\hspace{5pt}$\blacksquare$\hspace{5pt}$\blacksquare$\hspace{5pt}$\blacksquare$\hspace{5pt}$\square$ & 

    $\blacksquare$\hspace{5pt}$\blacksquare$\hspace{5pt}$\blacksquare$\hspace{5pt}$\blacksquare$\hspace{5pt}$\blacksquare$\hspace{5pt}$\blacksquare$\hspace{5pt}$\blacksquare$\hspace{5pt}$\square$ & 

    $\blacksquare$\hspace{5pt}$\square$ & 

    $\blacksquare$\hspace{5pt}$\square$ & 

    $\blacksquare$\hspace{5pt}$\square$ & 
    
    $\blacksquare$\hspace{5pt}$\square$ & 
    
    $\blacksquare$\hspace{5pt}$\square$\hspace{5pt}$\square$ 

			\\

			\hline
			
		\end{tabular}
	}
\end{sidewaystable}

\begin{challenge}
\textbf{\textit{Since attacks and defenses are primarily tested against image-based datasets. Further work is required to establish the viability of the application field to real case scenarios like IoT.}}
\end{challenge}

According to the gathered data, attacks and defenses proposals mainly focus on IID. environments, which is not a real case scenario in decentralized settings.
\begin{challenge}
\textbf{\textit{Many of the proposed attacks and defenses are proposed under IID environments. Further studies that consider both IID and non-IID distributions will need to be undertaken.}}
\end{challenge}

As mentioned above, image-based tasks enable specially designed attacks and defenses. GANs-based attacks and defenses are mainly applied in 2D environments and lack applicability in 1D environments, such as IoT. Therefore, adapting GANs or Variational Autoencoders for 1D would extend the attacks and defenses application field.
\begin{challenge} \textbf{\textit{We recommend establishing the viability of GAN-powered attacks and defenses in 1D environments like IoT.}}\end{challenge}

Lastly, our research discovered that attacks and defenses are only proposed for supervised learning.
\begin{challenge}\textbf{\textit{Since FL threats and countermeasures are developed for supervised learning, we suggest evaluating their usage in unsupervised, regression learning and FTL.}}\end{challenge}











%% file: 09-RelatedWork.tex
\section{Related Work}
\label{sec:related}

The security and privacy of ML have been extensively evaluated~\cite{huang2011adversarial, biggio2018wild, kurakin2016adversarial}. However, due to FL's novelty, such aspects are still under early-phase evaluation. The diversity of scenarios makes performing an FL's security and privacy evaluation as a whole difficult. For that reason, existing works solely focused on specific threats or use-cases for auditing FL. This effect is directly visible in the latest overviews for FL's security and privacy, presented in Table~\ref{tab:SuveysComparison}. 

The authors ~\cite{mothukuri2021survey, kairouz2019advances} evaluated FL's security and privacy issues by reviewing the SoTA attacks and defenses. While both works considered different types of FL architectures,~\cite{mothukuri2021survey} solely reviewed data partitioning types. In contrast,~\cite{kairouz2019advances} slightly studied data distribution and its implications. Other works~\cite{lyu2020threats, xia2021survey, zhang2018survey, li2020preserving, liu2022threats} mainly focused on Poisoning and Inference attacks.
Existing works~\cite{yang2019federated, li2020federated, yin2021comprehensive, lim2020federated, kaissis2020secure} evaluated privacy implications and issues on Inference attacks regarding privacy. Other authors~\cite{yin2021comprehensive} further studied privacy implications in FL by proposing a taxonomy for privacy-preserving defenses.
Regarding security, the authors~\cite{goldblum2020dataset} analyzed the security implications of datasets. Similarly, recent work~\cite{lim2020federated} examined some SoTA attacks under a security lens and proposed a taxonomy for that.

In our work, we thoroughly evaluate the SoTA attacks and defenses under the scope of security and privacy. We further study different FL architectures, data distributions, partitioning, evaluation metrics, and frameworks. Leveraging the gathered data, we design the first holistic taxonomy for attacks and defenses while considering security and privacy. 



Other works~\cite{lyu2020threats, xia2021survey, zhang2018survey, li2020preserving, jere2020taxonomy} apart from Poisoning attacks also investigated Inference attacks and their countermeasures. Similarly, the authors in~\cite{yang2019federated, li2020federated, yin2021comprehensive, lim2020federated, kaissis2020secure} focused their research solely on Inference threats and their privacy considerations. They also evaluated defense mechanisms, but only~\cite{li2020federated, lim2020federated} assessed privacy-driven frameworks for FL. Contrary, the authors in~\cite{li2019survey} concentrated their research on evaluating different defenses. They focused on Server Cleaning, CM, Differential Privacy, and Robust FL Algorithms.

%% file: 10-Conclusions.tex
\section{Conclusions}
\label{sec:conclusions}

The active area of FL's security and privacy research is heterogeneous in proposals. This work proposes a threat model that contains the attack surface, adversarial actors, capabilities, and goals. Under its scope, we analyze the SoTA attacks and defenses and classify them accordingly, and develop two novel taxonomies for attack and defenses, respectively. 
We discuss properties of attacks and defenses relations, data distribution, evaluation metrics, and dataset effects. Overall, FL merged different research areas towards a common objective of securing it. Moreover, we recognize promising challenges that should be addressed in future work. The taxonomy provides context for existing and emerging proposals, slowly contributing further knowledge on the FL's security and privacy.